\documentclass[12pt,preprint]{aastex}
\usepackage{longtable}
\usepackage{booktabs}
\usepackage{emulateapj5}

\begin{document}

\shortauthors{Welikala et al.}

\title{Pixel-z: Studying substructure and stellar populations in
  galaxies out to $z\sim3$ using pixel colors I. Systematics
}

\author{N. Welikala\altaffilmark{1},  A.M. Hopkins\altaffilmark{2},
  B.E. Robertson\altaffilmark{3},  A.J. Connolly\altaffilmark{4},
  L. Tasca\altaffilmark{5},  A.M. Koekemoer\altaffilmark{6},
 O. Ilbert\altaffilmark{5}, S. Bardelli\altaffilmark{7},
 J.P. Kneib\altaffilmark{5}, A.R. Zentner\altaffilmark{8} 
       }

\altaffiltext{1}{Insitut d'Astrophysique Spatiale, B\^atiment 121, Universit\'e Paris-Sud XI \& CNRS, 91405 Orsay Cedex, France, niraj.welikala@ias.u-psud.fr}
\altaffiltext{2}{Australian Astronomical Observatory, PO Box 296,
  Epping NSW 1710, Australia }
\altaffiltext{3}{Department of Astronomy, University of Arizona, 933
  North Cherry Avenue, Tucson, AZ 85721, U.S.A }
\altaffiltext{4}{Department of Astronomy, University of Washington,
  Box 351580, Seattle, WA 98195, U.S.A}
\altaffiltext{5}{Laboratoire d'Astrophysique de Marseille, CNRS \&
  University Aix-Marseille, 38, rue Fr\'ed\'eric Joliot-Curie, 13388
  Marseille Cedex 13, France }
\altaffiltext{6}{Space Telescope Science Insititute, 3700 San Martin
  Drive, Baltimore, MD 21218, U.S.A}
\altaffiltext{7}{INAF Osservatorio Astronomico di Bologna, Bologna, Italy}
\altaffiltext{8}{Pittsburgh Particle Physics, Astrophyics and
  Cosmology Center (PITT PACC), Department of Physics and Astronomy,
  University of Pittsburgh, Pittsburgh, PA 15260, U.S.A}

\begin{abstract}
We perform a pixel-by-pixel analysis of 467 galaxies in the GOODS-VIMOS survey to study systematic effects
in extracting substructure and properties of stellar populations (age, dust,
metallicity and star formation history) from the pixel colors using
the pixel-$z$ method. Several systematics are examined in this paper, including the effect of
the input stellar population synthesis models whose SEDs are fitted to
each pixel's colors, the effect of passband limitations and
differences between the individual SED fits to pixels and global SED
fitting to a galaxy's colors. We find that with optical-only colors
($bviz$), the systematic uncertainties due to differences among
stellar population synthesis codes are well constrained. The largest
impact on the stellar population age and SFR e-folding time estimates in the pixels arises from differences 
between the Maraston~(2005) models on one hand and the Bruzual \& Charlot~(2003) and Charlot \&Bruzual~(2007) models on the other, when optical-only ($bviz$)
colors are used. This results in systematic differences larger than
the $2\sigma$ uncertainties in over 10 percent of all pixels in
the galaxy sample. The effect of varying the number and choice of
available passbands is more
severe. In 26 percent of the pixels in the full sample, these
limitations result in systematic biases in the age 
determination which are larger than the $2\sigma$ uncertainties in
the measurements. Robust results can, however, still be obtained with a
minimum of 3 optical filters provided they span the 4000\,\AA \ break. Near-IR data is also added to a
subsample of 46 galaxies from the GOODS-NICMOS survey and
systematics arising from model differences are again
investigated. Differences among the models in their predicted rest-frame red/NIR colors manifest themselves as follows. For $z > 1$ galaxies the observed optical/NIR
colors span the rest frame UV-optical SED, and the use of different population
synthesis models does not significantly bias the estimates of the stellar
population parameters compared to using optical-only colors. However,
for $z < 1$, where the rest-frame NIR is still probed, there is a larger
discrepancy between models when using optical-only colors compared to optical/NIR
colors. This affects in particular the age determination in the
pixels. With this characterization of the systematic errors, we illustrate how pixel-$z$ can be applied robustly to make
detailed studies of substructure in high redshift objects such as (a) radial
gradients of properties such as age, SFR and dust and (b) the
distribution of these properties within subcomponents such as spiral arms and clumps. 
Finally, we show preliminary results from applying pixel-$z$ to
 galaxies in the CANDELS survey, illustrating how the new HST/WFC3 data can be exploited 
to probe substructure and stellar populations in $z\sim1-3$ galaxies.
\end{abstract}


\section{INTRODUCTION
\label{sec:intro}}

\subsection{Clumpy Disks: Resolved Stellar Populations in $z\sim 1-3$ Objects as a
  Probe of High-z Disk Galaxies}

Many galaxies at $z\sim1-3$ have morphologies that are very different
from the galaxy population at $z\sim0$ (Lotz~et~al.~2004; Lotz~et~al.~2006;
Law~et~al.~2007; Scarlata~et~al.~2007; Elmegreen~et~al.~2007; Pannella~et~al.~2009). However, the morphologies of these objects only tell one part of the story. At $z>1$, rest-frame UV wavelengths,
which are most impacted by dust obscuration and star formation,
can be traced by high resolution optical
imaging (Toft~et~al.~2007; Overzier~et~al.~2010; Cameron~et~al.~2010). The location and distribution of
stellar populations within these galaxies, particularly sites of active
star formation, is likely to be affected by various physical and dynamical
mechanisms which are present in earlier epochs in the galaxy's
evolution. The pixel-$z$ technique (Conti~et~al.~2003; Welikala~et~al.~2008; Welikala~et~al.~2009) can measure spatially resolved stellar
populations for large samples of $z\sim1-3$ galaxies
using their multi-band HST images. With this approach, one can study
asymmetries that exist in the star formation, age and dust distribution across these
galaxies. In parallel with the detailed kinematical studies of these
objects from spatially resolved spectroscopy, this will enable
insights into the formation mechanisms of these galaxies.

One such mechanism that could be responsible for the formation of these clumps on
kiloparsec scales is disk fragmentation in Toomre-unstable gas-rich disks (Elmegreen~et~al.~2007, Elmegreen~et~al.~2009). There is now direct kinematic evidence that high-z 
disks are much more turbulent than their counterparts at $z \sim 0$
(Bournaud~et~al.~2007; Bournaud~et~al.~2008; Genzel~et~al.~2008;
Genzel~et~al.~2011). For example, high-redshift disks have been found to
have local intrinsic gas velocity dispersions of $20-90
\,$km\,s$^{-1}$ as well as high gas-to-total baryonic mass
fractions (Daddi~et~al.~2010; Tacconi~et~al.~2010). 
Numerical simulations of gas-rich turbulent disks indicate that
massive kpc-sized clumps can form in-situ through gravitational
instabilities, a phenomenon known as a `clump-cluster'
phase. (Noguchi~et~al.~1999; Immeli~et~al.~2004a,b; Bournaud et
al.~2007; Elmegreen et al.~2008; Dekel, Sari, \& Ceverino~2009;
Agertz~et~al.~2009; Ceverino~et~al.~2010; Aumer et~al.~2010; Genzel~et~al.~2010). 
According to these simulations, the clumps can migrate towards the gravitational center 
as a result of both their mutual interactions and dynamical friction
against the host disk, so that they can eventually coalesce into a
young bulge on the order of a few dynamical times. 

Alternatively, these high-redshift clumpy disk and irregular galaxies
may form via mergers. In particular, numerical simulations
of gas-rich mergers result in remnant disks that show a low velocity
field asymmetry that also satisfies the criteria
necessary to be classified as a high-redshift disk galaxy
observationally. For example, Robertson \& Bullock (2008) compared one
of the merger remnants to the bulk properties of the well-studied $z=2.38$ galaxy
BzK-15504 and showed that it has a star formation rate (SFR), gas
surface density, and a circular velocity-to-velocity dispersion ratio
that is in excellent agreement with BzK 15504. Such numerical
simulations suggest that gas-rich mergers can play a prominent role in
the formation of disk galaxies at high redshift. 


 Recent and upcoming larger IFU programs such as the SINS survey
 (F\"{o}rster-Schreiber et al.~2006, Genzel et al.~2006), the SAURON
 project (Bacon et al.~2001) and the MASSIV survey (Contini et al.~2011; Epinat et al.~2011; Queyrel et al.~2011) are expected to shed further light on these formation
 mechanisms by performing detailed studies of the gas kinematics of $z \sim 2$
galaxies. Other large IFU programs include the CALIFA project (Sanchez
et al.~2011), which will target approximately 600
objects with PMAS IFUs on the Calar Alto Telescope, 
and the SAMI program (Croom et al.~2011) which is a planned low
redshift survey of serveral thousand galaxies. In addition to
kinematic studies, a complementary study of the
spatial distribution of the stellar populations within
these galaxies can provide additional insights into whether these
objects are formed through mergers or are, in fact, disks formed
in-situ. F\"{o}rster-Schreiber et al.~(2011a, b) used a combination of
near-infrared integral field spectroscopy from the SINFONI instrument
on the VLT, combined with deep high resolution HST NIC2/F160W imaging
of six $z \sim 2$ star-forming galaxies to characterize the properties
of kpc-scale clumps and their contribution  to the rest-frame optical emission. 
In our next papers, we aim to extend these previous studies and make
detailed spatially-resolved measurements of the properties of large
samples of disk galaxies. These studies will also make use of new
fully reduced NIR data from the HST Wide Field Camera 3 (WFC3) which will become available for the
GOODS-South field in early 2012.

\subsection{Spatially Resolved Star Formation and The Role of Environment in Galaxy Evolution}

The relation between galaxy environment and galaxy properties have
been extensively studied in the local Universe (Gomez et al.~2003,
Blanton et al.~2005, Hogg et al.~2004). 
A number of studies have observed that the high-$z$ ($z\gtrsim 1$) SFR-density relation is
either reversed or weaker at $z \sim 1$ than that seen locally
(Cucciati et al.~2006; Cooper et al.~2007; Elbaz et al.~2007; Ilbert et al.~2006; Poggianti et al.~2006; Ideue et
 al.~2009; Salimbeni et al.~2009; Scodeggio et al.~2009; Tran et
 al.~2010; Gr\"{u}tzbauch et al.~2011). These examples are consistent
 with the scenario that galaxies in dense environments form stars
 rapidly at early times, quickly building up mass and becoming
 quiescent, while galaxies in less dense environments form stars at a
 more sedate pace but over longer timescales, a phenomenon we refer to
 as {\em in-situ\/} evolution (Wijeshinge et al.~2011). This is distinct from the ‘‘infall and
quench’’ scenario where the properties of galaxies are impacted directly
by their local environment through physical mechanisms such as
ram-pressure stripping and galaxy harassment. In addition, Peng at
al.~(2010) makes a distinction between this type of quenching which is directly related to the environment of
galaxies and {\em mass quenching} which dominates at high
masses and early cosmic times.

A few recent studies have, however,  continued to observe the same SFR-density relation upto
$z\sim1$ as seen locally (Patel et al.~2011). 
A few studies have investigated the relation beyond $z\sim1$. Quadri
et al.~(2011)
showed that galaxies with quenched SF tend to reside in dense environments
out to at least $z \sim 1.8$ and that the the SFR-density relation
holds even at fixed stellar mass. Their density estimates, however, 
have relatively high uncertainties as they are derived from photometric redshifts.
While the SFR-density relation at high redshifts is a matter of
some debate, it is clear that an {\em in-situ\/} evolution makes a distinct
prediction compared to the ‘‘infall and quench’’ models. The latter
suggests that galaxies in dense environments should show a SFR
distribution that is progressively suppressed from the outside in, as
the outer regions are those which will be affected first by their
rapidly changing environment. The former, on the contrary, suggests
that the suppression should either happen uniformly as a galaxy
ages, or that the inner regions should be suppressed first. By studying the spatial distribution of
SFR in star-forming galaxies as a function of environment, we should
be able to distinguish clearly between these two scenarios.

Studies by Welikala et al.~(2008; 2009)  at $z
\sim 0.1$ confirm the global 
SFR-density relation observed locally but also show that the SFR
at $z \sim$ 0.1 is centrally concentrated relative to the outskirts, 
and in more dense environments it is this central SFR that is
supressed, favouring the  {\em in-situ\/} evolution scenario.
Park et al.~(2007)  also studied the color gradients of galaxies as a
function of the local galaxy density and galaxy morphology in the
SDSS. They found no environmental dependence of the color gradient for
early-type galaxies at a given luminosity, and only a weak
dependence on environment for faint late-type galaxies which are
observed to become bluer in their outskirts (relative to the galaxy center) in low
density environments. It thus becomes extremely interesting to 
determine whether the radial variation of star formation with
environment seen locally is also in place at high redshifts.
The development of MOS IFU instruments and 
surveys with large photometric and spectroscopic samples, such as
the Galaxy And Mass Assembly (GAMA) survey (Driver et al.~2011), will allow such a statistical analysis 
for a large sample of galaxies.
There are two complementary approaches to this. 
While spatially resolved spectroscopy provides detailed measures of
the current spatial distribution of physical properties within galaxies, pixel-$z$
adds the ability for the star formation history to be probed. While pixel-$z$ also
provides the current SFR and other properties of the stellar
populations in subcomponents of galaxies, these measurements will have
higher uncertainties than spectroscopic measures, and cannot provide
detailed kinematic information that spectroscopic observations
can. However, until the advent of highly-multiplexed MOS IFUs, even IFU
surveys will be limited to several thousands of targets, while
pixel-$z$ can be applied to tens of thousands and up
to millions of objects in order
to determine the radial variation of stellar population parameters
within them and the evolution of this radial variation with redshift.
We therefore aim to extend studies performed by Welikala et al.~ (2008; 2009) which
applied the pixel-$z$ method to $\sim300,000$ galaxies at $z\sim0.1$ in the SDSS and
determined the radial variation of star formation for this sample, 
the scatter in this relation, and its dependence on galaxy environment
and galaxy morphology. There are, however, inherent systematic effects that may limit the utlity of pixel-$z$ in
probing spatially resolved stellar populations in galaxies. The
current analysis explores these in order to quantify the impact of such
systematics on the measurements made by pixel-$z$.


\subsection{Stellar Populations Extracted from Pixel
  Colors: Assumptions and Limitations}

A number of studies have sought to extract stellar populations for
spatially resolved colors. These compare the colors in individual
pixels to a library of SEDs generated by stellar population
synthesis models. The majority of these utilised small samples
of galaxies at $z \sim 0$ and $z \sim 1$ (Abraham et al.~1999; Johnston
et al.~2005; Kassin et al.~2003; Lanyon-Foster et al.~2007; Zibetti et
al.~2011). 
Given a multi-band image of a galaxy, the pixel-$z$ method (Conti et
al.~2003; Welikala et al.~2008, 2009) computes the stellar population properties such 
as stellar population age, SFR e-folding timescale $\tau$, dust obscuration
through $E(B-V)$ and
metallicity $Z$ in each pixel in the galaxy, as well as the associated
uncertainties in these parameters. It does this by taking a library of
SEDs generated from stellar population synthesis (SPS) codes, redshifting
each of these SEDs and convolving them with the available
passbands. The simulated fluxes are then compared to the observed fluxes
in each pixel and a best-fitting SED is computed for that
pixel. The statistical errors in each parameter are calculated by
marginalizing the likelihood function from the fit over the remaining parameters, 
and are conservative. 

The pixel-$z$ approach makes two fundamental assumptions: 
\begin{itemize}
\item{\textbf{Independently evolving pixels: } The pixels in the galaxy image are assumed to be evolving
    independently of each other. This ignores any mixing of stellar
    populations between neighboring pixels which is expected to 
    occur in real galaxies. Neighbouring pixels will also not truly be
  independent, due to a finite Point Spread Function (PSF), and this
  effect is also neglected.}
\item{\textbf{Single stellar populations with exponentially
      declinining SFH: } The stellar population underlying each pixel
    is assumed to be one whose spectrum is derived from
    integrating over single stellar populations (SSPs) of a given age,
    weighted by a SFR which is characterized by an initial burst followed by an exponential
    decrease in SFR. The decrease is parameterized by a range of allowed values of e-folding
    times $\tau$. In reality, galaxies are expected to have more complicated SFHs, including multiple bursts.} 
\end{itemize}

In addition to these two fundamental assumptions, there is a
limitation in the form of degeneracies among
the SED models which could impact the accuracy with which pixel-$z$ 
determines stellar population parameters in the pixels. SPS models show
    that age, metallicity and dust all tend to affect spectra in
    similar ways (Bruzual \& Charlot~2003). Multiple SEDs each generated for
    different populations (different ages, metallicites and SFHs) can,
    as a result, still predict very similar colors of a galaxy or its
    subcomponents. Some of these
degeneracies arise from well known physical correlations, such as those between age and
metallicity of the stellar population. Worthey~(1994) constructed detailed models of old and
intermediate stellar populations and their absorption indices to show
that if $\delta_{age}$/$\delta{Z} \approx 3/2$ for two populations, they would appear identicial
in most absorption indices. The age-metallicity degeneracy can also
affect the photometric evolution of the SSPs since optical and
near-infrared (NIR) colors reflect the relative contribution of hot
and cool stars to the integrated light. Bruzual \& Charlot~(2003)
followed the evolution of optical and NIR colors and the stellar
mass-to-light ratio $M/L$ in their model, and found that at fixed age,
increasing metallicity tends to redden the colors and increase the
$M/L$ ratio. This is explained by the fact that increasing the
metallicity (at fixed stellar mass) causes (a) stars to evolve at lower
effective temperatures and lower luminosities (Girardi et al.~2000)
and (b) changes the relative number of red and blue supergiants, which 
strongly impacts the color evolution of the SSP. They also found that increasing metallicity at
fixed age had a similar effect to increasing age at fixed
metallicity. The age-metallicity degeneracy is therefore inherent in the SSP model
used. There could also be other degeneracies, including those between dust
obscuration and metallicity as well as between age and dust
obscuration. The redshift-evolution of the obscuration-metallicity
degeneracy in the pixels was characterized by Conti et al.~2003.

This work does not test the two fundamental assumptions above. As for
degeneracies, exploring the different correlations between the parameters on a pixel-to-pixel
basis and propagating the resulting errors are beyond the
scope of this paper. Welikala et al.~2008 characterized some of these
correlations using the likelihood function from the fit to the pixels
in SDSS galaxies. Since the severity of the degeneracies depend on the number and type of available
broadband colors, the SPS model and the chosen grid of stellar
population parameters, we focus on these systematics in this paper.
The validity of both the two assumptions and the impact of
degeneracies will be explored in a future paper that compares
 the spatial distribution of stellar populations inferred from spatially
resolved spectroscopy obtained from large IFU programs (Croom et
al.~2011)  with those inferred from applying the pixel-$z$ technique
to the same galaxies. Such a detailed comparison with IFU programs
will enable a robust calibration and testing of pixel-$z$, leading
eventually to a possible refinement of these assumptions. In this
paper, we focus on the following systematics that could potentially
bias our results:

\begin{itemize}

\item{\textbf{The SPS model: } An inevitable bias is the choice of the stellar population
    synthesis model.  Different  stellar population synthesis
   models predict different colors.
In particular, there are significant differences in the
prescriptions for Thermally Pulsating Asymptotic Giant (TP-AGB)
stars among the following models: Maraston 2005 (M05), Bruzual \& Charlot 2003
(BC03) and Charlot \& Bruzual 2007 (CB07). In this paper, we compare the outcomes of using 
these different models on a pixel-by-pixel basis, for all galaxies in
the sample. Similar comparisons of population synthesis models have been
performed previously on the integrated colors of galaxies but these have focused
primarily on the effect on their stellar mass (Maraston et
al.~2006). In this work, we focus on the effect of these differences
on the age, SFR e-folding time and obscuration estimated in the pixels.
}
\item{\textbf{The available passbands: }The accuracy with which the pixel-$z$ parameters are determined in
    the pixels depends on the the available passbands.  This paper
    explores the systematic effects arising from passband
limitations, using both optical and near-IR colors. 
}
\item{\textbf{The parameter space searched by pixel-$z$ and the number
      of SEDs per model: } Pixel-$z$ searches a grid of
ages, e-folding times, dust attenuation values and
    metallicities. In general, a large as possible range of values is
    chosen while physically unmeaningful values will be associated
with high statistical uncertainties. Despite this, a large parameter
space can result in degeneracies among the various SEDs searched by pixel-z which in turn may bias the
    pixel-$z$ estimates. Some priors are already built in to stop pixel-$z$ searching
    unphysical parameters e.g., the age of a stellar population in a
    pixel cannot exceed the age of the Universe at a given
    redshift. Here, we explore the systematic error arising from
    increasing the range of parameter values allowed on the pixel-$z$
    grid. In particular, we focus on the impact of increasing the
    metallicity range. This has two predicted effects: (a) it
    increases the severity of the age-metallicity degeneracy and would
    bias the pixel-$z$ estimates in quantities such as age,
    $\tau$ and dust and (b) it introduces an uncertainty
    associated with the models: while spectral synthesis models have
    been well established for solar metallicity stellar populations in
    optical photometry, the models are less well established for very
    sub-solar or super-solar populations. Increasing the metallicity
    range will test the impact of both these effects on the derived pixel-$z$ parameters.
    }
\item{\textbf{Global SED-fitting versus pixel-$z$: } We examine any
    biases resulting from measuring stellar population quantities,
    particularly the SFR, in individual pixels compared to the
    same quantities derived from aperture photometry and
    SED-fitting. These biases arise because individual pixels can have
    different colors to the integrated color of the galaxy. 
    This bias could give rise to a difference between (1) the mean
properties of the galaxy obtained by integrating the fluxes in an
aperture and SED-fitting to the total fluxes and (2) the mean properties derived from 
fitting to the individual pixels within the same aperture and summing
the resulting fits.}

\end{itemize}


\section{DATA AND METHODS}
\label{sec:analysis}

We use a magnitude-limited ($R<25$) sample of galaxies in the GOODS-South
survey (Giavalisco et al.~2004) with imaging in four ACS bandpasses,
$B435$, $V606$, $i775$ and $z850$. We refer to the full set of these four ACS filters
as $bviz$ throughout this paper. We make use of the version 2.0 of the publicly available ACS
source catalog \footnote{\url{http://archive.stsci.edu/pub/hlsp/goods/catalog\_r2}}
as well as a photometric redshift galaxy catalog for CDFS (Cardamone et al.~2010). Cutout images of each objects are obtained from the reduced,
calibrated, stacked and mosaiced ACS images\footnote{\url{http://archive.stsci.edu/pub/hlsp/goods/v2/}}.
The pixel scale in these final drizzled images is $0.03\arcsec$.

In the pixel-$z$ analysis, the redshift in each pixel is fixed to that
of the host galaxy and we fit only to the SED type. Using spectroscopic redshifts 
provides a more accurate determination of the stellar population
properties in each pixel, compared to using photometric redshifts.
We thus select only galaxies which have been spectroscopically measured in the GOODS-VIMOS
campaign (Popesso et al.~2009; Balestra et
al.~2010). GOODS-VIMOS consists of two surveys which target galaxies in
different redshfit ranges. The VIMOS Low Resolution Blue (LR-Blue) is aimed at observing
galaxies mainly at $1.8<z<3.5$ while the Medium Resolution (MR) orange
grism is aimed mostly at galaxies at $z<1$ and Lyman Break Galaxies
(LBGs) at $z>3.5$. Our sample has measured
spectroscopic redshifts from both grisms.
 

A final selection is done on galaxy size to ensure that there are a
sufficient number of pixels for the pixel-$z$ analysis. 
Galaxies are chosen to have a half light radius $R_{50} > 0.3\arcsec$ according to 
simulations in Welikala et al.~(2011) that estimate the
minimum galaxy sizes needed to recover color gradients within galaxies
with a fractional error smaller than 0.1. This implies that the
half light diameter should be at least 20 pixels across, corresponding
to 5.5 times the PSF FWHM. 
This leaves us with a sample of 467 galaxies with four-band imaging ($bviz$). 
Out of this sample, we select a subsample of 46 objects with NIR
colors which have been detected by the GOODS-NICMOS survey (Bouwens et
al.~2011; Magee et al~2007; Conselice et al.~2011). This provides deep
($> 26.5$ mag at $5 \sigma$) NICMOS data with the NIC3 camera over the GOODS North and South fields. These studies have found
several `first-light' ($z \sim 7$) galaxy candidates but the survey has
also detected several $z\sim1-3$ galaxies that we use in this work.

\begin{figure}
\includegraphics[width=0.8\textwidth]{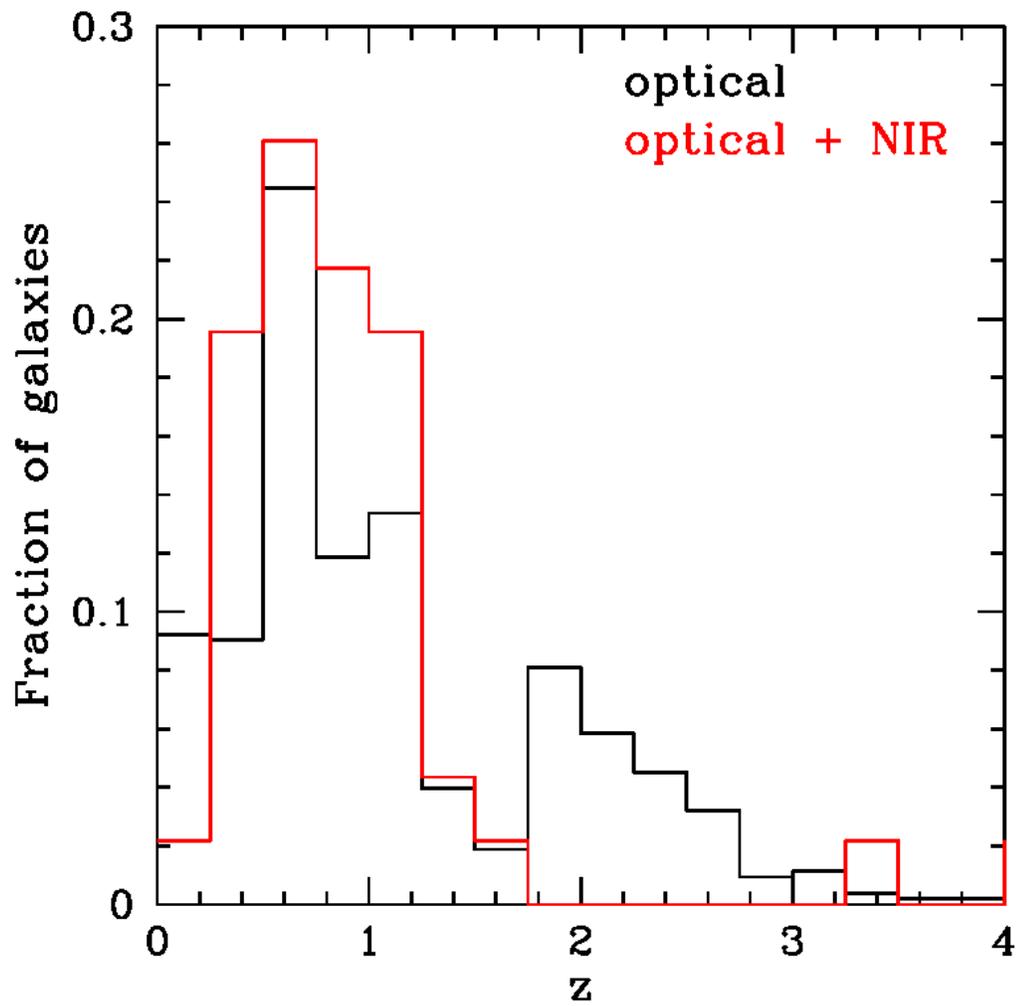}
\tiny{
\caption{The redshift distribution of (a) the optical sample of 467
  galaxies with $bviz$ imaging (black)
  and (b) the subsample of 46 galaxies with near-infrared imaging from NIC3
  (red).}
\label{fig:redshift_distribution}}
\end{figure}

The redshift distribution of both the optical and the NIR sample is
shown in Figure~\ref{fig:redshift_distribution}. The median redshift
of the optical sample is $z=0.95$, while that of the NIR sample is
$z=0.81$. For the optical-only sample, there are 145 galaxies in $1.0 < z
\le 2.0$ and 77 galaxies in $2.0 < z \le 3.0$. For the NIR sample, 
there are 22 galaxies in $0.5 < z \le 1.0$, 12 galaxies in $1.0 < z
\le 2.0$ and one object found at $z\sim3$.  Using the Sersic indices
($n_{Sersic}$) for these objects obtained from a publicly available morphology catalog for the GOODS-South
field\footnote{\url{http://www.ugastro.berkeley.edu/$\sim$rgriffit/Morphologies/catalogs}}, 
we find that the optical-only sample consists of approximately 4 times
more late-type galaxies compared to early-types: 299 late-type
galaxies ($n_{Sersic} < 1.5$) and 81 early-type galaxies ($n_{Sersic}
> 2.5$). This study is not concerned with tracing a particular
population of galaxies, which would require a volume-limited sample of
objects. Rather, we investigate systematics of a particular technique
which can be applied to galaxies in a range of redshifts,
luminosities, morphological types and inclinations as long as they are
well-enough resolved for the pixel-$z$ analysis. Nevertheless, we
repeat our analysis on subsets of the full galaxy sample in order to test the robustness of the results against selection effects.

Finally, we use preliminary NIR data from the first 6 epochs of the
CANDELS survey (Grogin et al.~2011; Koekemoer et al.~2011) using the WFC3 $F125W$ and $F160W$ filters in the GOODS-South
field\footnote{\url{http://archive.stsci.edu/pub/hlsp/candels/goods-s/gsd06/v0.5/}},
in order to illustrate the power of combining this high resolution optical and NIR data from ACS and WFC3
with the pixel-$z$ method in order to probe substructure in $z\sim1-3$
galaxies and to determine the properties of the stellar populations in the
pixels. A more thorough investigation of substructure in galaxies using the final CANDELS
data product will be performed in a future paper.

\subsection{Obtaining Fluxes and Variances in the Pixels}

ACS and NICMOS science images are in counts/sec, while the weight maps
are inverse variance maps (i.e., $1/\sigma_{f}^2$).
Flux calibration of the ACS and NICMOS science and weight postage stamp
images of each source is peformed using the ACS\footnote{\url{http://archive.stsci.edu/pub/hlsp/goods/v2/h\_goods\_v2.0\_rdm.html}}
and NICMOS zeropoints\footnote{\url{http://archive.stsci.edu/prepds/goodsnic/release/index.html}}. 
These contain the background variance from the sky noise, read-out noise and dark
currents. 

Further, because the pixels are correlated, the actual noise is higher than the
theoretical noise in the weight files (Casertano et al.~2000). We account for this by computing the noise in empty patches (whose size is larger than the
correlation length scale) within the science image and we calculate the ratio of this measured noise value to the
theoretical noise value. The noise in the weight images is then rescaled
by this ratio to give a more accurate estimate of the noise in the
pixels due the background and instrument. The Poisson uncertainty from
pixels in the source itself is then added in quadrature.

These final ACS flux and variance images are resampled to the coarser
0.1 arcsecond pixel scale of the NIC3 images using a Lanczos4 kernel.  We then perform
PSF-matching of the ACS images to the NIC3 images using 
the \textit{ip\_diffim}\footnote{\url{http://dev.lsstcorp.org/trac/}}
image mapping software currently in the pipeline of the Large Synoptic
Survey Telescope (LSST). The \textit{ip\_diffim} code is an implementation of the 
Higher Order Tranform of Psf and Template Subtraction code
(\textit{Hotpants}\footnote{\url{http://www.astro.washington.edu/users/becker/hotpants.html}}), 
and uses the algorithm of Alard~(1999) and Alard \& Lupton~(1998) to find the PSF-matching kernel that maps the
input (resampled) ACS images to the NIC3 ones. The kernel is
decomposed into a linear set of basis functions, which are
 Gaussians of varying FWHM. The process is linear
and ends up matching the PSFs of the two input images. 
A similar process is used to PSF-match the
variance images. Finally, for the analysis on the CANDELS data, the
WFC3 images from the first 6 HST epochs are
stacked by performing a weighted sum of the images in the different epochs
and dividing by the sum of the weight
maps\footnote{\url{http://archive.stsci.edu/pub/hlsp/candels/goods-s/gsd06/v0.5/}}
in order to increase the signal-to-noise in the final flux images. 
The existing ACS images are resampled to the pixel scale of the  WFC3
ones and PSF-matching is performed using the same method.

\subsection{Deriving SSPs from Colors with pixel-$z$}
\label{sec:sspsfromcolors}

The pixel-$z$ method (Conti et al.~2003; Welikala et al.~2008;
Welikala et al.~2009) fits SEDs from a
library of stellar population synthesis models to colors of individual
pixels. The redshift is fixed to the spectroscopic redshift of the
galaxy, and the fitting is done for the SED only. 
A best-fitting SED is identified for each pixel using a maximum
likelihood method and an age, $\tau$, dust obscuration and metallicity is
then inferred for each pixel. RMS errors are found for each of these
parameters by marginalizing the likelihood function in each pixel
over the remaining parameters. A signal-to-noise threshold $SNR > 5$  in the
$i$ band is set to ensure that the fitting is done to pixels in the galaxy
and not to the sky pixels. Low SNR pixels in the outskirts of a
galaxy and in any residual sky pixels that have been artificially fitted
generally have the largest uncertainties in their pixel-$z$
parameters. 

We examine three distinct stellar population synthesis models which are
currently widely used: Bruzual \& Charlot 2003 (BC03), Charlot \&
Bruzual (CB07) and Maraston 2005 (M05). For each model, we generate
spectra derived from single stellar populations with an exponentially
declining star formation history. In order to test differences
between the codes, the parameter space has to be made identical (or if
not, as similar as possible). The grid of stellar
population parameters ecompasses 2178 SEDs:

\begin{itemize}

\item{11 values of stellar population age: this is the time elapsed since the
    intial burst of stars. The stellar populations range
    from extremely young (0.001, 0.01, 0.1,
    0.5 Gyr), to middle age (1.0, 3.0, 5.0 Gyr) to very old (7.0, 9.0,
    11.0, 12.0 Gyr). In addition, the pixel-$z$ algorithm rejects ages
    which are greater than the age of the Universe at the redshift of
    each galaxy.}
\item{11 values of $\tau$: 0.1, 0.25, 0.50, 0.75,
    1.0, 1.5, 2.0, 3.0, 4.0, 5.0, 10.0 Gyr.}
\item{6 values of $E(B-V)$ from the Calzetti al.\ 1998 attenuation curves:  0.0, 0.1,
  0.2, 0.3, 0.5, 0.9}.
\item{3 values of metallicity Z: 0.01 (0.5 $Z_{\odot}$), 0.02
    ($Z_{\odot}$) and 0.04 (2 $Z_{\odot}$)}.

\end{itemize}

In the case of metallicity, stellar populations generated from the M05
models allow only 3 values. We have adopted 3 values of 
Z across all the models compared here (in BC03 and CB07, these are 0.008, 0.02
and 0.05). To test the effect of this restricted metallicity
range, we perform a control test over the entire population of
galaxies and all the associated pixels, using the BC03 models. The
allowed number of metallicities is increased to include significantly sub-solar
values ($Z=0.0001, 0.0004, 0.004$). The resulting systematic error, 
i.e., the difference in each pixel-z parameter (age, $\tau$, $E(B-V)$)  between using 6 values and 3 values of
$Z$, is computed pixel-by-pixel, and we examine the
distribution of this systematic acrosss the entire galaxy sample.
The effect of this relatively restricted range of metallicity is that the systematic difference in
the $Z$ values recovered from pixel-$z$ between the models is usually no
more than 1 $\sigma$, indicating that the majority of pixels in the
galaxy sample are described by metallicites which are close to solar.
 Permitting only metallicities that are close to solar
is therefore a reasonable approximation. The effect on the other
parameters (age, $\tau$ and dust obscuration) is examined below. This
control test is also a test of the impact of degeneracies in the models used. 
 Increasing the input metallicity range would be expected to increase the severity of these
degeneracies when fitting to the pixel colors, and therefore increase
the systematic error in the pixel-$z$ parameters. The results of this
test are presented in section ~\ref{subsec:paramspace}.

\subsection{Measuring the Systematic Effects}

\subsubsection{Measuring systematic effects in pixels for all galaxies
in the sample}
\label{subsubsec:methods_pixels}

In the tests below, we
compute the ratio of the systematic error in a particular pixel-$z$ quantity (age, $\tau$, $E(B-V)$) 
 to the statistical error in that quantity, for each pixel in the galaxy: 

\begin{itemize}

\item{When testing the effect of the stellar population synthesis
    models, we measure for each pixel and for each parameter $q$ the quantity $f$, where:
\begin{equation}\label{first}
      f(m1,m2) = \frac{q(m1) - q(m2)} {\sqrt{ {\sigma_q(m1)}^2 +
        {\sigma_q(m2)}^2 } }\\
\end{equation}
where $m1$ and $m2$ are the 2 models being compared and $\sigma_q$ is
the rms uncertainty in the parameter $q$ for that pixel.
 }
\item{When testing the effect of limiting the available passbands, we
    measure for each pixel and for each parameter $q$ the quantity
\begin{equation}\label{first}
      f(p1,p2) = \frac{q(p1) - q(p2)} {\sqrt{ {\sigma_q(p1)}^2 +
        {\sigma_q(p2)}^2 } }\\
\end{equation}

where $p1$ and $p2$ are the filter set used. 
}

\item{The systematic error resulting from expanding the metallicity range
from 3 to 6 values is computed similarly.}

\end{itemize}

For each galaxy in the sample, we measure $f$ within $2R_{50}$, considering only pixels with $S/N > 5.0$ to remove sky pixels. 
Since the effect of each systematic could vary from galaxy to galaxy, 
we are interested in the distribution of a particular systematic error
across the whole galaxy sample. For each test of a particular systematic, we plot the distribution of $f$ for
all the pixels inside $2R_{50}$ for all the galaxies in the
sample. With such statistics, we are able to determine conclusively 
the effect of the particular systematic being tested. We check how robust the results are
against selection effects by repeating the analysis on several independent subsets of
the galaxy sample. In particular, we find that, for the optical-only
sample, the results are robust against the effect of redshift. We thus
present our results for the full sample for the optical-only case. 
When considering only the 46 galaxies of the NIR sample and 
comparing the results of using only the optical data for this sample with those
obtained using the optical and NIR data combined, we find
a small difference in the results for galaxies with $z<1$ and $z>1$. 
For the NIC3/NIR sample, therefore, we present our results for $z<1$ and $z>1$ separately.

\subsubsection{Estimating the bias from using pixel-$z$ relative to
  global SED fitting} 

We estimate the bias in the inferred stellar population parameters
from using pixel-$z$ relative to the global quantities inferred from
aperture photometry. In particular, we investigate whether the sum of
the SED fits to individual pixels in a galaxy is consistent with the
fit to the total flux from the same pixels. These differences can
arise because individual pixels can have colors that differ from the mean
color of the galaxy. As a result, individual pixels can have stellar
population quantities that can differ significantly from what is
derived from the integrated colors of galaxies. The result of summing
the fits to the individual pixels in a galaxy may, in principle, be significantly
different from the global quantity measured from SED-fitting to the
total fluxes of the same pixels. Our aim is to investigate these
differences and quantify their magnitude.

 We focus on the SFR since it is a direct
observable in spectroscopic (including spatially resolved
spectroscopic) measurements. 
In both approaches, the current SFR is derived using the best-fit star
formation history, normalized using the same scale factor required to
normalize the SED to match the observed broad-band luminosity, in
either the pixel (for the first approach) or the entire galaxy (for
the second). We perform the comparison of
the two approaches for the same pixels by considering only pixels with
$S/N>5$ and within a fixed aperture of radius $0.5\arcsec$. 
We use the same library of SED templates throughout the
comparison: 2178 SEDs generated by the BC03 models.
The residual difference between the sum of the SED fits to the
invidual pixels and the SFR derived from SED-fitting to the total flux
in the same pixels is then computed in
terms of the statistical error in the SFR measurement.

\subsubsection{Substructure measurements: radial stellar population gradients, spiral arms and clumps}

As described in Section~\ref{sec:intro}, we aim to use pixel-$z$ to study (a) the spatial distribution of SFR
and dust and its dependence on the galaxy environment and (b) clumpy
star-forming disk galaxies at $z\sim2$. We use selected face-on disk galaxies
to demonstrate how pixel-$z$ can be used to measure large-scale features such as radial stellar population gradients and to
compute the mean properties of spiral arm structures and
clumps. We also wish to know how
the different systematic errors computed for the individual pixels, as
described in section~\ref{subsubsec:methods_pixels}, translate to systematics in
these large-scale features in the galaxy. 

\begin{itemize}
\item{\textbf{Radial gradients of stellar populations:}  We compute
    the mean luminosity-weighted age, $\tau$ and E(B-V) of successive radial annuli in the
    galaxy. The luminosity $L_{i}$ in each pixel is computed directly from the normalization
    of the best-fitting SED template to the observed fluxes. The mean pixel-$z$ quantity $\langle q_j\rangle$ computed in
    annulus $j$ is thus:

\begin{equation}
\langle q_j \rangle = \frac{\sum_{i=0}^N   {q_i \times L_i}} {\sum_{i=0}^N L_i}    
\end{equation}

where $N$ is the number of pixels in the annulus. 

From the age, $\tau$ and $L$ in each pixel, two futher quanties, the
$\langle SFR \rangle$ and the mean specific star formation rate
($\langle sSFR \rangle$) are also
derived for each annulus. The gradients of these quantities are then
computed for (a) varying stellar population synthesis models and (b) varying passbands. 
The size of the smallest annulus is chosen such that it is larger than the full width half
maximum (FWHM) of the PSF in the $i$ band. The radius in $kpc$ is computed
directly from the spectroscopic redshift of each galaxy and the ACS
angular scale of $0.03\arcsec/$pixel, and assuming $\Omega_{\Lambda}=0.728$, $\Omega_{\rm M}=0.27$,
and $H_{0}=70.2\,{\rm km\,s^{-1}\,Mpc^{-1}}$.
}
   
\item{\textbf{Mean properties of spiral arms and clumps:}   We use the GALFIT
    code (Peng et al.~2002; 2010) to isolate components such as
    spiral arms and clumps for selected face-on disk galaxies in our
    sample. A disk galaxy model is subtracted from the galaxy image in
    the $i$ band and the residual image is then used as a mask to
    effectively isolate both the spiral
    arms and clumps in the galaxies. These structures of interest
    correspond to the tail of the flux distribution in the
    residual image, and by adjusting the threshold, we are able to
    isolate the components individually. The pixel-$z$ code is then
    run on the total fluxes of the pixels identified as being part of
    these substructures. The distribution of the age, $\tau$, $E(B-V)$, SFR
    and sSFR of all the pixels belonging to each clump/spiral arm is
    then calculated for the different cases of stellar population
    synthesis models. }
\end{itemize}

We perform these analyses on selected disk galaxies, an example of
which is shown in this work, in order to identify preliminary trends, and to refine our
approach for a more thorough investigation. A statistical study of radial gradients, spiral arms and clumps,
which will also require a more careful selection of galaxies, will be
explored in a future paper.

\section{RESULTS AND DISCUSSION}
\label{sec:results}

\subsection{Effect of SSP Model Differences using Optical-only Bands}
\label{subsec:model_tests}

Here, we illustrate the effect of using different stellar
population synthesis models on the values of the parameters that pixel-$z$ determines using the \textit{bviz} passbands. This is first
shown qualititatively by displaying the resulting maps, and second,
quantitatively, by measuring the systematic error in each parameter
and following its distribution. 

\begin{figure}
\centerline{\rotatebox{0}{
\includegraphics[width=0.35\textwidth]{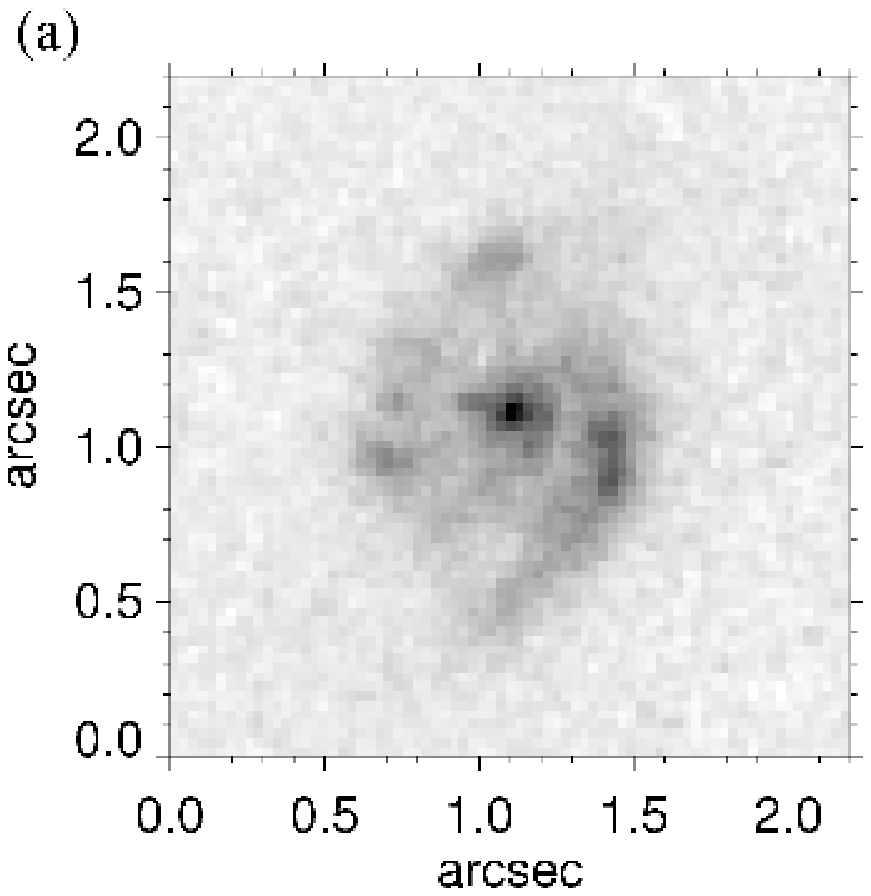}
}
}
\vspace{0.2cm}
\centerline{\rotatebox{0}{
\includegraphics[width=1\textwidth]{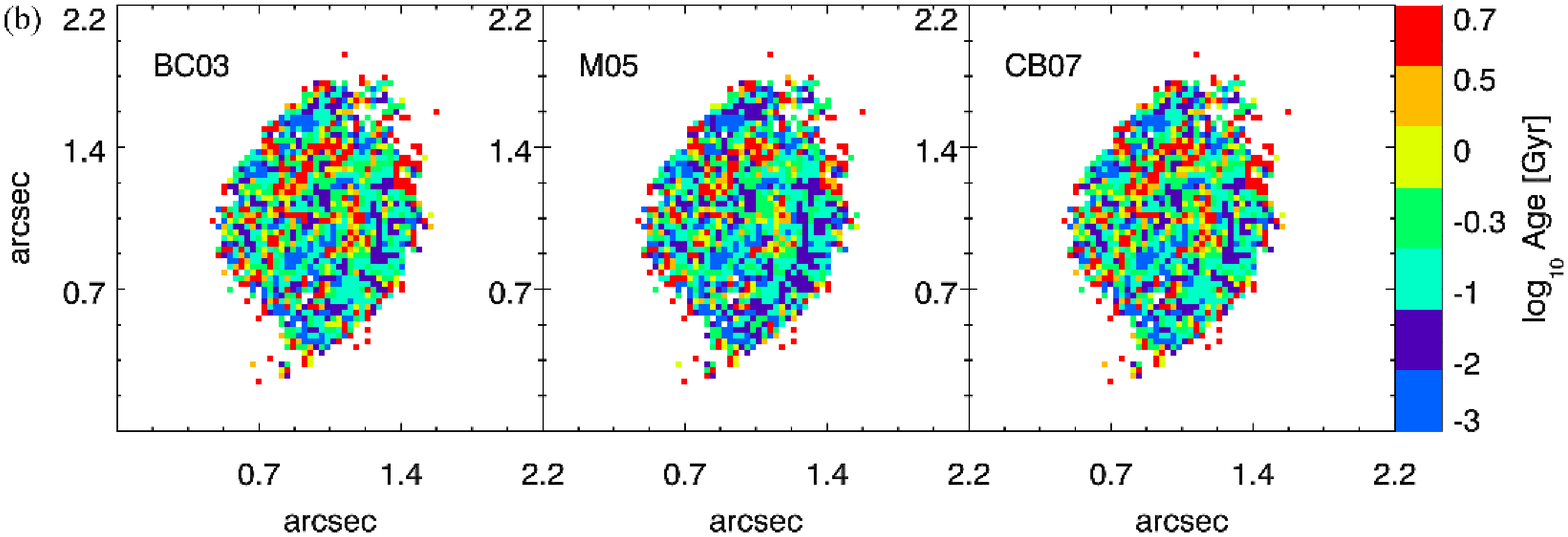}
}
}
\vspace{0.2cm}
\centerline{\rotatebox{0}{
\includegraphics[width=1\textwidth]{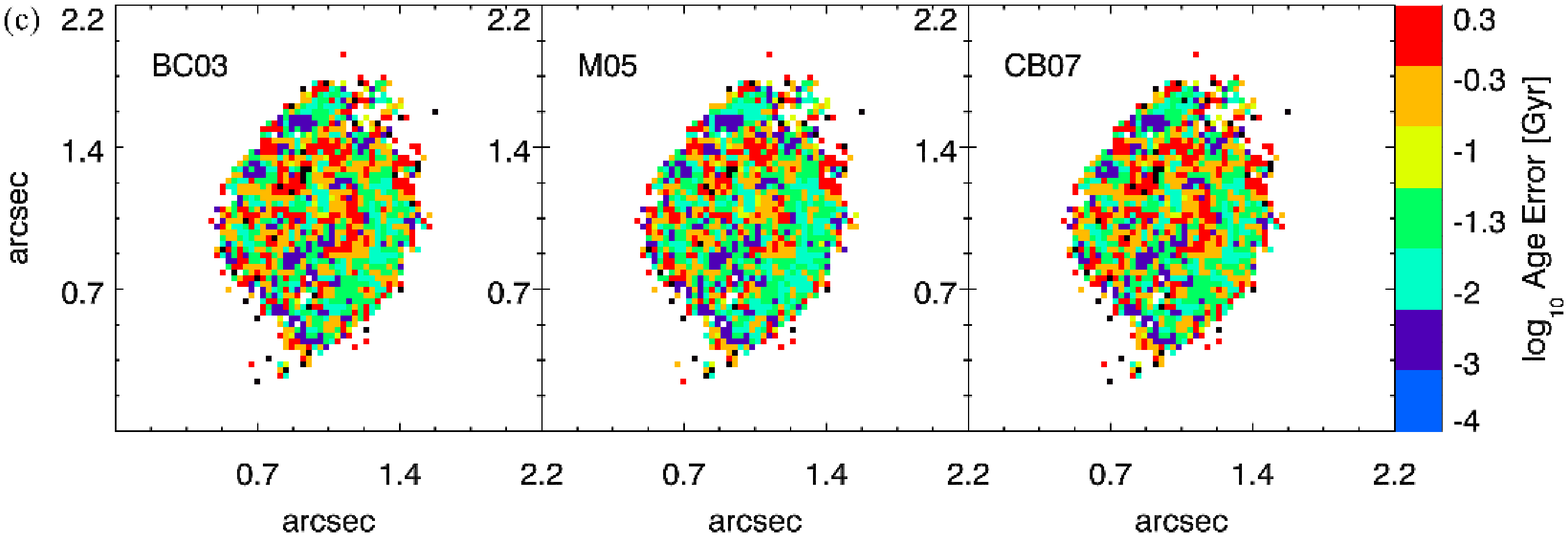}
}
}
\tiny{
\caption{
Illustration of the small effect of the chosen stellar
 population model in extracting stellar population properties from
  the pixel colors. Four optical bands ($bviz$) were used in this galaxy at
  $z=1.05$. (a) \textit{i} band image (b) stellar population age
 decomposition for BC03, CB07 and M05 population synthesis models
 (2178 spectra in each model) (c) Error in the stellar population age for the same models.
}
\label{fig:models_optical_age}}
\end{figure}

In the top panel of Figure~\ref{fig:models_optical_age}, we illustrate the
effect of the chosen stellar populations model on the age maps. 
There is a clear correlation between the pixel-$z$ maps
of this galaxy and the $i$ band image. The qualitative differences from pixel-to-pixel between
the various models(M05,BC03,CB07) are small but there is a difference
along the right spiral arm between BC03 and CB07
on the one hand and M05 on the other. BC03 and CB07 both predict that the
majority of pixels along the arm have ages $\sim0.1$Gyr, while M05
shows a larger proportion of pixels showing younger ages ($\sim0.01$Gyr).
Otherwise, both age values and their
associated uncertainties in the pixels are qualitatively
similar for all 3 models. The red pixels in the inter-arm
regions indicate much older populations but since these are also typically lower
signal-to-noise pixels, the rms error is also
significantly worse for these pixels. Pixels belonging to
the sky which are artificially fit are also assigned large uncertainties.

\begin{figure}
\centerline{\rotatebox{0}{
\includegraphics[width=0.35\textwidth]{fg2aarxiv.eps}
}
}
\vspace{0.3cm}
\centerline{\rotatebox{0}{
\includegraphics[width=1\textwidth]{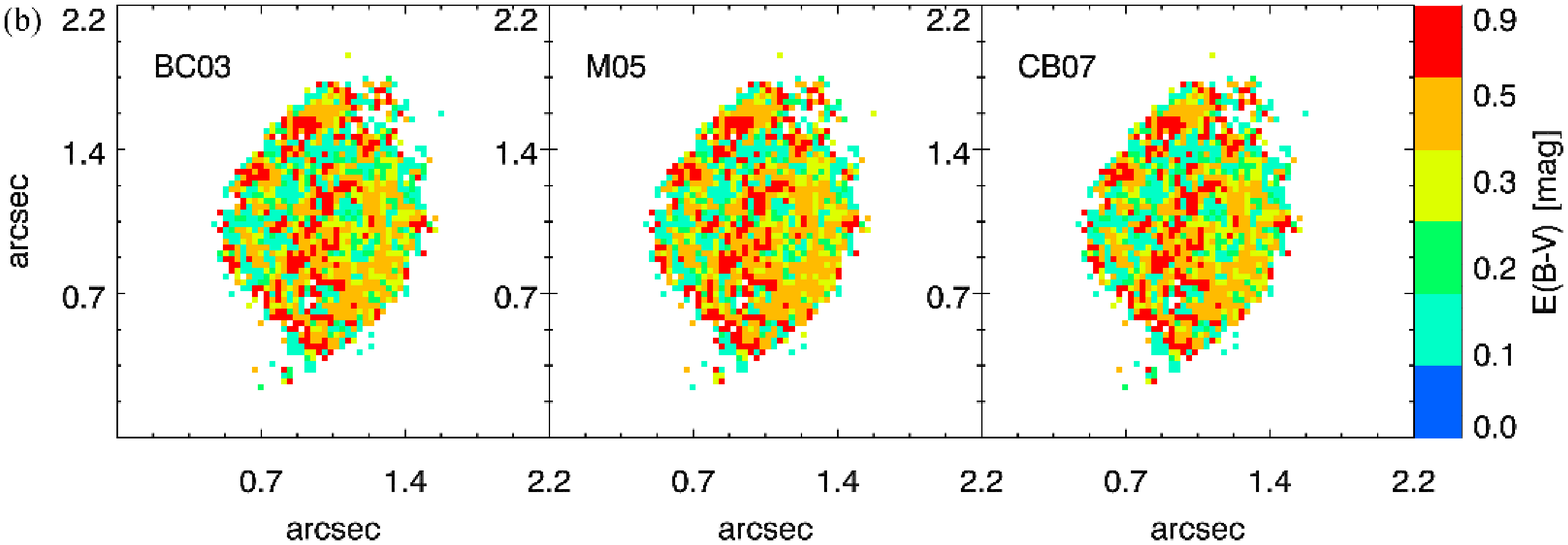}
}
}
\vspace{0.2cm}
\centerline{\rotatebox{0}{
\includegraphics[width=1\textwidth]{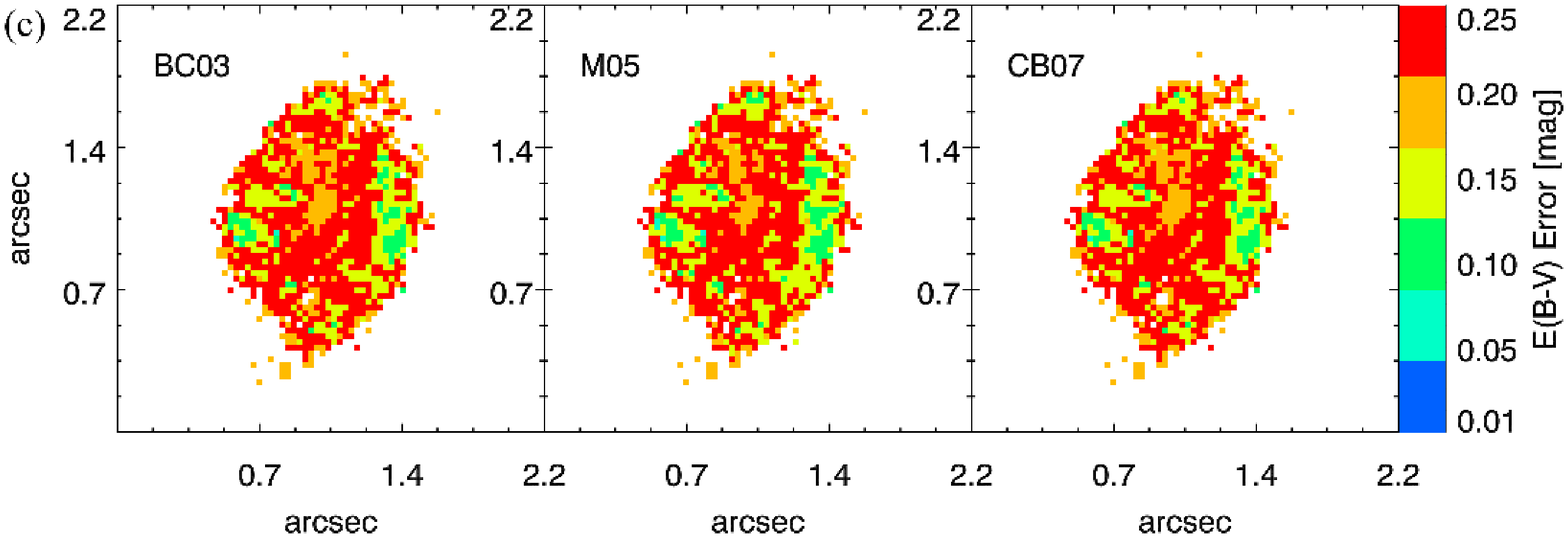}
}
}
\tiny{
\caption{
Illustration of the small effect of the chosen stellar
 population model in extracting stellar population properties from
  the pixel colors. Four optical bands ($bviz$) were used in this galaxy at
  $z=1.05$. (a) \textit{i} band image. (b) Dust obscuration
  decomposition (given in terms of $E(B-V)$) for BC03, CB07 and M05 population synthesis models
 (2178 spectra in each model) (c) Error in $E(B-V)$ for the same models.
} 
\label{fig:models_optical_ebv}}
\end{figure}

The dust obscuration, shown in Figure~\ref{fig:models_optical_ebv}
is also correlated with the spatial distribution of the stellar
age. Unlike the disagreement seen in the age values of the spiral arm between the
M05 models on one hand and the BC03 and CB07 models on the other hand,
the obscuration maps demonstrate a high level of agreement among the various models. In all the models, the arm regions, which were predicted to have the
youngest stellar populations, also have a moderately high level of dust
obscuration, with $E(B-V)\sim 0.5$. The uncertainties in the dust obscuration
in the pixels are similar among the various models as well. 
These uncertainties correlate
strongly with morphological features in the image as in the case of
the stellar population age. The knots on the spiral arms, corresponding to star-forming HII regions, are clearly displayed in the obscuration
uncertainty map with the lowest errors of the entire galaxy. These
statistical uncertainties are similar among all the models
considered. The inner arms and the
outermost edges of the galaxy, which are the regions with the lowest
signal-to-noise in the entire galaxy, have, predictably, the highest rms
uncertainties in their pixel-$z$ quantities. 

\begin{figure}
\centerline{\rotatebox{0}{
\includegraphics[width=0.55\textwidth]{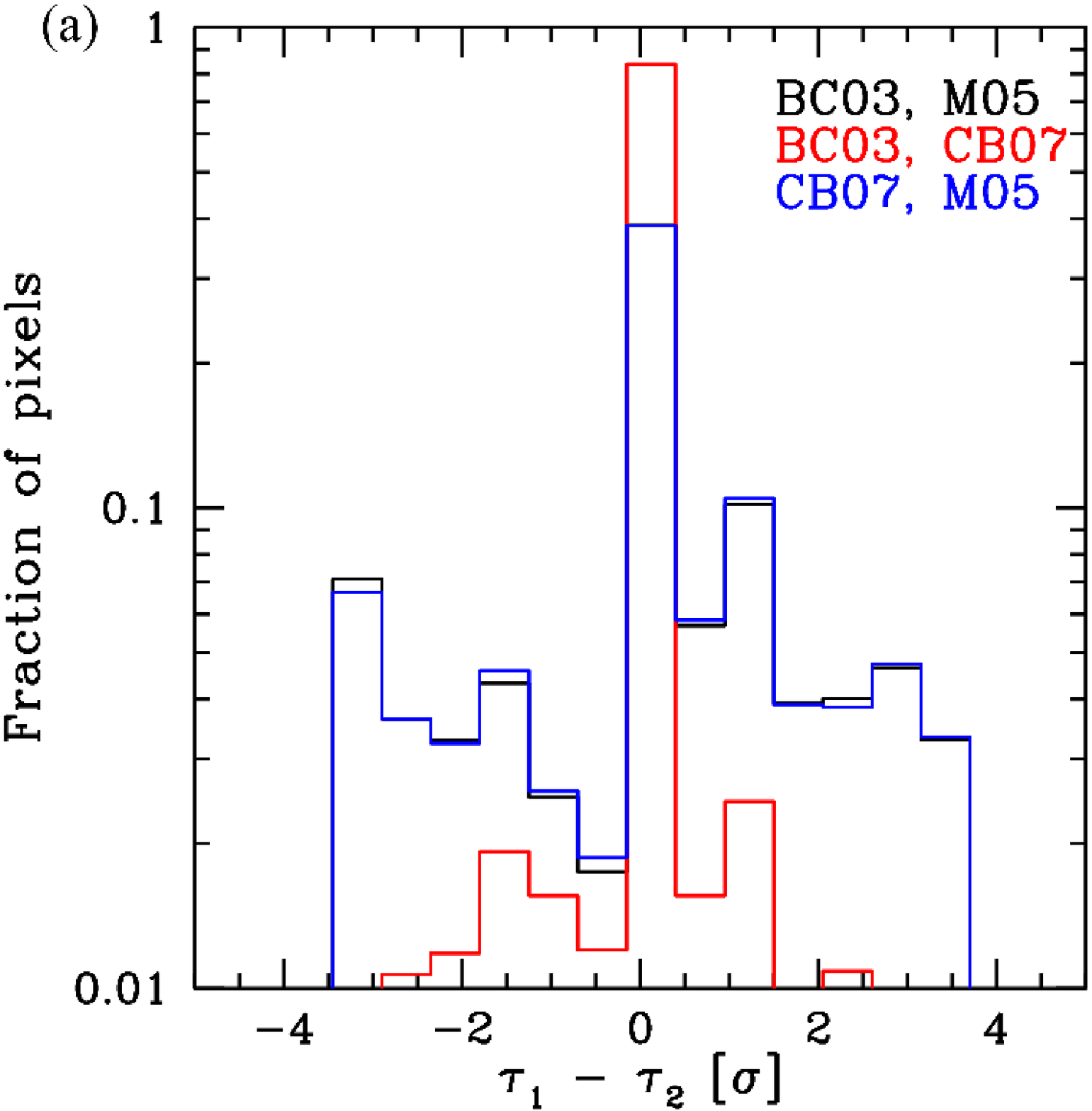}
\hspace{0.1cm}
\includegraphics[width=0.55\textwidth]{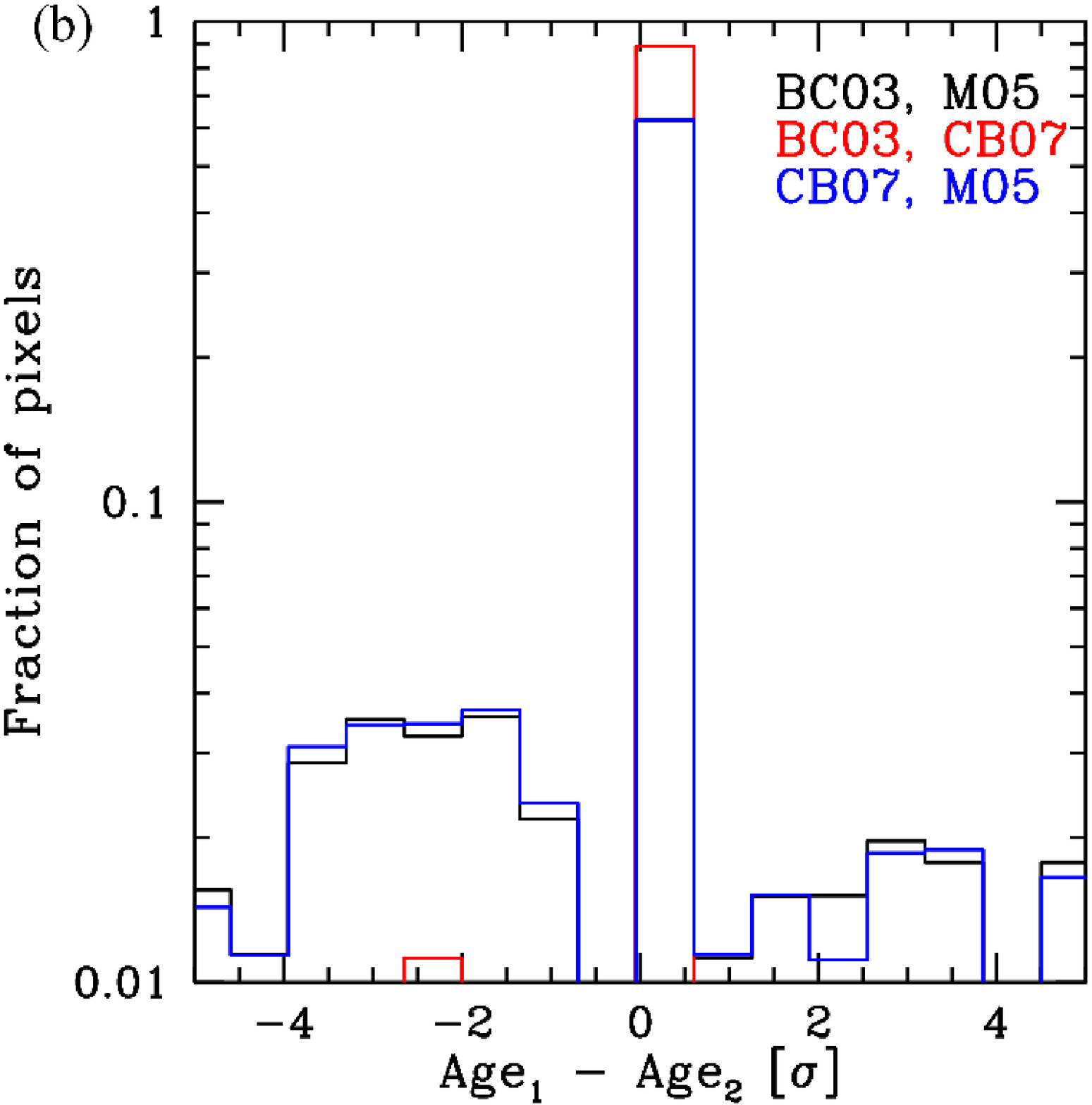}
}}
\centerline{\rotatebox{0}{
\vspace{0.2cm}
\includegraphics[width=0.55\textwidth]{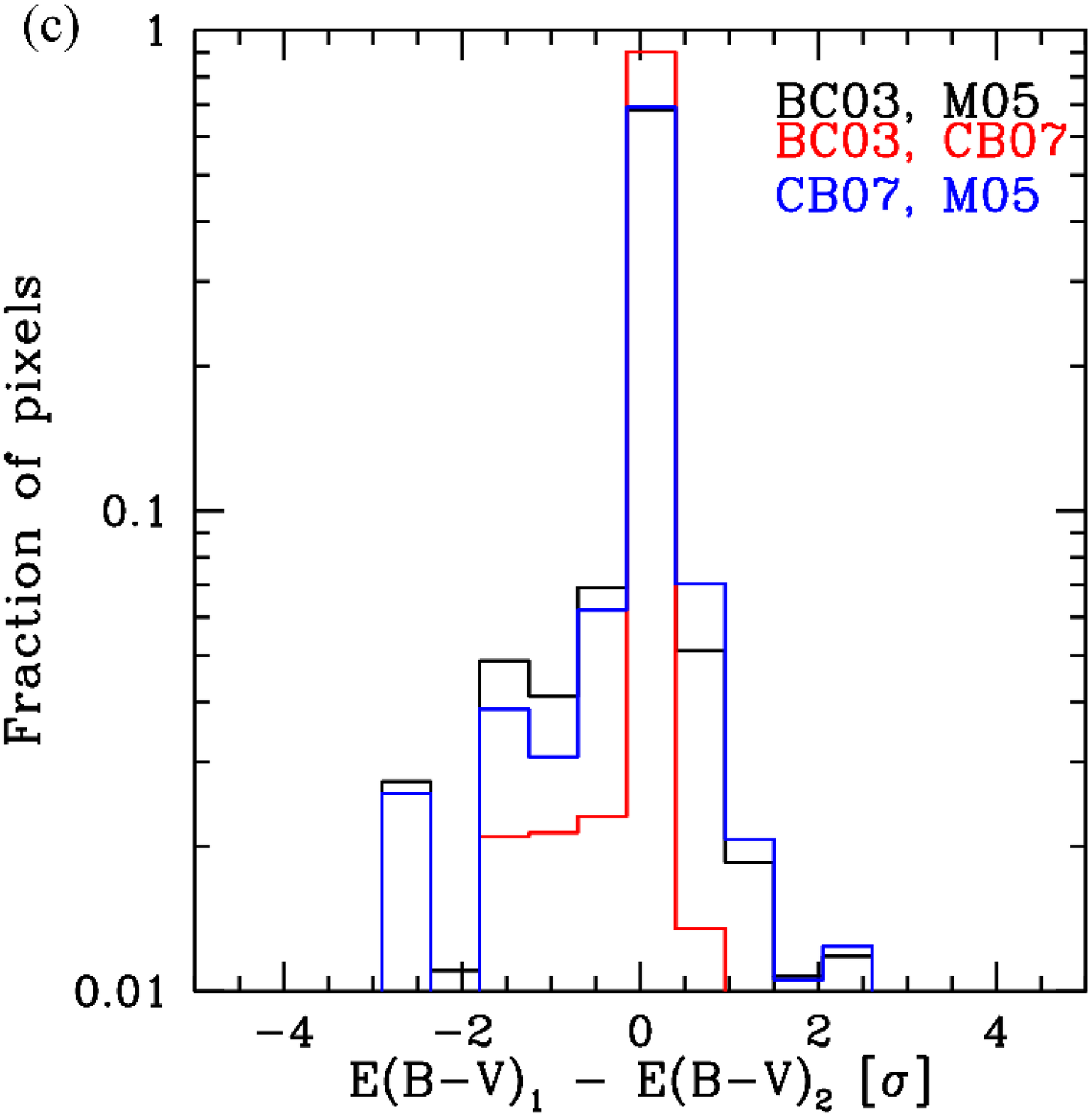}
}
}
\tiny{
\caption{
Quantifying the effect of stellar population synthesis models (in
terms of the statistical error) for all the pixels in the 467 galaxies, selected to have half light
radii $R_{50}> 0.5\arcsec$ and measured
  spectroscopic redshifts. We plot in each panel the distribution of the systematic
  error in the pixel-$z$ parameter (due to model differences) in terms of
  the statistical error ($\sigma$) in that parameter, for all the pixels in the sample. (a) SFR e-folding time, (b) age of
  the stellar population and (c) dust obscuration shown in terms of $E(B-V)$. The pairs of models
  compared in each panel are: BC03 and M05 (black), BC03 and CB07 (red) and CB07
  and M05 (blue). Only pixels within $2R_{50}$ are
  considered. 
}   
\label{fig:models_test_optical}}
\end{figure}

In order to quantify these effects, we use pixels from all 467
galaxies in our sample. The systematic error due to different
population synthesis models is measured according to equation 1, with
the resulting distribution shown in
Figure~\ref{fig:models_test_optical}. 
The distributions are generally non-Gaussian, peaking around zero
systematic error, and showing a relatively small fraction of outliers. The shape of the
distribution is a result of the quantized nature of the problem, since
the pixel-$z$ parameters lie on a discrete grid of values.
The level of agreement and the fraction of outliers in this
distribution for every pixel-$z$ parameter is summarized in Table 1 for each of the pairs of models
being compared. The two models that have the highest level of agreement in
all the parameters concerned (age, $\tau$ and E(B-V)) are BC03 and
CB07. In the $\tau$ parameter, 94 percent of pixels within the
galaxies are within $1\sigma$ and the fraction of outliers
($>2\sigma$) is 0.04. The high level of agreement between BC03 and CB07 is due to the fact that
the continuum in the rest-frame UV is very similar between
these two models. There is a larger systematic difference in the pixel-$z$
parameters between BC03 and M05, and between CB07 and M05. In
particular, the $\tau$ parameter is most sensitive to 
differences between these models. The fraction of pixels with a systematic difference in 
$\tau$ less than $1\sigma$, is 0.71 and 0.80
 for the [BC03,M05] and [CB07,M05] model pairs respectively. 
The fraction of outliers in $\tau$ ($>2\sigma$) is 0.15 and 0.17 respectively for the
 same model pairs. 

The impact of model differences between 
BC03/CB07 and M05 is much smaller for the age and obscuration. 
Dust obscuration is least sensitive to differences between any of the
3 models, with only $1-3$ percent of pixels having systematic
differences in $E(B-V)$ which are larger than $2\sigma$. For the inferred stellar
age, the fraction of pixels having systematic differences in their
stellar population age of $<1\sigma$ is 0.86 and 0.87 for [BC03,M05] and [CB07,M05] 
respectively, while the fraction of outlier pixels in the age
distribution for the same pairs of models is 0.11. For [BC03,CB07],
0.94 of pixels agree in their inferred stellar population age to within
$1\sigma$, while the fraction of outlier pixels for the same models is
0.04. The comparison of the stellar population synthesis models for the
metallicity are not shown in Table 1 because the small number of values of
metallicity being sampled means that all fits are within $2\sigma$. 

Out of the three pixel-$z$ parameters tested, it is thus in the SFR e-folding time where the
largest discrepancy lies between M05 on one hand and BC03/CB07 on the other.  This discrepancy
is unlikely to be due only to the differing treatment of TP-AGBs stars
in these models, since if that were the case, we would expect better
agreement between CB07 and M05 as contrasted against BC03. The discrepancy
observed in $\tau$ is likely related to fundamental
differences between M05 on one hand and BC03/CB07 on the other when
the SED fitting is performed using optical colors only.
In section 3.5.2, we use some example galaxies to investigate where the
discrepancy in the predicted values of $\tau$ (between M05 and
BC03/CB07) occurs spatially in the galaxies. Finally, we test the
effect of redshift on these results by repeating the above analysis for
subsamples of galaxies in redshift intervals of 0.5 from $z=0.5-3$ and
find that our results for the optical-only passbands are not
significantly impacted by the redshift of the galaxy.

\subsection{Effect of Parameter Space and SED Degeneracies}
\label{subsec:paramspace}

In this test, we examine the effect of changing the parameter space on the pixel-$z$ estimates
 using the BC03 models. The number of metallicity values used in the pixel-$z$ grid of stellar
 population parameters is increased from 3
to 6, to include sub-solar metallicities, as detailed in Section~\ref{sec:sspsfromcolors}. The
increase in the metallicity range means we are comparing the original 2178 SEDs
with an expanded set of 4356 SEDs fitted to each
pixel.

\begin{figure}
\centerline{\rotatebox{0}{
\includegraphics[width=.5\textwidth]{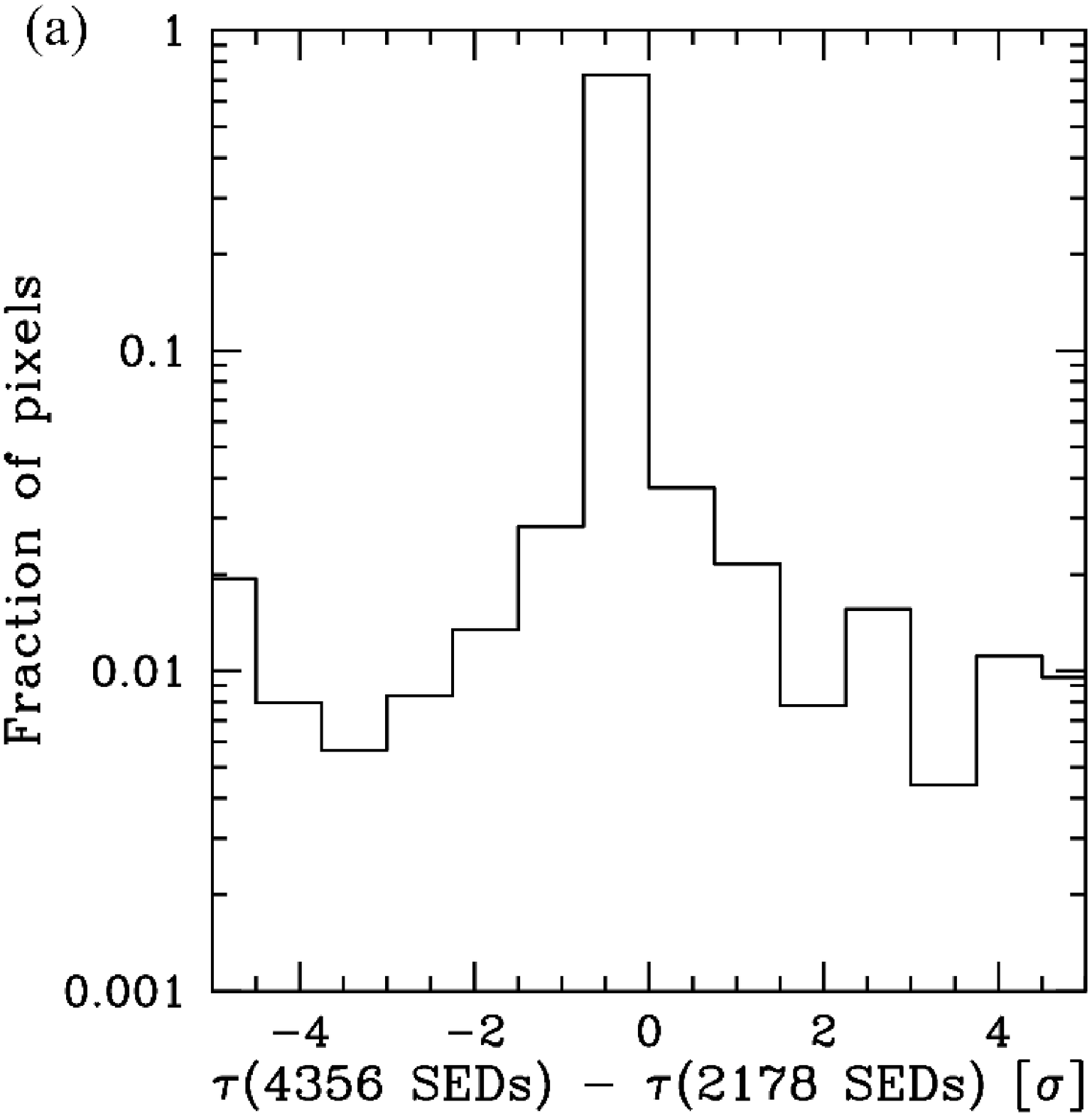}
\hspace{0.1cm}
\includegraphics[width=.5\textwidth]{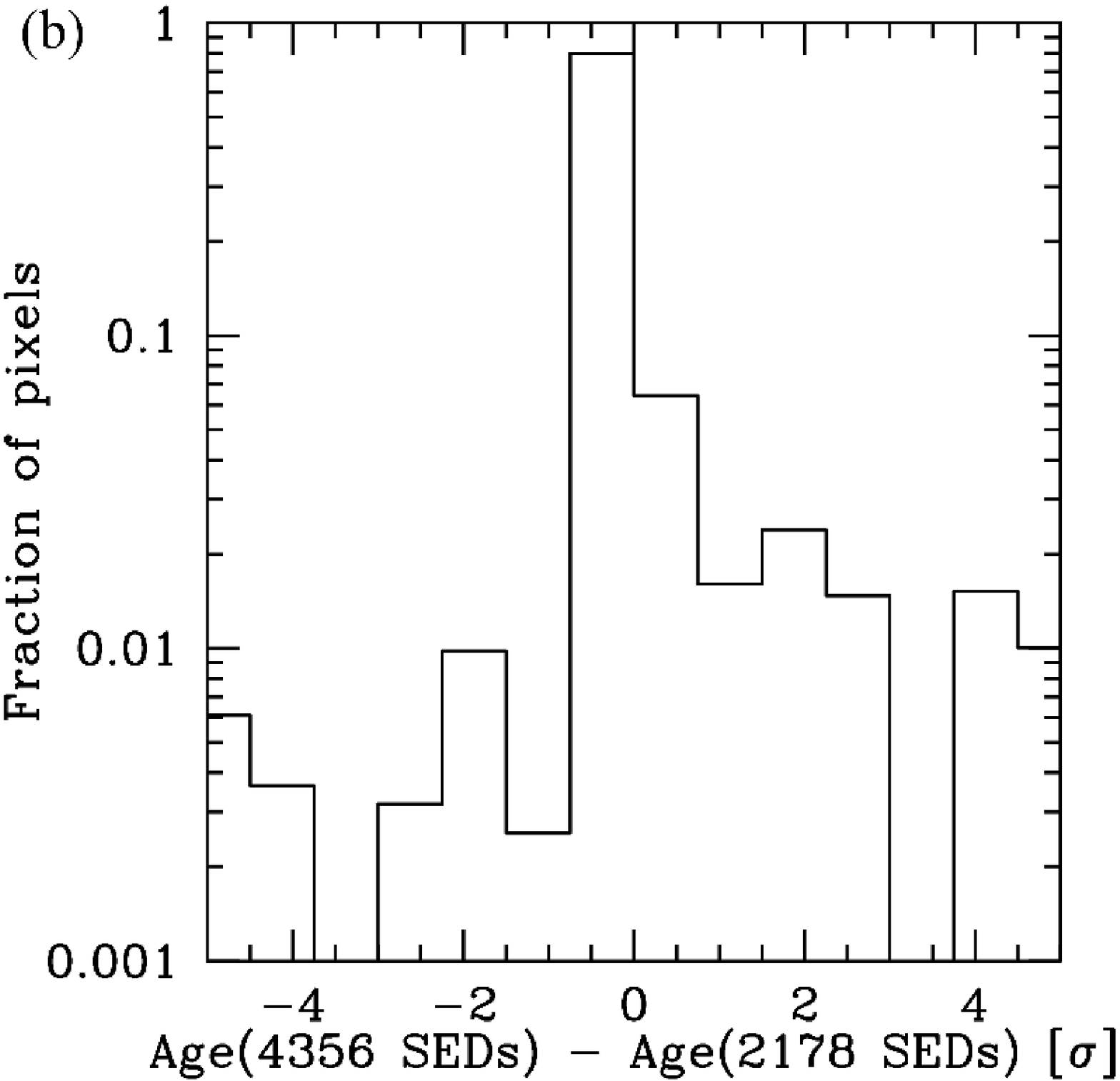}
}}

\vspace{0.2cm}
\centerline{\rotatebox{0}{
\includegraphics[width=.5\textwidth]{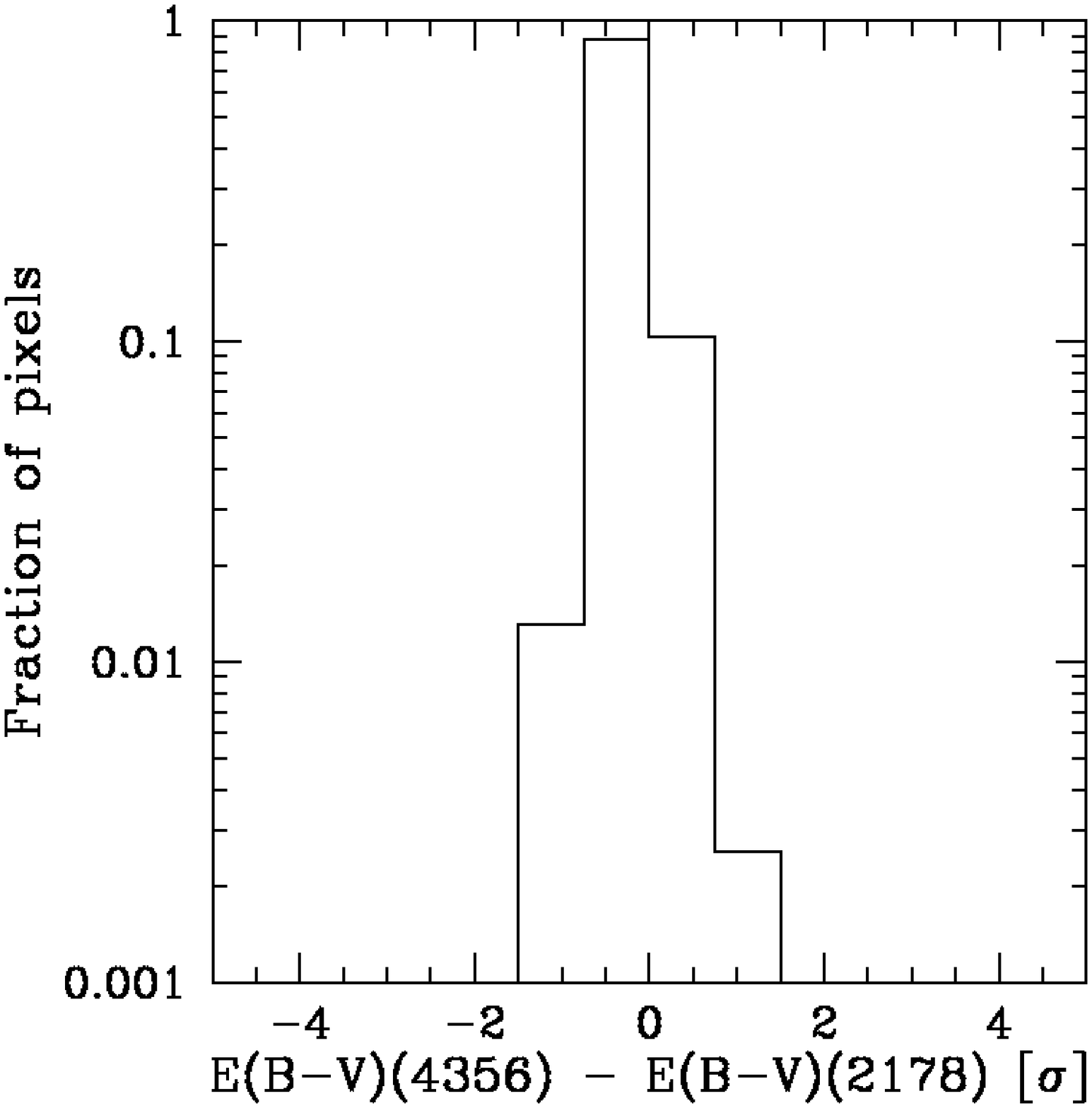}
}
}

\tiny{
\caption{
Effect of increasing the range of metallicity values (and hence the
number of SEDs) allowed
  for the BC03 models for all pixels in the entire galaxy sample of 467
  objects. Here we fix the model (BC03) and examine the
  effect of increasing the range of metallicity values from 3 allowed
  metallicities (resulting in a total of 2178 SEDs) to 6 allowed values (4356 SEDs). Shown in each panel is
  the distribution of the systematic error in the recovered pixel-$z$
  parameter in terms of the statistical
  error ($\sigma$) in the parameter, for all pixels in the sample: (a) SFR e-folding time (b) stellar population age and (c) dust obscuration given in
  terms of $E(B-V)$. Only pixels within $2R_{50}$ are
  considered. 
 }

\label{fig:metallicity_test}}
\end{figure}

The distribution of the systematic error (as a function of the statistical error) 
introduced by extending the allowed metallicity
range into the sub-solar regime is shown in
Figure~\ref{fig:metallicity_test} for all pixels in all the galaxies
in the sample. 
The fraction of pixels with small systematic 
errors ($<1\sigma$) and the fraction of outliers in each parameter is
summarized in Table 2. The largest effect of a broadening the metallicity range
is, predictably, on the stellar population age but the effect is minimal: the fraction of outliers in the age systematic is only 0.06.
Around 6 percent of pixels have a bias in their $\tau$ that is larger
than the $2\sigma$ uncertainties. The obscuration is the parameter which is least
affected ($< 0.01$ of pixels) by the introduction of sub-solar metallicities. 
This is in agreement with the findings of Conti et al.~(2003) who
examined the evolution of the obscuration-metallicity degeneracy in
pixels in galaxies in the HDF-North, and found that there is a strong
degeneracy in galaxies at low redshifts but that the relation flattens
out at $z>1$.

The conclusion of this test is that while the stellar population age
and $\tau$ are marginally affected by the increased metallicity range, 
as expected from the age-metallicity degeneracy, the resulting
systematic differences in all the pixel-$z$ quantities are still well
within the uncertainties between different stellar population synthesis models. 

\subsection{Effect of Passband Limitations (Optical-only)}

\begin{figure}
\centerline{\rotatebox{0}{
\includegraphics[width=1.1\textwidth]{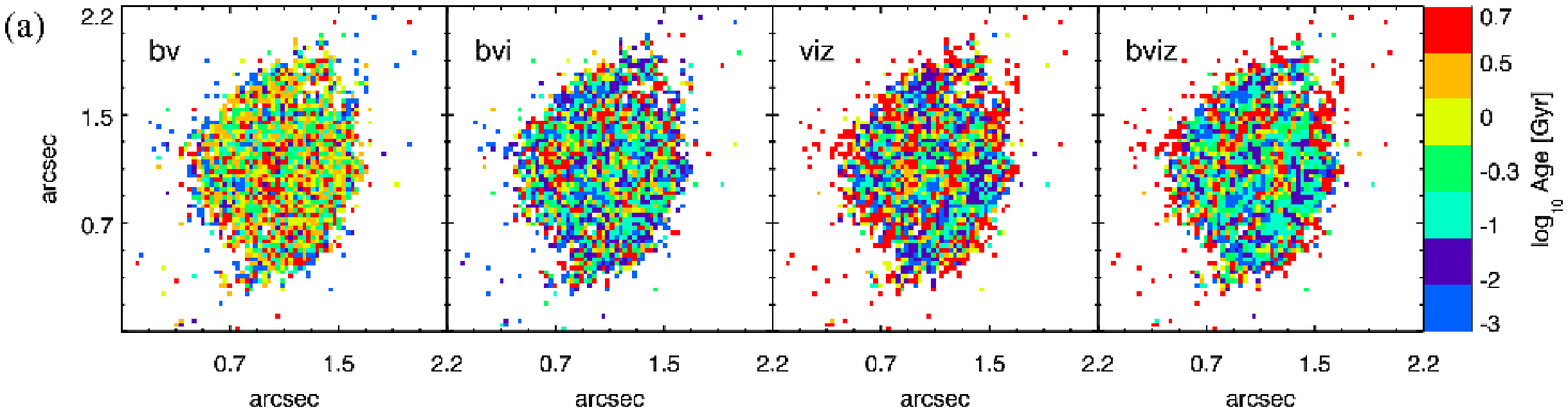}
}}
\vspace{0.2cm}
\centerline{\rotatebox{0}{
\includegraphics[width=1.1\textwidth]{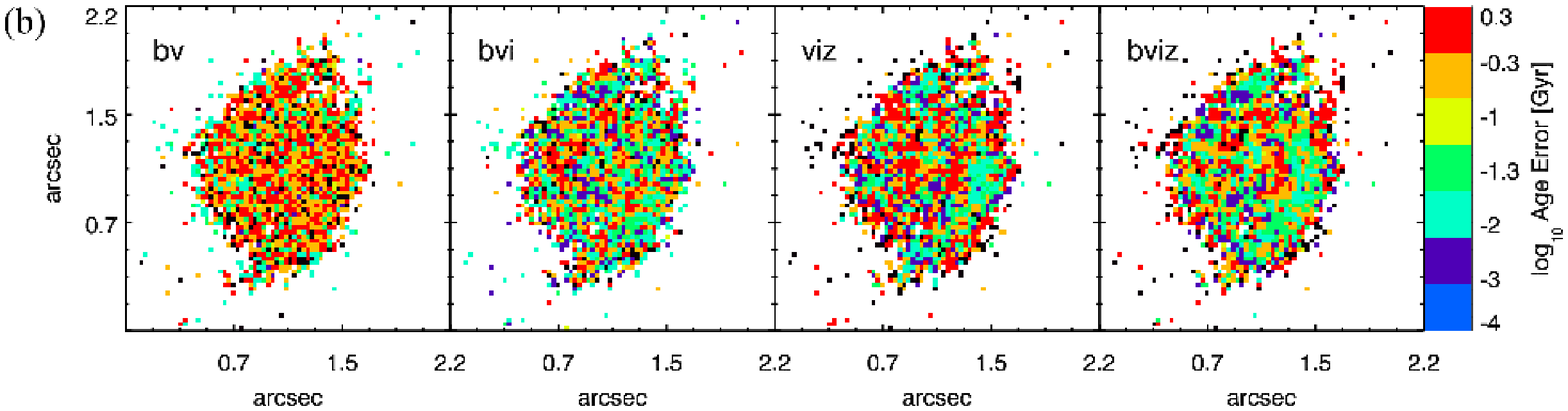}
}}
\tiny{
\caption{
Systematic effect of passband changes on the stellar population age
map for the galaxy shown in Figures 2 and 3. The stellar population
synthesis model used here is BC03.  (a)
Map of the stellar population age for the galaxy for the following passband combinations: 
(b+v), (b+v+i), (v+i+z), (b+v+i+z). (b) Map of the error in the
stellar population age for the same passband combinations.
}
 \label{fig:passbands_optical_age}}
\end{figure}

Here we test the effect of the passband limitations on the estimates of the
pixel-$z$ parameters. The stellar population synthesis model is fixed to BC03 in this test and
passbands are added and removed and the pixel-$z$
output is compared to that produced using the full complement of 4 optical
bands. This test is initially performed for selected individual galaxies to identify the
likely trends, before a full investigation on all pixels in the full
galaxy sample is performed. Figure~\ref{fig:passbands_optical_age} illustrates the test for one example
galaxy, which is the same as shown in
Figures~\ref{fig:models_optical_age} and~\ref{fig:models_optical_ebv}. It shows the
impact of passband variations on the age values in the
pixels and their associated uncertainties. Starting with only 2 optical bands ($bv$), the
effect of different permutations of bands are tested. 
From the stellar population age maps, it is clear that 2 passbands are insufficient to
resolve many of the features seen in the light distribution of
the galaxy. For the case of 3 bands, the use
of the redder $viz$ combination of filters comes closer to the result
obtained with the full $bviz$ filters than does the $bvi$ set.  The
$viz$ combination produces an age map that comes closer to mirroring
the galaxy morphology than does the $bvi$ set, separating 
the arm and inter-arm regions more clearly. For this example galaxy at $z\sim1$, most
of the relevant color information for the pixel-$z$ therefore comes from the optically redder passbands.

\begin{figure}
\centerline{\rotatebox{0}{
\includegraphics[width=1.1\textwidth]{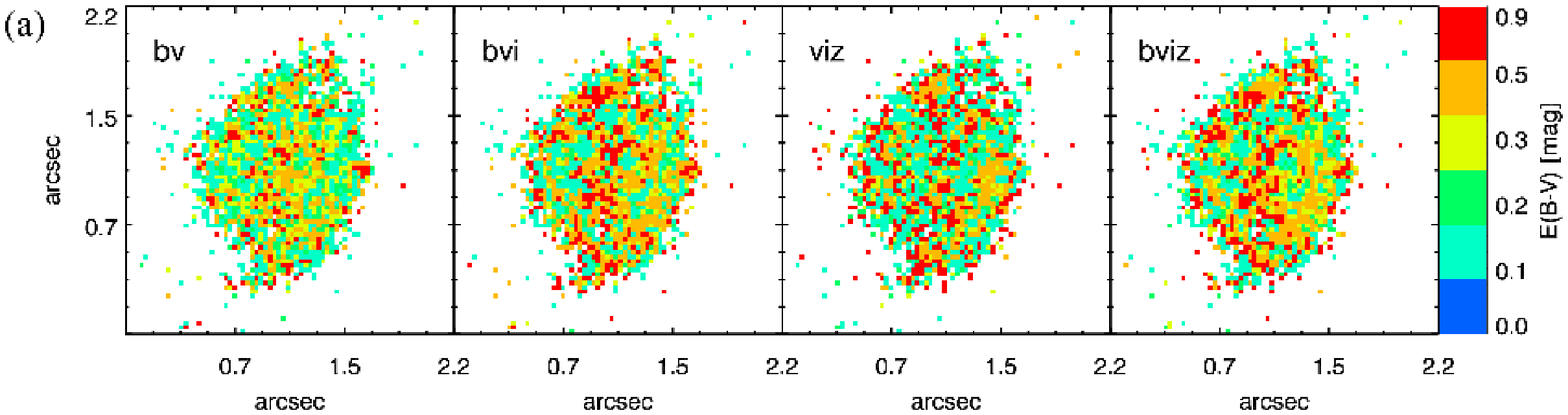}
}}
\vspace{0.2cm}
\centerline{\rotatebox{0}{
\includegraphics[width=1.1\textwidth]{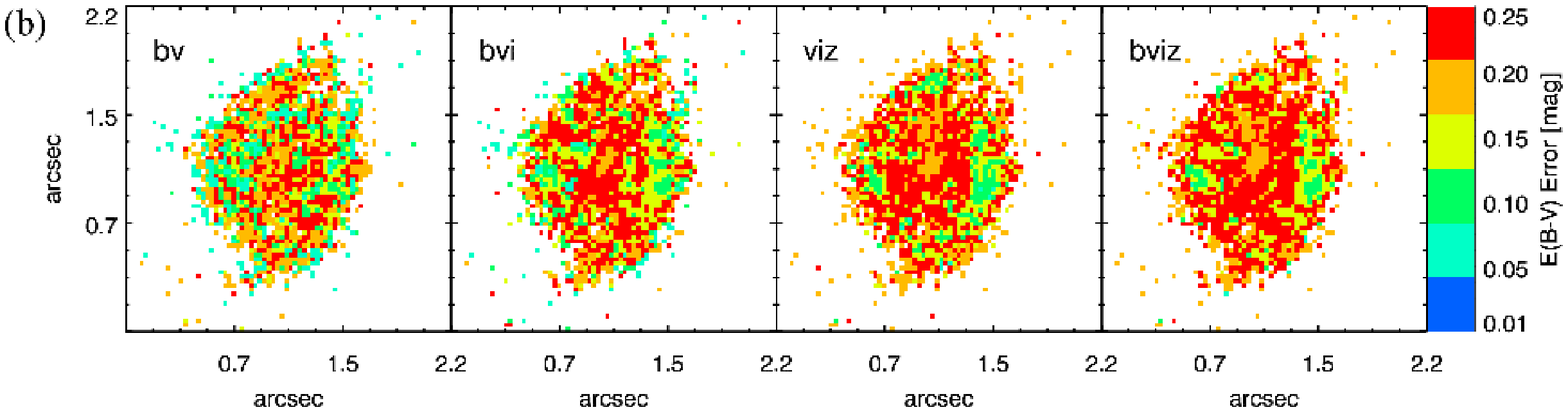}
}}
\tiny{
\caption{
Systematic effect of passband changes on the dust obscuration map of
the galaxy shown in Figures 2  and 3. The stellar population
synthesis model used here is BC03.  (a)
Map of the $E(B-V) $ for the galaxy for the following passband combinations: 
(b+v), (b+v+i), (v+i+z), (b+v+i+z). (b) Map of the error in $E(B-V)$ for the same passband combinations.
}
 \label{fig:passbands_optical_ebv}}
\end{figure}

A similar conclusion can be drawn from the map of rms values in the age
parameter. With only 2 optical bands, high uncertainties 
throughout the disk of the galaxy reflect the inability of pixel-$z$ to
find the correct SED with only 1 color, thus resulting in higher
statistical errors in the ages inferred for the pixels. With 3 passbands, again it is
the $viz$ combination that seems to be closest to the $bviz$ 
values. However, the maps of dust obscuration in
Figure~\ref{fig:passbands_optical_ebv} show that the $bvi$ combination
coming closest to the result predicted from the full complement of
bands. Nevertheless, the $bvi$, $viz$ and $bviz$ combinations all agree that the arms have intermediate
levels of obscuration ($E(B-V)\sim0.4$) and low associated RMS errors.

\begin{figure}
\centerline{\rotatebox{0}{
\includegraphics[width=.55\textwidth]{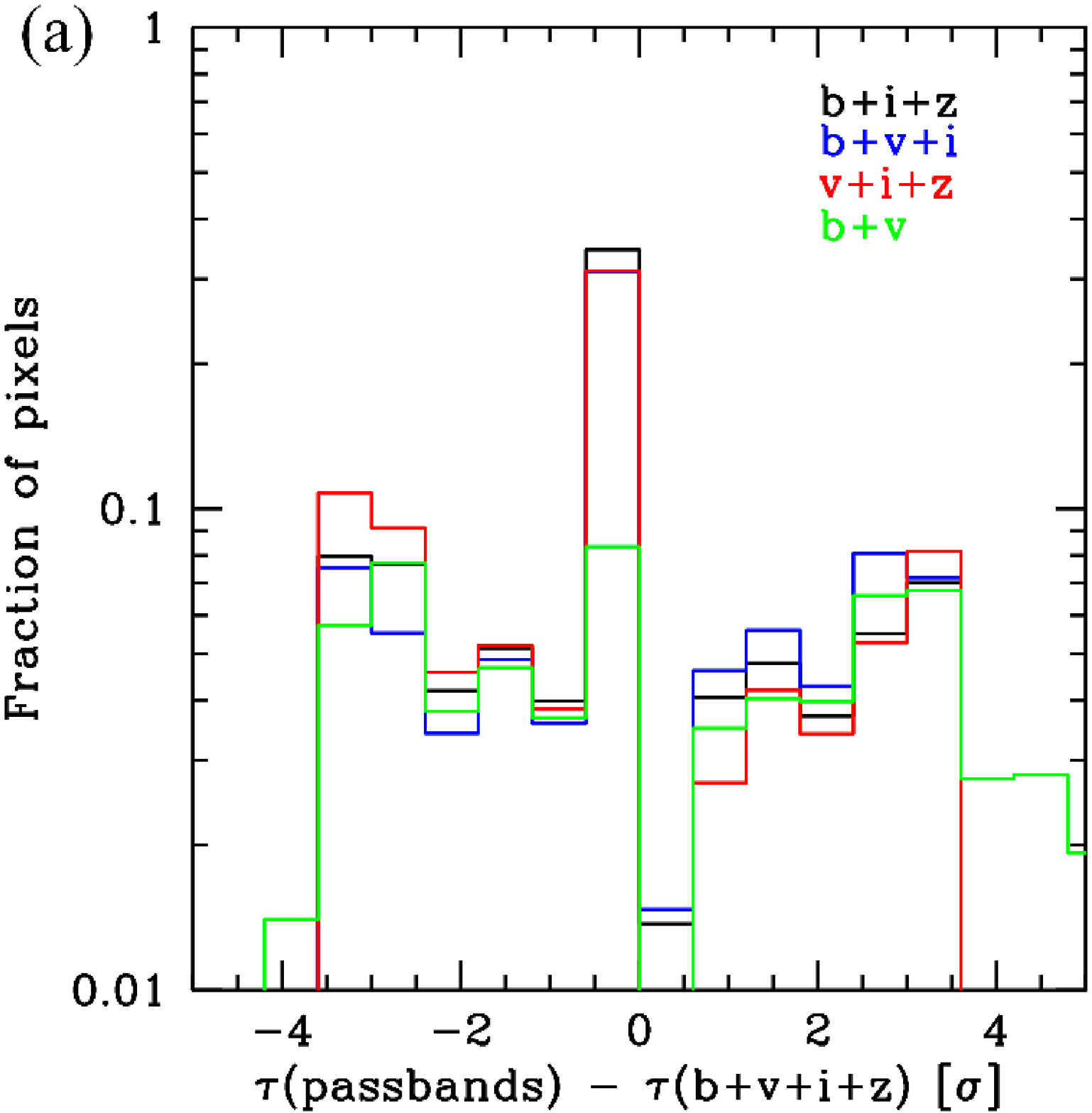}
\hspace{0.1cm}
\includegraphics[width=.55\textwidth]{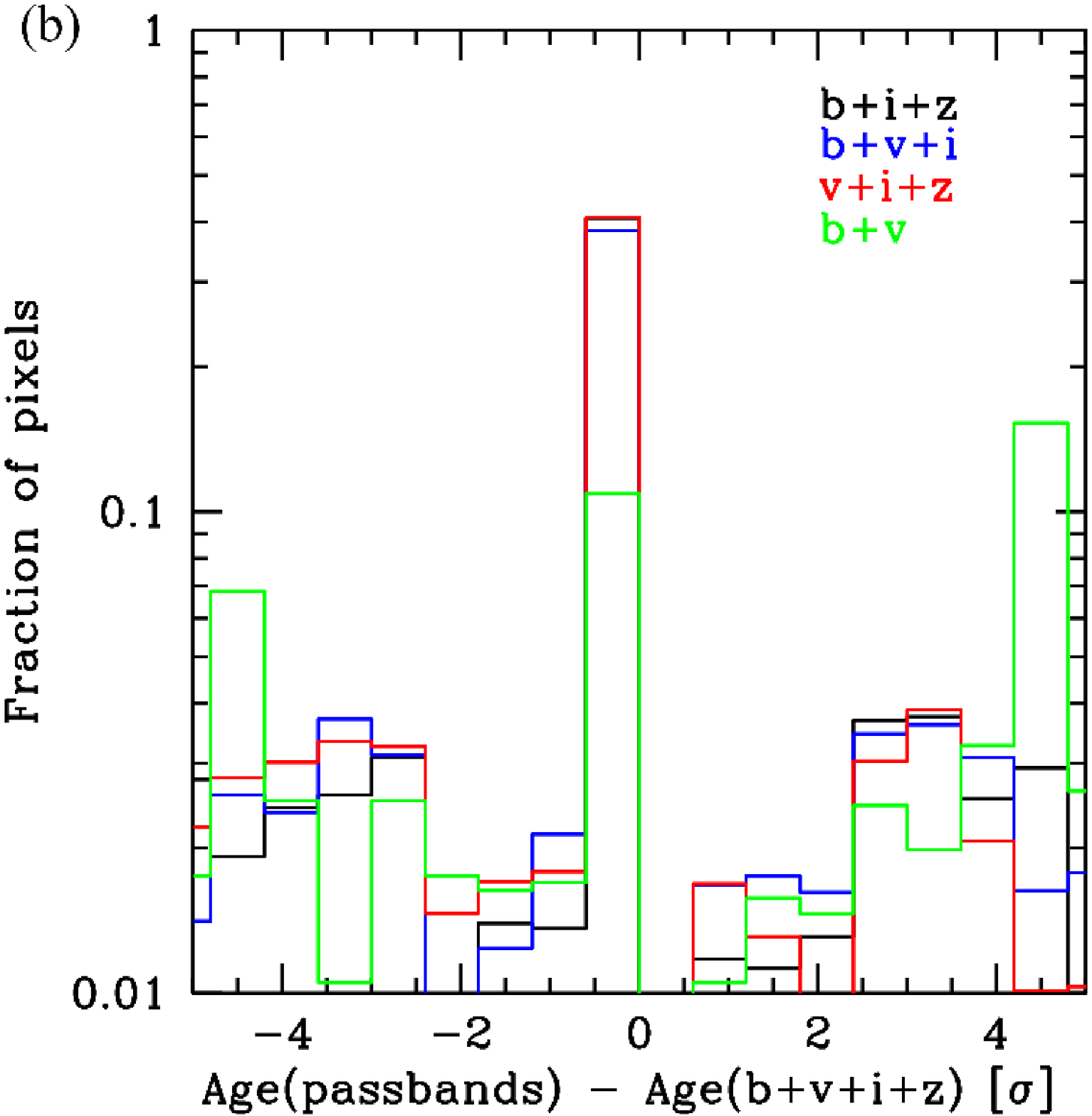}
}
}
\vspace{0.2cm}
\centerline{\rotatebox{0}{
\includegraphics[width=.55\textwidth]{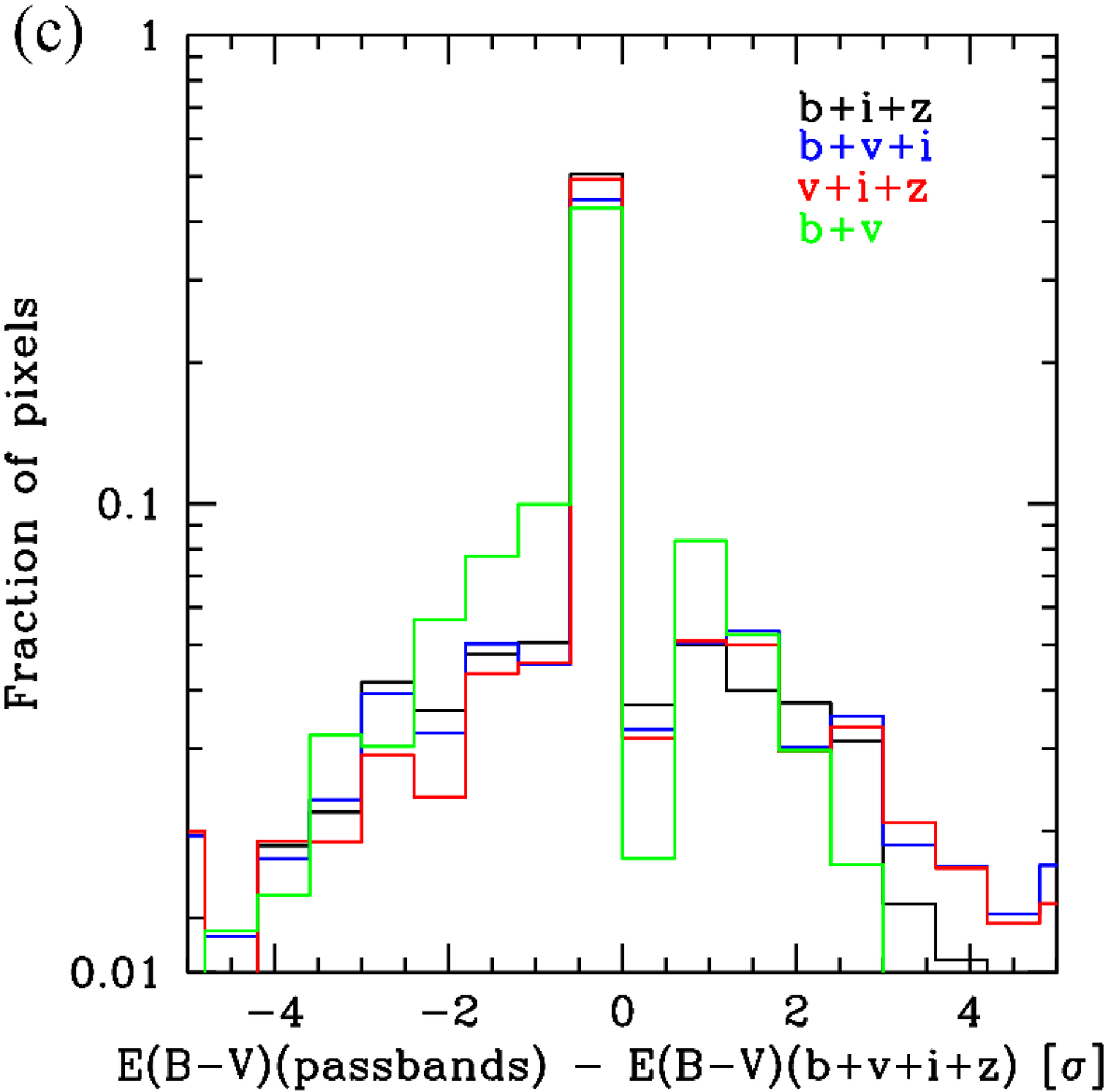}
}
}
\tiny{
\caption{
Systematic effect of passband changes for all the pixels 
  in the entire galaxy sample, with a fixed stellar population
  synthesis model (BC03). Shown is the distribution of the systematic
  error in the parameters (given in terms of the statistical error $\sigma$) that arise from omitting passbands relative
  to the full complement of optical bands ($bviz$). The systematic error is
measured for (a) the SFR e-folding time (b) stellar population age and (c)
the dust obscuration given in terms of $E(B-V)$. For each parameter,
the passbands considered are: $biz$ (black), $bvi$ (blue) and $viz$ (red). For illustrative purposes, we also
plot, for only the single galaxy shown in Figures 2 and 3, the
distribution of the systematic error resulting from using only the 2
passbands $bv$ (green).
}   
\label{fig:passbands_optical_test}}
\end{figure}

Moving to the full sample of all pixels in all galaxies, we use
equation 2 to compute for each parameter and for each pixel, the systematic error (again in
terms of the statisitical error in the parameter) due to the difference 
between a given filter permutation and the full optical filter set
($bviz$). The stellar population model is fixed to BC03 as before. 
Figure~\ref{fig:passbands_optical_test} shows the distribution of this systematic error for all the pixels in the full galaxy
sample. It is evident that the systematic error distribution has larger tails and a lower peak ($<1\sigma$) than the systematic errors
resulting from stellar population model differences. As for the
example galaxy in Figures~\ref{fig:passbands_optical_age}, it is clear also that 2 filters
($bv$) result in a large fraction of outliers ($>3\sigma$).
For the 3-filter permutations ($biz$, $bvi$ and $viz$), across the
entire sample of galaxies, we find that there is only a small 
difference in the distribution of the systematic error among the
filter permutations.  

We show the fraction of outliers in each case in Table 3. It is
clear that the SFR e-folding time ($\tau$) and the stellar population age are the most impacted of the 3
parameters by passband limitations. In $\tau$ and in the stellar
population age, over 20 percent of the
pixels become outliers ($>2\sigma$) when one band of the full filter
set $bviz$ is dropped. There is not a very significant difference
between the 3 sets of filters, but omitting the $z$ band results in an
outlier fraction for the $\tau$ parameter of 0.26 compared to 0.20 for
the other bands. For the stellar population age, the outlier fraction varies from
$0.22-0.28$ relative to the full filter set ($bviz$). The dust obscuration is
least impacted by passband changes, as long as a minimum of 3 are
used. For each of the 3 filter sets, about 80 percent of pixels in the
full galaxy sample have an $E(B-V)$ which agrees with the $bviz$
result to within $1\sigma$. Finally, as in Section~\ref{subsec:model_tests}, we test the
effect of redshift on these results by repeating the above analysis for
subsamples of galaxies in different redshift intervals from $z=0.5-3$ and
find that our results for the optical-only passbands are not
significantly impacted by the redshift of the galaxy.

\subsection{Effect of Adding the Near-IR Data}

\begin{figure}
\centerline{\rotatebox{0}{
\includegraphics[width=.55\textwidth]{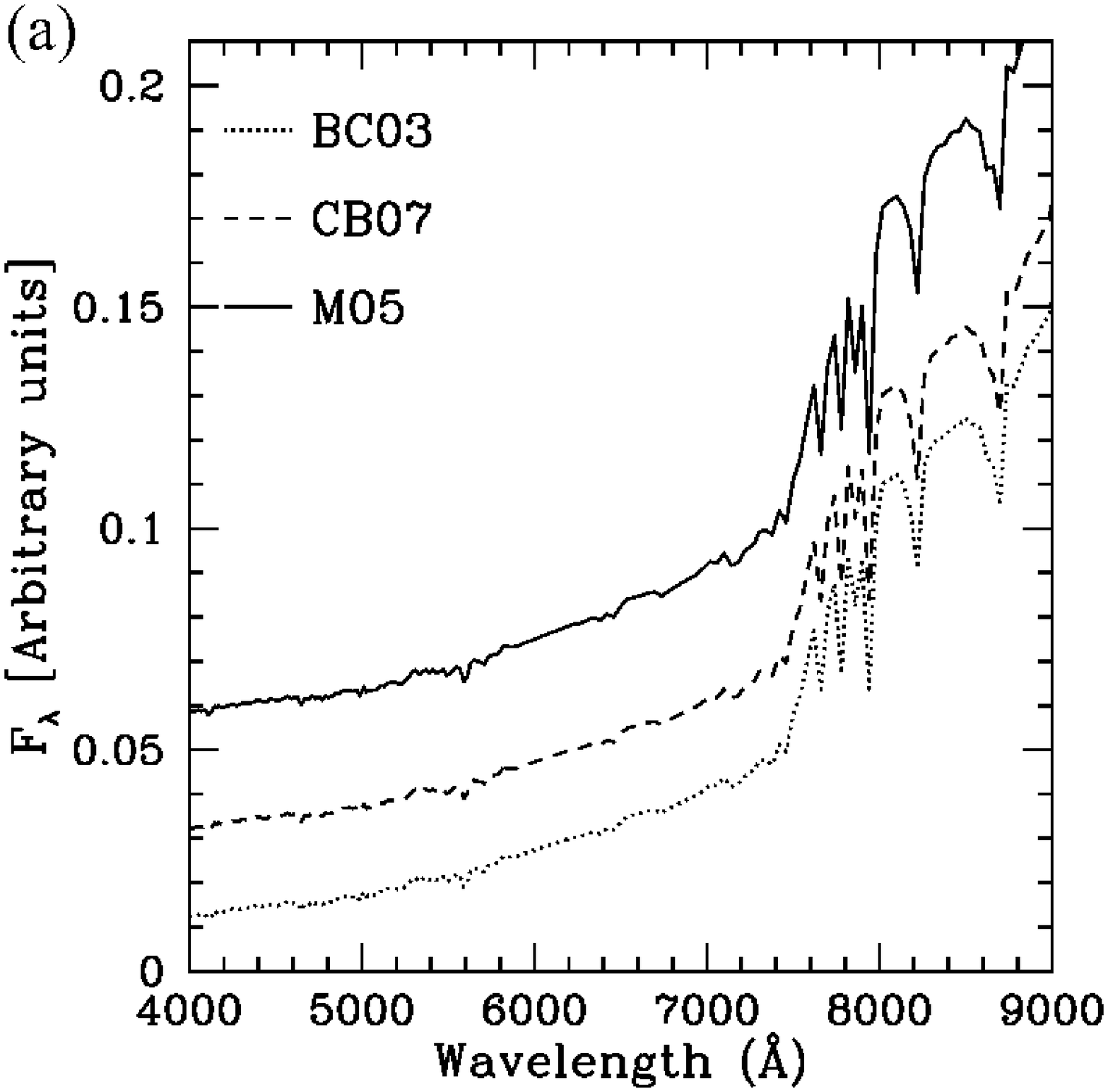}
\includegraphics[width=.55\textwidth]{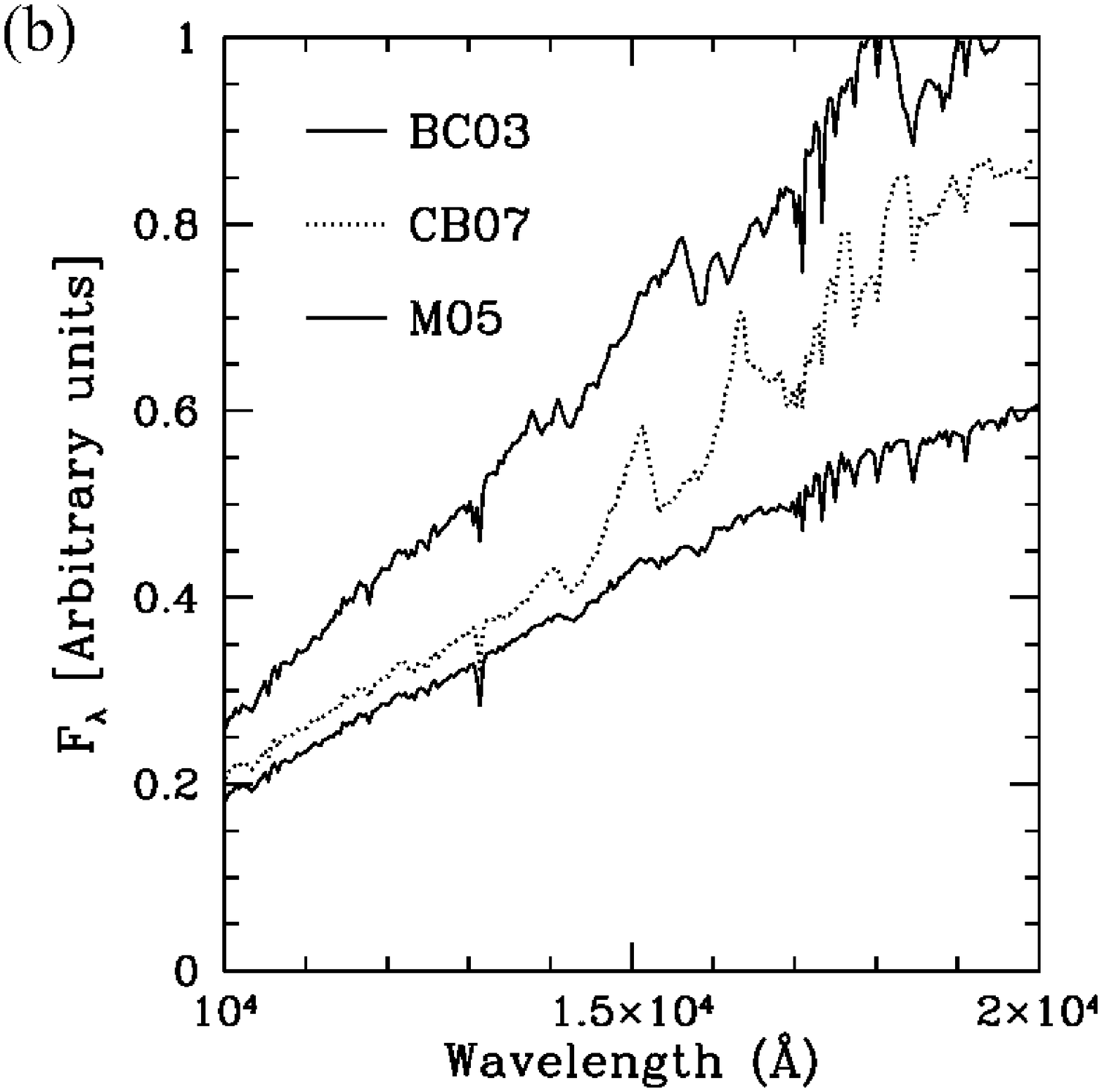}
}
}
\tiny{
\caption{
(a) Example SEDs in the optical spectral range of a galaxy at
  redshift=1.0, age = 3 Gyr, tau = 10 Gyr, E(B-V)=0.9 and Z = 0.008.
There is very little difference in the shape of the continuum among
the BC03, CB07 and M05 stellar population synthesis models in the
optical. (b) The SEDs of the same galaxy in the NIR, where
differences in the shape of the continuum emerge among the three
models. These will give rise to different predicted NIR colors.
}
 \label{fig:spectra}}
\end{figure}

In this section, we explore the effect of adding NIR passbands from NIC3 to the existing
 optical data for a subsample of 46 galaxies. Adding the NIR data allows us to explore systematic
 differences between various population synthesis models more fully, since they can predict
 different NIR colors. In particular, it is known that BC03 and M05
differ significantly in their treatment of TP-AGB stars, consequently
resulting in different predicted NIR colors. This is illustrated in
Figure~\ref{fig:spectra} which shows an example SED from all three models
considered in this work, for a galaxy at $z=1.0$ containing a stellar population 
that is 3 Gyr old, with $\tau=10$ Gyr, $E(B-V)=0.9$ and
$Z=0.008$. There is very little difference between the shape of the
continuum between the the models in the optical. However,
the same galaxy in the NIR shows significant differences in the
shape of the continuum among the three models. These will give rise to different predicted
NIR colors in the pixels.

\begin{figure}
\centerline{\rotatebox{0}{
\includegraphics[width=.55\textwidth]{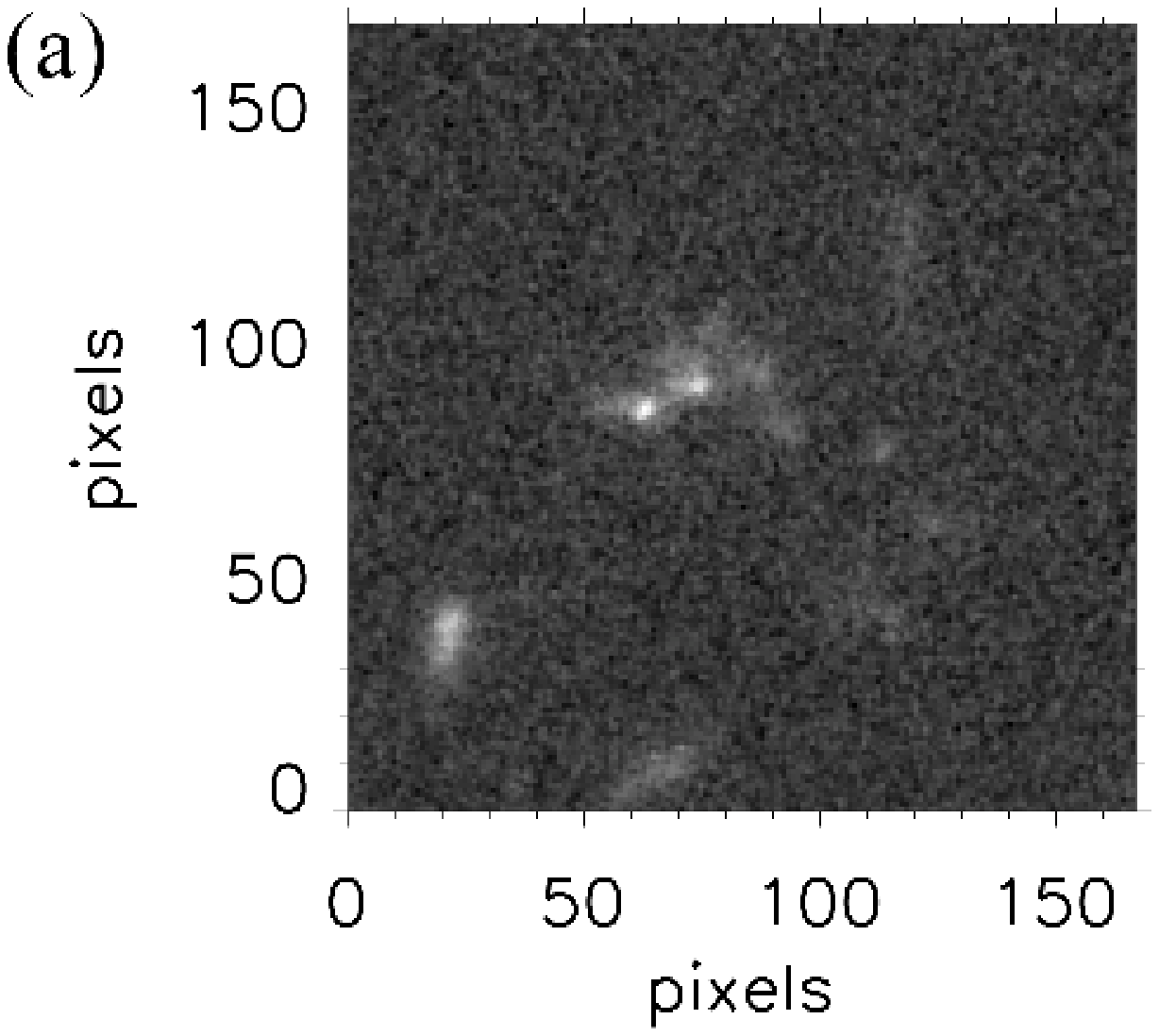}
\hspace{0.1cm}
\includegraphics[width=.55\textwidth]{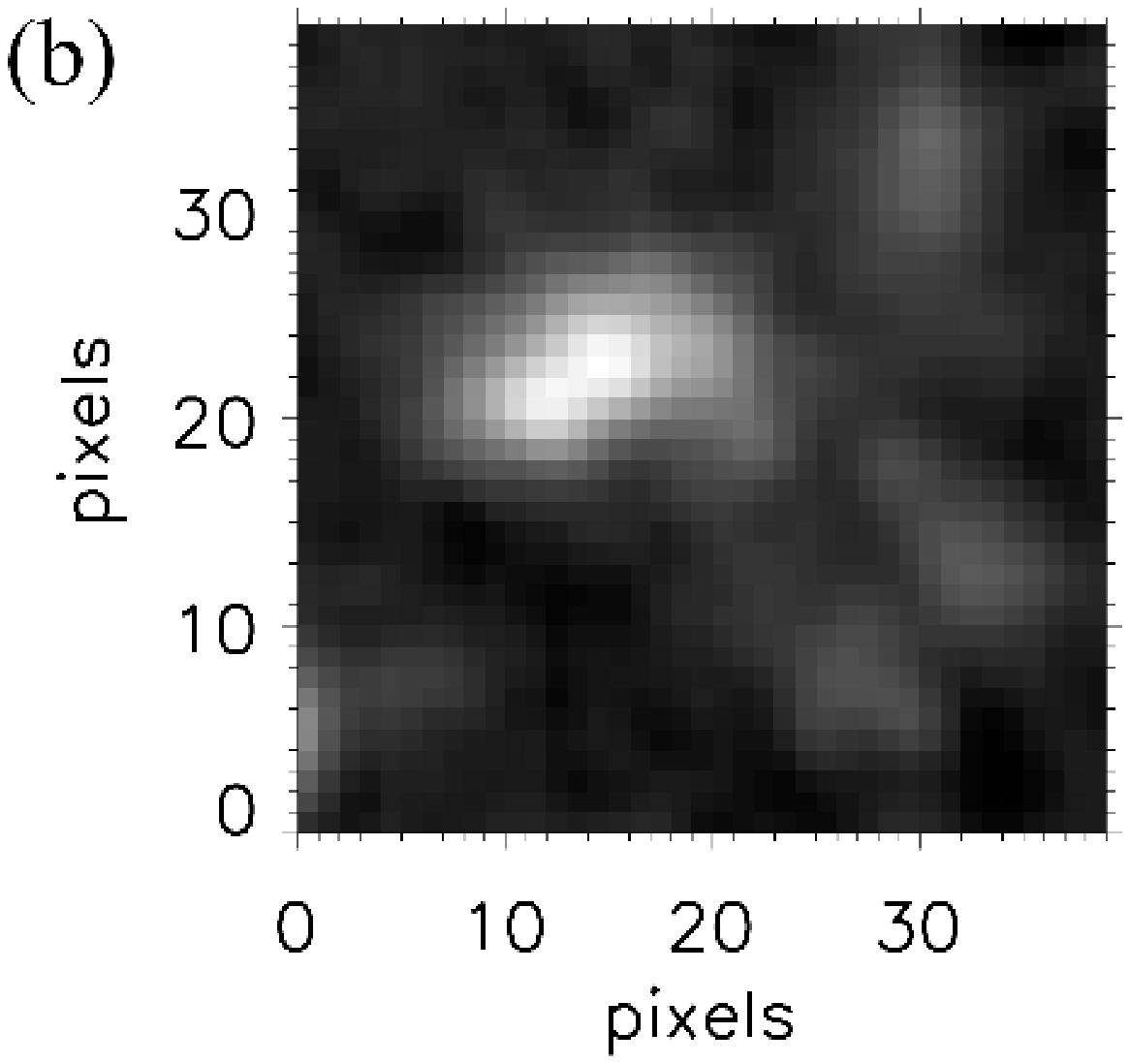}
}
}
\centerline{\rotatebox{0}{
\includegraphics[width=1\textwidth]{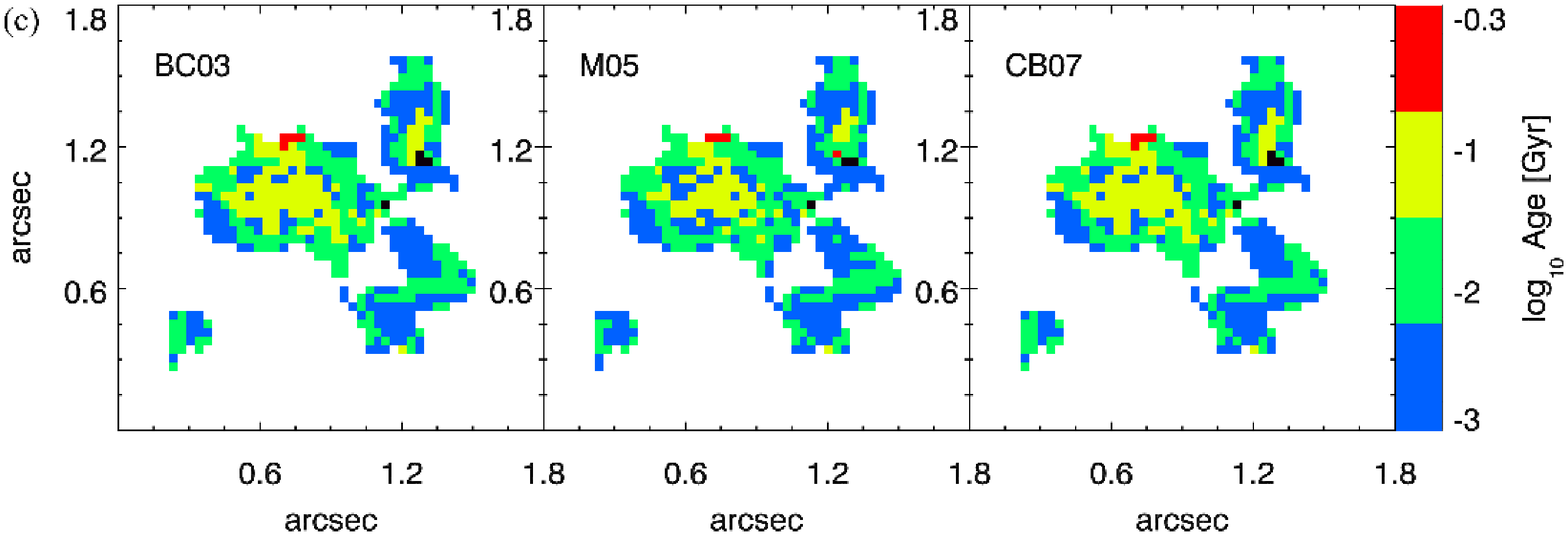}
}
}
\vspace{0.1cm}
\centerline{\rotatebox{0}{
\includegraphics[width=1\textwidth]{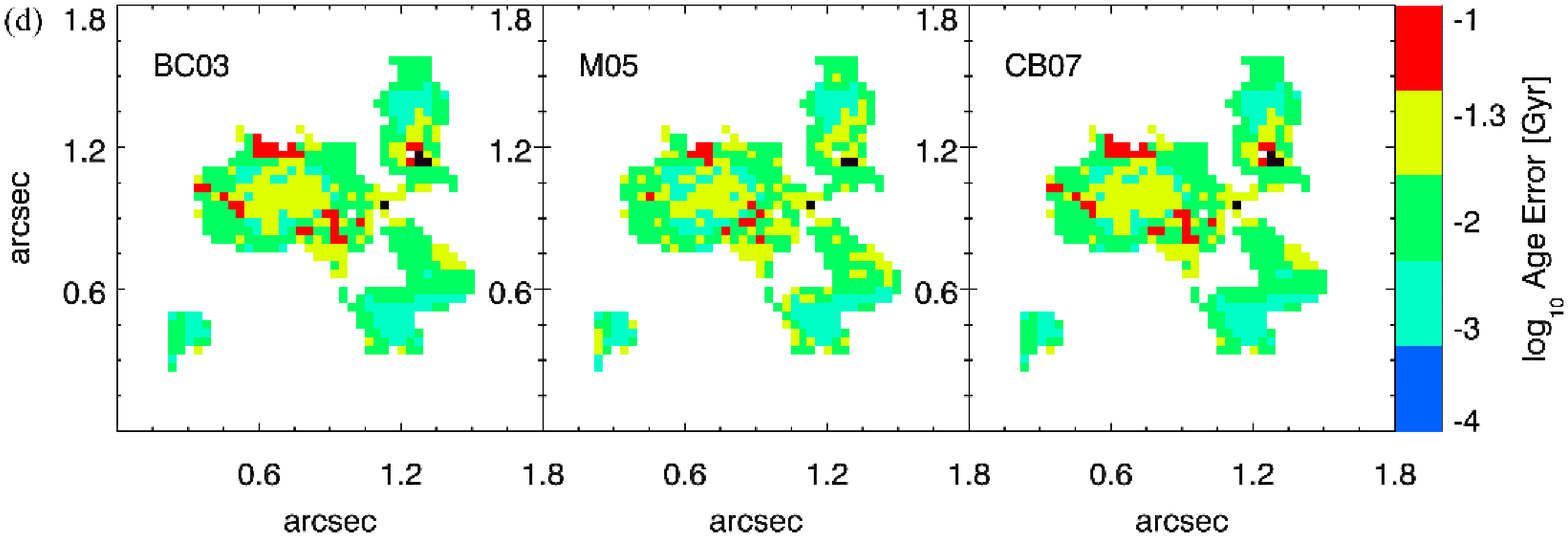}
}
}

\tiny{
\caption{
Spatial distribution of the stellar population age in a $z=2.5$
 galaxy after adding NIR colors to the pixel-$z$ analysis. 
  The galaxy was detected in both GOODS-South and the
  GOODS-NICMOS survey. It has an irregular morphology and has two
  associated companions on the right of the image. In total, 6
  passbands ($bvizJH$) are used. (a) the original $i$ band image of the
  galaxy (b) $i$ band image of the galaxy after resampling and PSF-matching to the
  $F160W$ image of the galaxy (c) map showing the distribution of the
  stellar population age inferred from the BC03, M05 and CB07 stellar
  population synthesis models (d) map showing the the error in
  the age for the same models.
}   
\label{fig:age_nir_example}}
\end{figure}
\begin{figure}
\centerline{\rotatebox{0}{
\includegraphics[width=.55\textwidth]{fg10aarxiv.eps}
\hspace{0.1cm}
\includegraphics[width=.55\textwidth]{fg10barxiv.eps}
}
}
\centerline{\rotatebox{0}{
\includegraphics[width=1\textwidth]{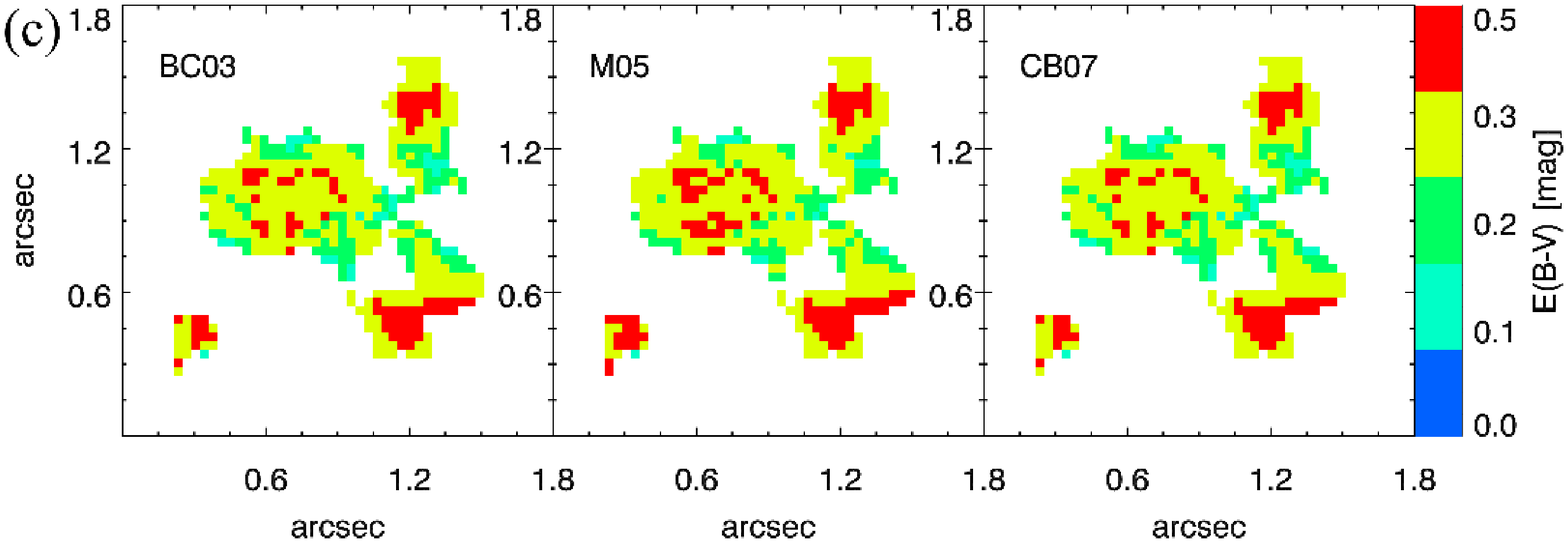}
}
}
\vspace{0.1cm}
\centerline{\rotatebox{0}{
\includegraphics[width=1\textwidth]{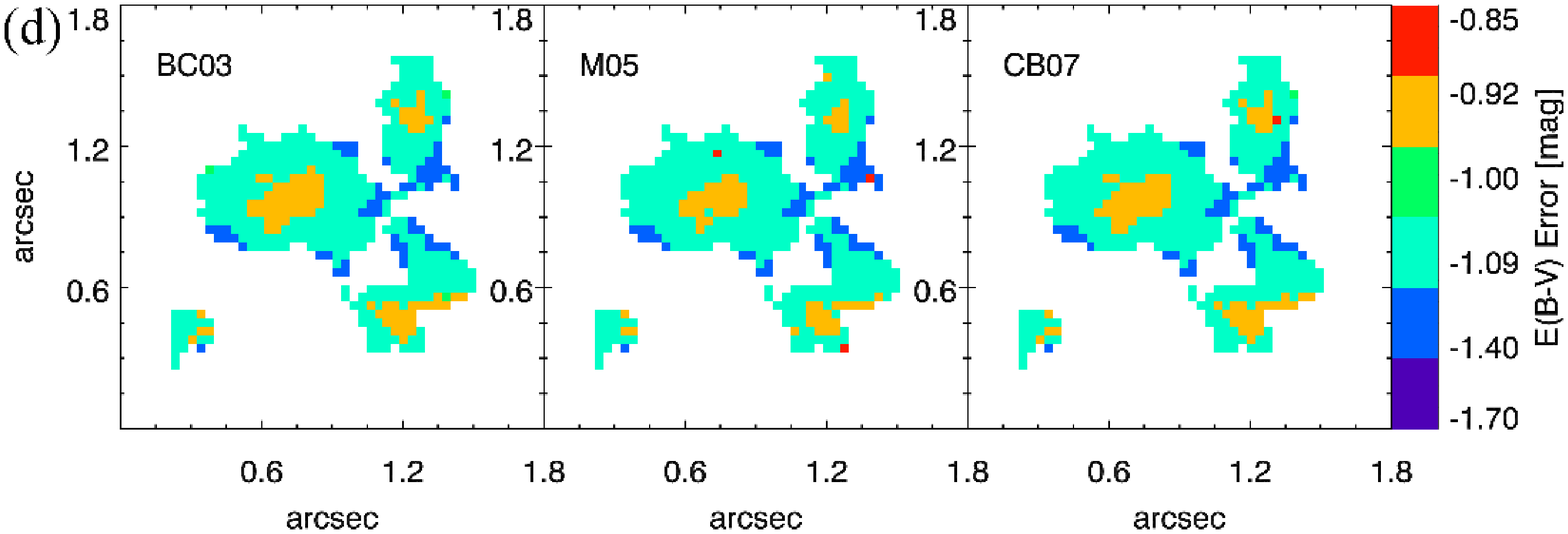}
}
}

\tiny{
\caption{
Spatial distribution of dust obscuration in a $z=2.5$
  galaxy and its companions after adding NIR colors to the pixel-$z$ analysis. 
The galaxy was detected in both GOODS-South and the
  GOODS-NICMOS survey.  It has an irregular morphology and has two
  associated companions on the right of the image. In total, 6
  passbands ($bvizJH$) are used (a) the original $i$ band image of the
  galaxy (b) $i$ band image of the galaxy after resampling and PSF-matching to the
  $F160W$ image of the galaxy (c) map showing the distribution of the
  dust obscuration (in terms of $E(B-V)$) inferred from the BC03, M05 and CB07 stellar
  population synthesis models (d) map showing the the error in
  the $E(B-V)$ for the same models.
}   
\label{fig:ebv_nir_example}}
\end{figure}

 Figures~\ref{fig:age_nir_example} and ~\ref{fig:ebv_nir_example} show
 examples of applying pixel-$z$ to a galaxy at $z=2.5$ with $bviz$ and $J$ and $H$
 band imaging from NIC3. There is a good agreement among the
 three models compared here in both their stellar population age maps
 and in the age rms map. All three models predict a very young population of
 stars throughout ($t<0.2$ Gyr). 
All models considered predict the core of the
 galaxy to have an older population of stars ($\sim0.1$ Gyr) than the
 outskirts ($t\sim0.01$Gyr). The one notable difference is that the
 M05 models predict a slightly larger proportion of very young stellar
 populations ($t\sim0.001$Gyr) surrounding the nucleus. The models also agree very well on the age
 distribution within the companions of this galaxy (to the right). The models also agree well in the uncertainty maps in the
 stellar population age. Figure~\ref{fig:ebv_nir_example} shows the obscuration
 maps of the same galaxy as predicted by the three models. Again, there
 is excellent agreement among the three models in the distribution of
 dust and its associated uncertainty. In general,
 all three models predict that the galaxy has moderate levels (E(B-V)$\sim0.3$
 magnitudes) of dust, but they also show pockets of high obscuration
 (E(B-V)$\sim0.5$ magnitudes) within the main galaxy and
also in its two companions. The location of these pockets are the same
in three models. This example therefore implies that model differences do not introduce large
 systematic biases in the pixel-$z$ parameters for $z\sim2.5$
 galaxies with such irregular morphologies. 

\begin{figure}
\centerline{\rotatebox{0}{
\includegraphics[width=1.\textwidth]{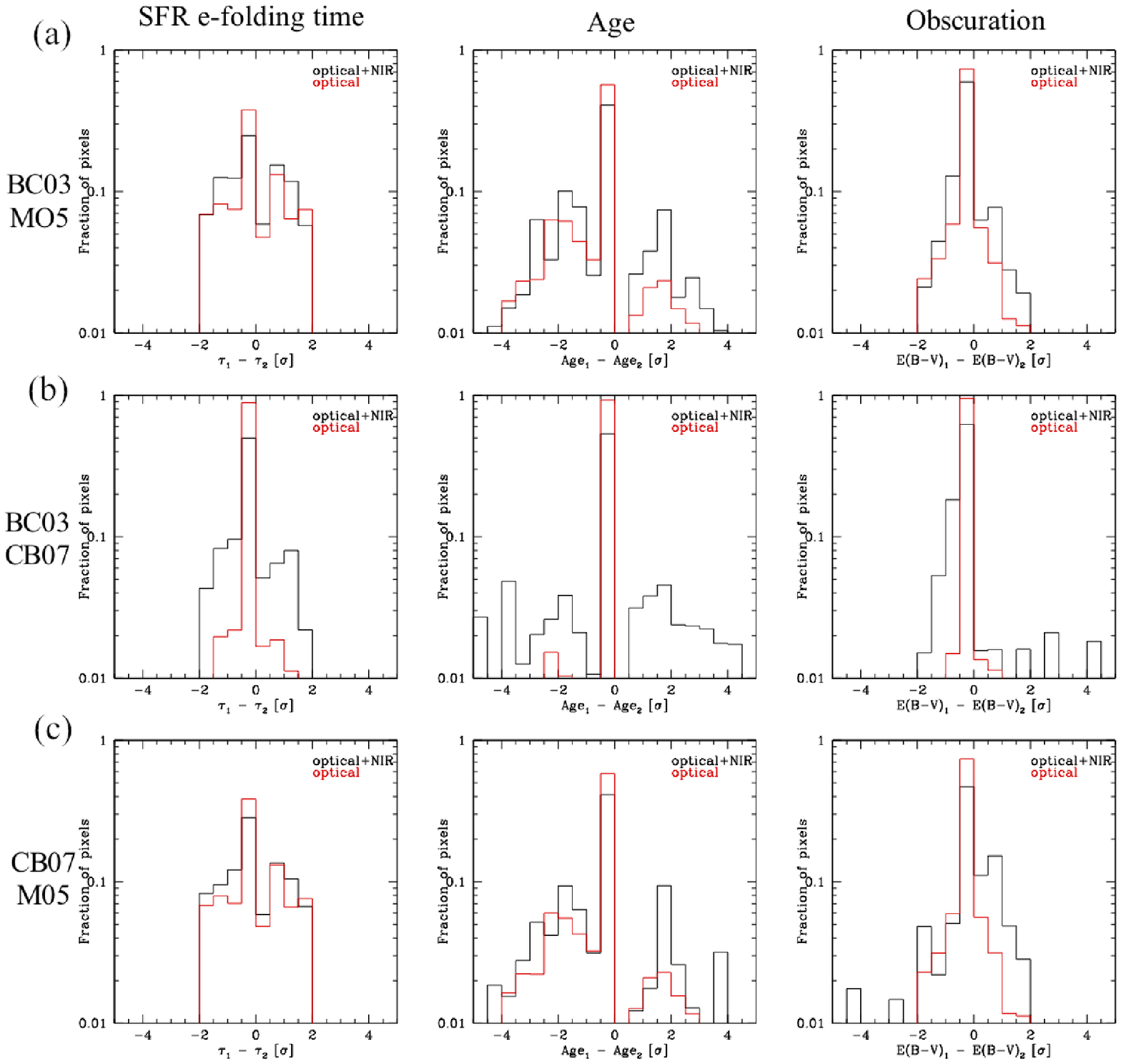}
}
}

\tiny{ 
\caption{ 
The effect of adding NIR passbands on the systematic differences
between stellar population synthesis models, for pixels in a subsample
of $z\le1$ galaxies. Shown are the distributions of systematic
differences in the pixel-$z$ parameters (given in terms of the
statistical error $\sigma$) between pairs of population
synthesis models when using optical-only colors (in red) and optical and
NIR colors (in black). We consider all pixels in
this subsample of $z\le1$ galaxies which have $J$ and $H$ as well as
optical imaging. The distribution of these systematic differences
between each pair of models is investigated for the SFR e-folding time, stellar population age and
dust obscuration. The pairs of models being compared for each
pixel-$z$ parameter are: (a) BC03 and M05, (b) BC03 and CB07 and (c) CB07 and
M05. 
}   

\label{fig:nir_test_zle1}}
\end{figure}

 We investigate this hypothesis fully when we measure systematic differences across all
 pixels in the combined optical-NIR galaxy sample
 for all 3 models considered in this paper. For each parameter, we compare this
 distribution of the systematic differences with that obtained with
 using only the optical data. We account for the effect of redshift,
 particularly because for $z>1$, the
 4000\,\AA \ break shifts into the $J$ and $H$ passbands. We
 thus split the NIR sample into two: $z<1$ and $z>1$.
In Figure~\ref{fig:nir_test_zle1}, we show the
distribution of the systematic difference between models for the low
redshift sample. The fraction of pixels showing $<1\sigma$
uncertainties in their pixel-$z$ parameters for each pair of models
is given in Table 4 for both the optical-only colors and the optical
and NIR colors. For the low redshift sample, there is a very high
level of agreement between the BC03 and CB07 models in all the pixel-$z$ parameters,
with over $97$ percent of pixels showing less than $1\sigma$
deviation between the two models. With the addition of the NIR bands
however, the systematic differences between BC03 and CB07 models are
exposed.  This is a result of relatively small differences among the
models in their predicted rest-frame UV colors but much larger
differences among them in their predicted rest-frame optical (red) and
rest-frame NIR colors. This affects in particular the stellar population age (with
only 78 percent of pixels showing less than $1\sigma$ difference
between the two models) and $\tau$ (88 percent of pixels showing
agreement between the two models). The obscuration is less impacted by
the model differences. The addition of the NIR colors also exposes 
differences between BC03 and M05, and this impacts the stellar
population age most strongly, with the fraction of pixels showing
$<1\sigma$ difference in the age decreasing from 0.89 with
optical-only colors to 0.80 with NIR colors. The differences between
BC03 and M05 have less impact on $\tau$ and the obscuration when the
NIR colors are added. When
comparing CB07 with M05, it is again the stellar population age in the
pixels that is most affected by the addition of the NIR passbands,
with only 80 percent of pixels showing an agreement in the age to
within $1\sigma$ between the two models, compared to 89 percent for
the optical-only case.

\begin{figure}
\centerline{\rotatebox{0}{
\includegraphics[width=1.0\textwidth]{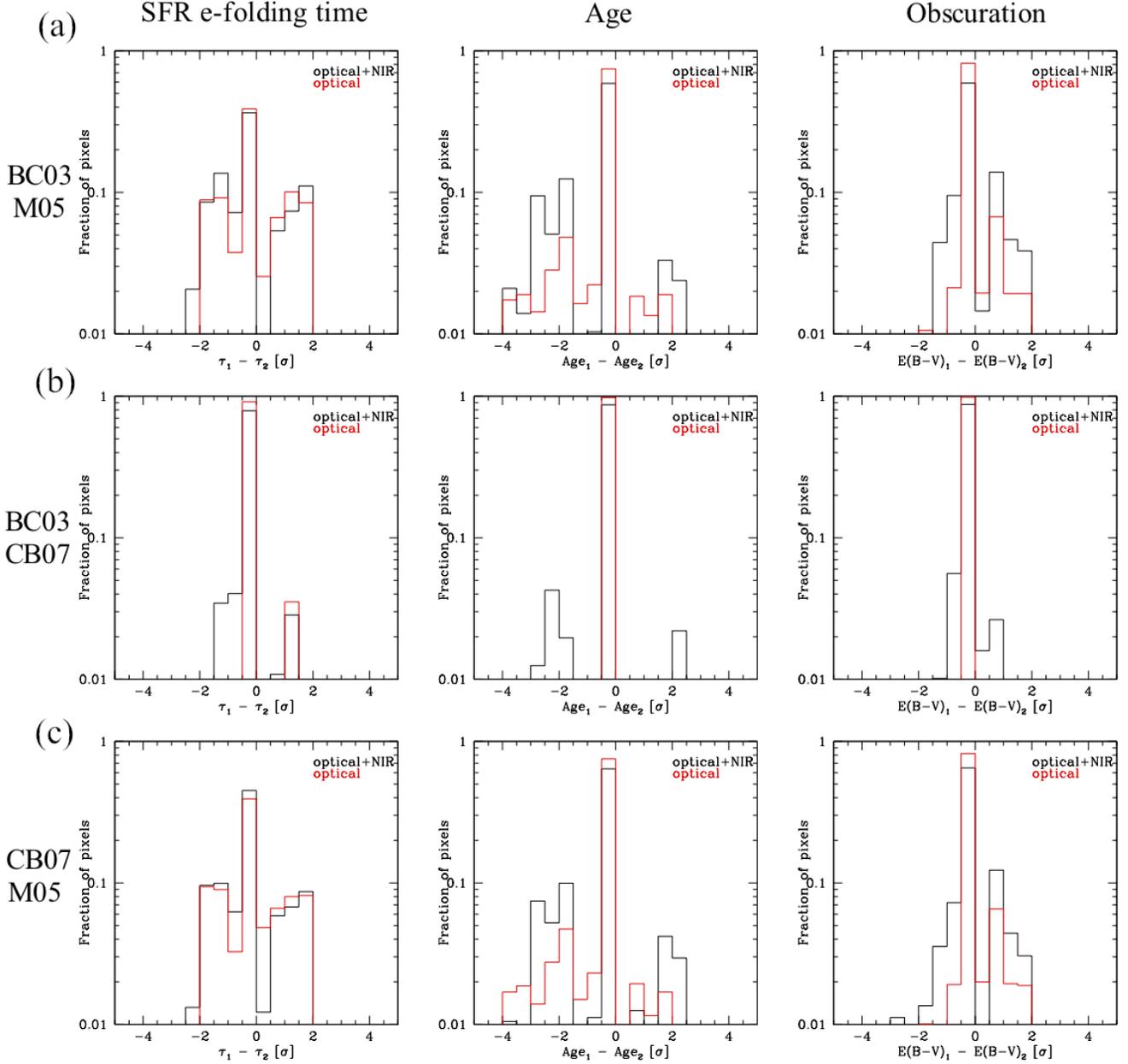}
}
}

\tiny{ 
\caption{ 
The effect of adding NIR passbands on the systematic differences
between stellar population synthesis models, for pixels in a subsample
of $z>1$ galaxies. Shown are the distribution of systematic
differences (given in terms of the statistical error $\sigma$) in the pixel-$z$ parameters between pairs of population
synthesis models when using optical-only colors (in red) and optical and
NIR colors (in black). We consider all pixels in
this subsample of $z>1$ galaxies which have $J$ and $H$ as well as
optical imaging. The distribution of these systematic differences
between each pair of models is investigated for the SFR e-folding time, stellar population age and
dust obscuration. The pairs of models being compared for each
pixel-$z$ parameter are: (a) BC03 and M05, (b) BC03 and CB07 and (c) CB07 and
M05. 
}   
\label{fig:nir_test_zgt1}}
\end{figure}

For the high-redshift ($z>1$) sample, the story is somewhat
different, as illustrated in Figure~\ref{fig:nir_test_zgt1} and Table
5. There is now a much higher level of agreement between the results
for the optical-only colors and the optical and NIR colors
combined, for all the parameters concerned. This is due to the fact that the rest-frame UV and blue
wavelength optical spectra (which are sampled by the optical and the NIR
filters for $z>1$) are similar among the models. This similarity 
in the predicted rest-frame UV/optical (blue) colors among the models
thus manifests itself as follows.  In the $z<1$
sample discussed above, the stellar population age showed a large
discrepancy between the $bviz$ and $bvizJH$ samples in terms of
the fraction of pixels that agree on their age for the different
models being compared. For the high redshift sample, the pixels in
both the $bviz$ and $bvizJH$ samples show similar levels of agreement
in the age parameter between the different models compared.  For
example, for the the BC03 and M05 models, the fraction of pixels with
systematic differences in their age that are smaller than $1\sigma$ is
0.95 in the optical-only case and 0.93 in the NIR case. 
The conclusion of these tests is that, for $z>1$ galaxies, differences
among the stellar population synthesis models in their predicted NIR colors do not significantly bias the 
pixel-$z$ estimates of the stellar population parameters in the
pixels, compared to using optical colors only.

\subsection{Global Galaxy Properties and the Effect of Systematics
}

Pixel-$z$ can measure global properties of structures
composed of many pixels within the same galaxy. These will be used to
determine the radial variation of stellar population parameters in
galaxies as a function of the galaxy environment, and also to allow a
detailed study of high redshift clumpy, disk galaxies, as outlined in
Section 1.1 and 1.2. Here, we investigate the effect of systematic
biases that can impact these global measurements, in particular of the
SFR since this will be a direct observable in spectroscopic measurements.

\subsubsection{Biases from using pixel-$z$ versus global SED-fitting} 

\begin{figure}
\centerline{\rotatebox{0}{
\includegraphics[width=0.6\textwidth]{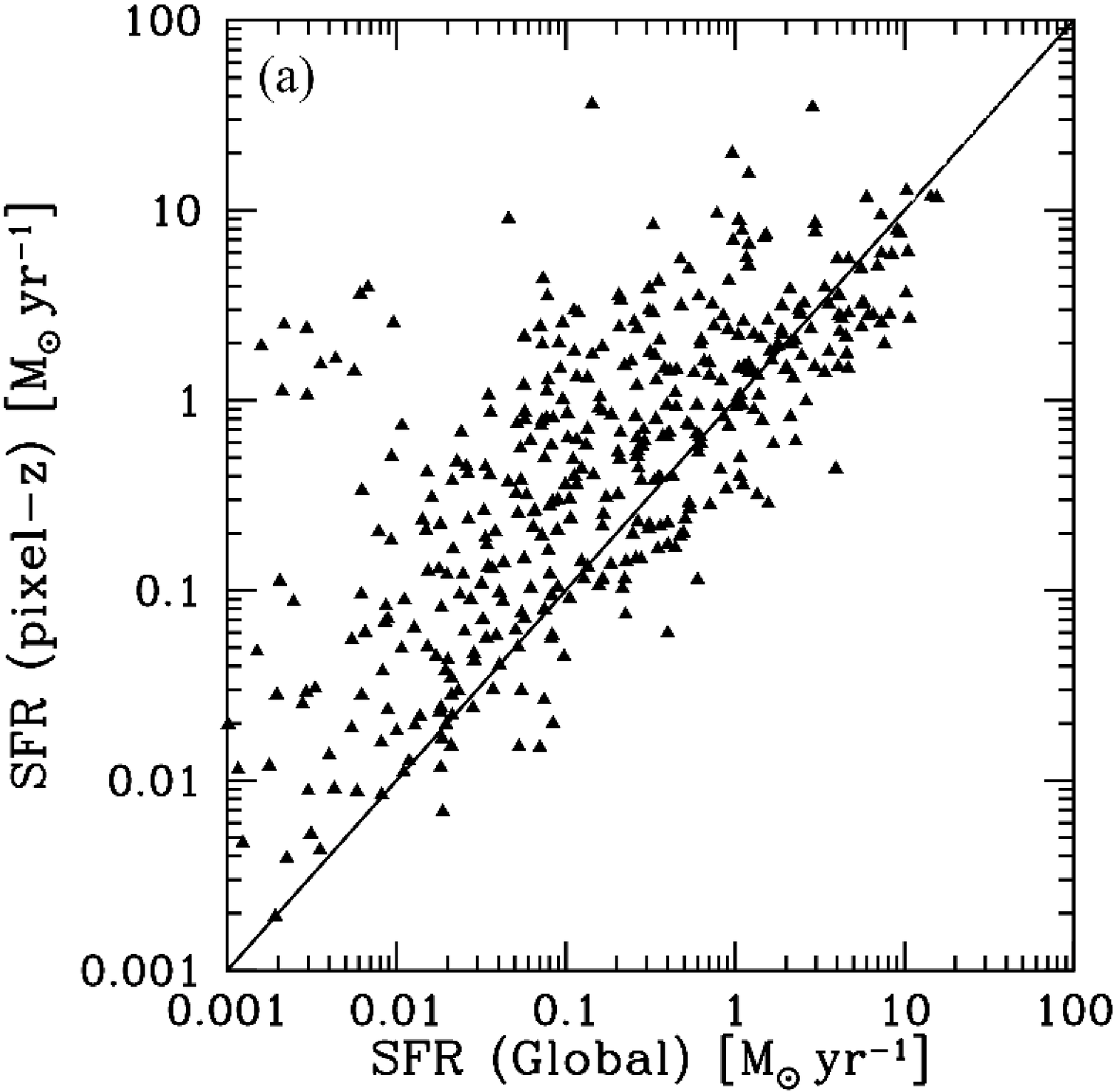}
}}
\vspace{0.1cm}
\centerline{\rotatebox{0}{
\includegraphics[width=0.6\textwidth]{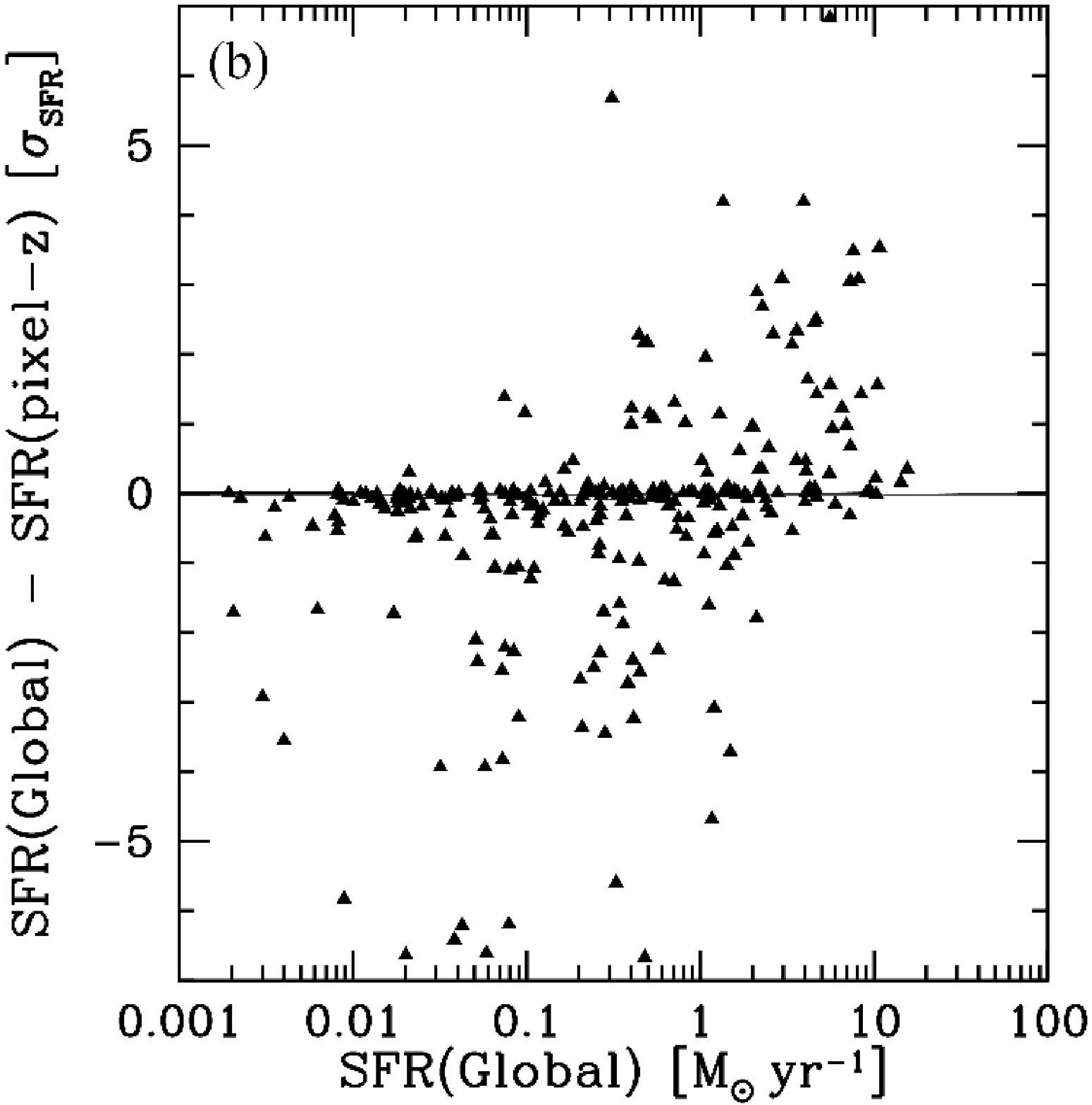}
}
}
\tiny{
\caption{
Comparison of the sum of SED fits to the individual colors in pixels
in each galaxy (as found by pixel-$z$) with the SFR derived from global SED-fitting 
to the integrated colors of the galaxy. For each galaxy, the same pixels within a
$0.5"$ aperture were used in both approaches. (a) The total SFR of galaxies
derived from pixel-$z$ as a function of the SFR derived from the global
SED fit to the integrated colors in the galaxies. (b) The residual
difference between the two approaches, given in terms of the statistical
error in the SFR measurement of the galaxy,  as a function of the SFR
derived from the global SED fit. 
}
 \label{fig:SFR_pixelz_global}}
\end{figure}

In Figure~\ref{fig:SFR_pixelz_global}, we show the
level of consistency between the sum of SED fits to individual pixels
in a galaxy and the global fit to the total flux in those same pixels,
as described in Section 2.3.2. We see a high level of consistency between the
two approaches in their estimates of the integrated SFR. 
We also quantify the magnitude of any deviation
between the two approaches in terms of the statistical uncertainty in
the SFR estimate for each galaxy from SED-fitting. We see that for the
majority of galaxies in the sample, the difference between the two SFR
estimates for the galaxies are within the statistical uncertainties of
the SFR measurement. However, there is a small population of outliers ($>2\sigma$) 
which are galaxies with low SFR ($<1 M_{\odot}\,yr^{-1}$) and which show a negative
residual i.e., the SFR estimates from pixel-$z$, obtained by summing
the results of the individual fits in each pixel, are higher than the
SFR estimates from global fits to the total flux in the pixels. In addition, there is a second small population of
outliers which are galaxies with high SFRs ($\sim10
M_{\odot}\,yr^{-1}$). In these, the SFR estimate from the global fit is significantly higher than that
estimated from the sum of the individual SED fits.

\subsubsection{Radial gradients of stellar populations: age, dust and star formation}

\begin{figure}
\centerline{\rotatebox{0}{
\includegraphics[width=.4\textwidth]{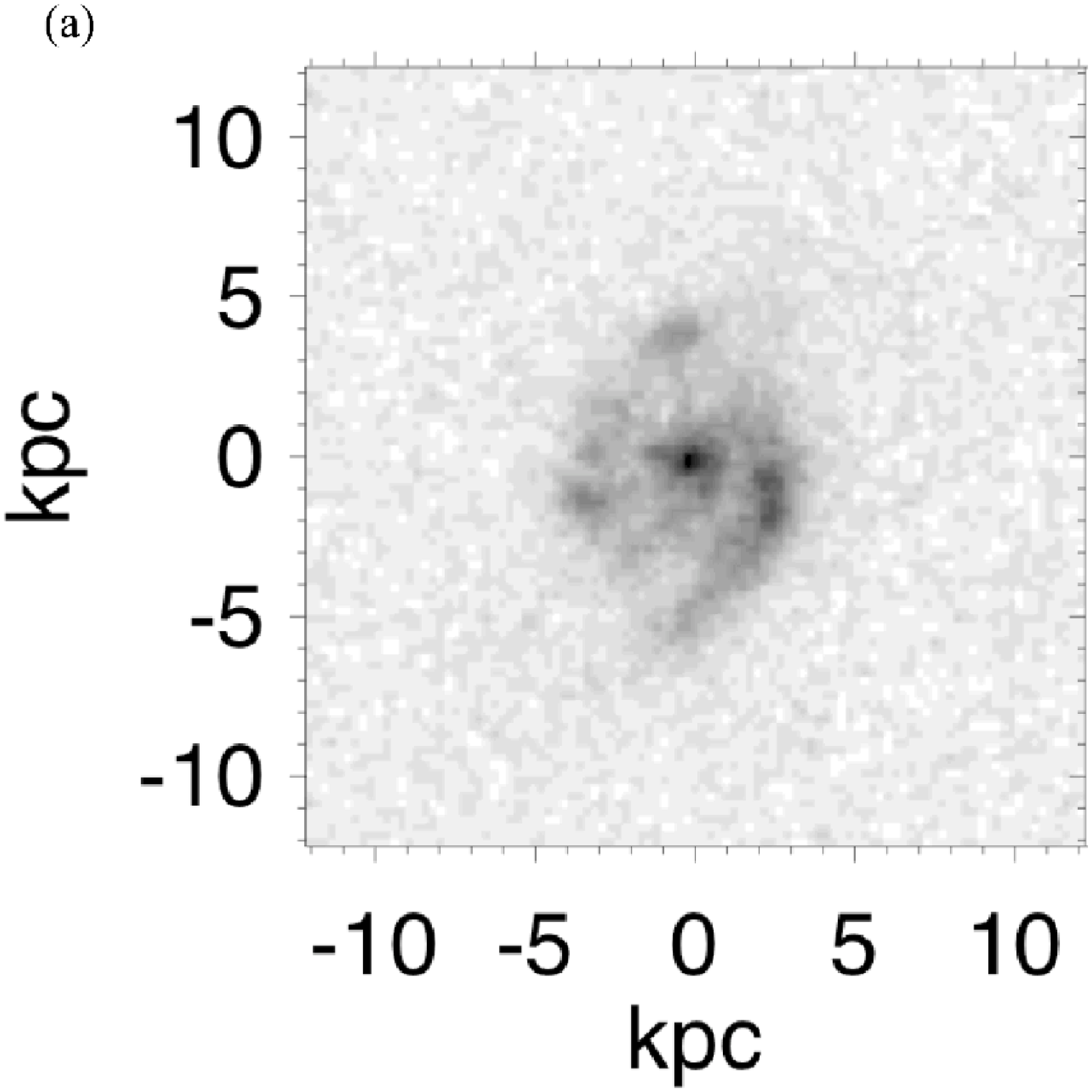}
}
}
\vspace{0.7cm}
\centerline{\rotatebox{0}{
\includegraphics[width=.35\textwidth]{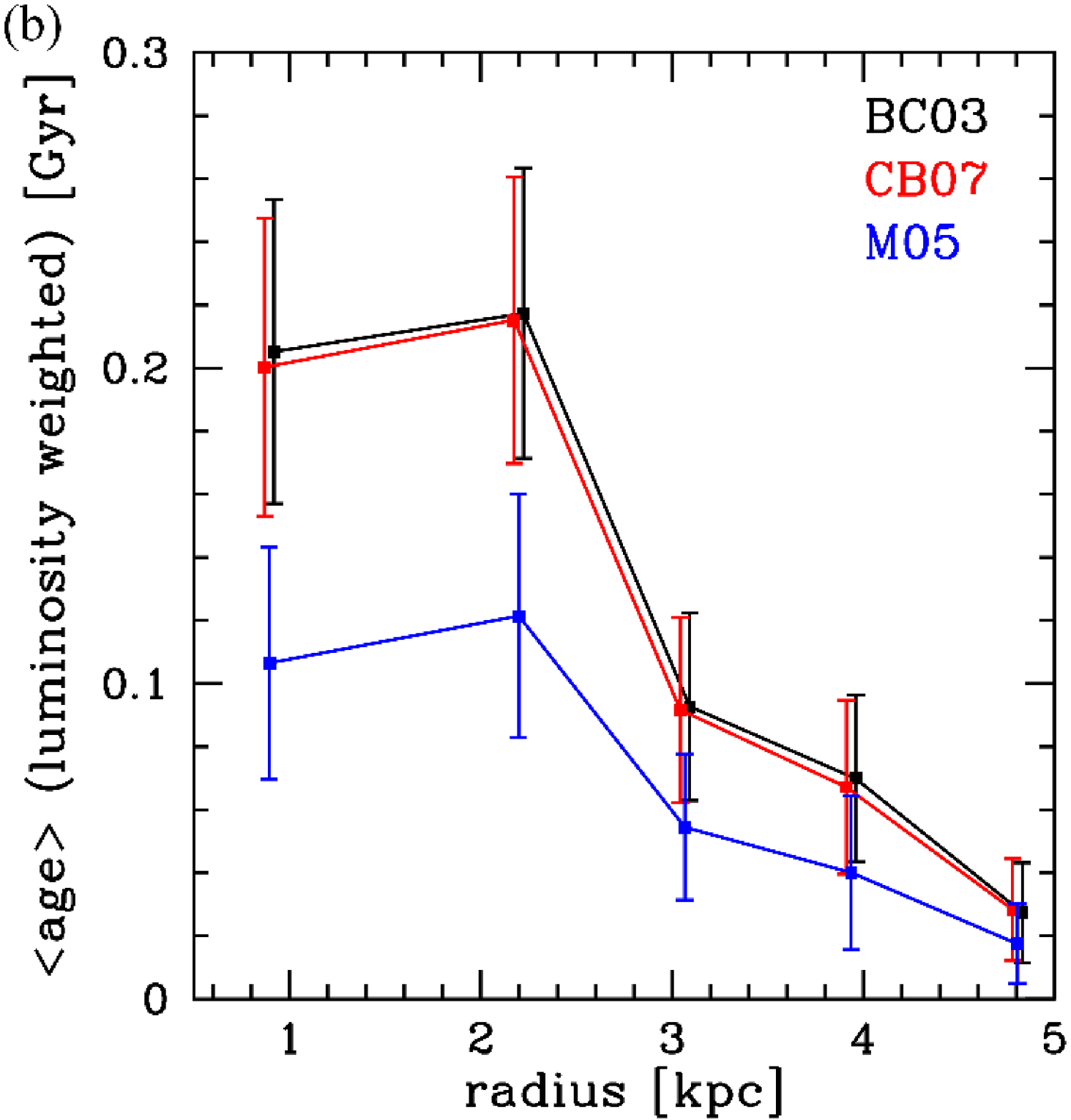}
\hspace{0.1cm}
\includegraphics[width=.35\textwidth]{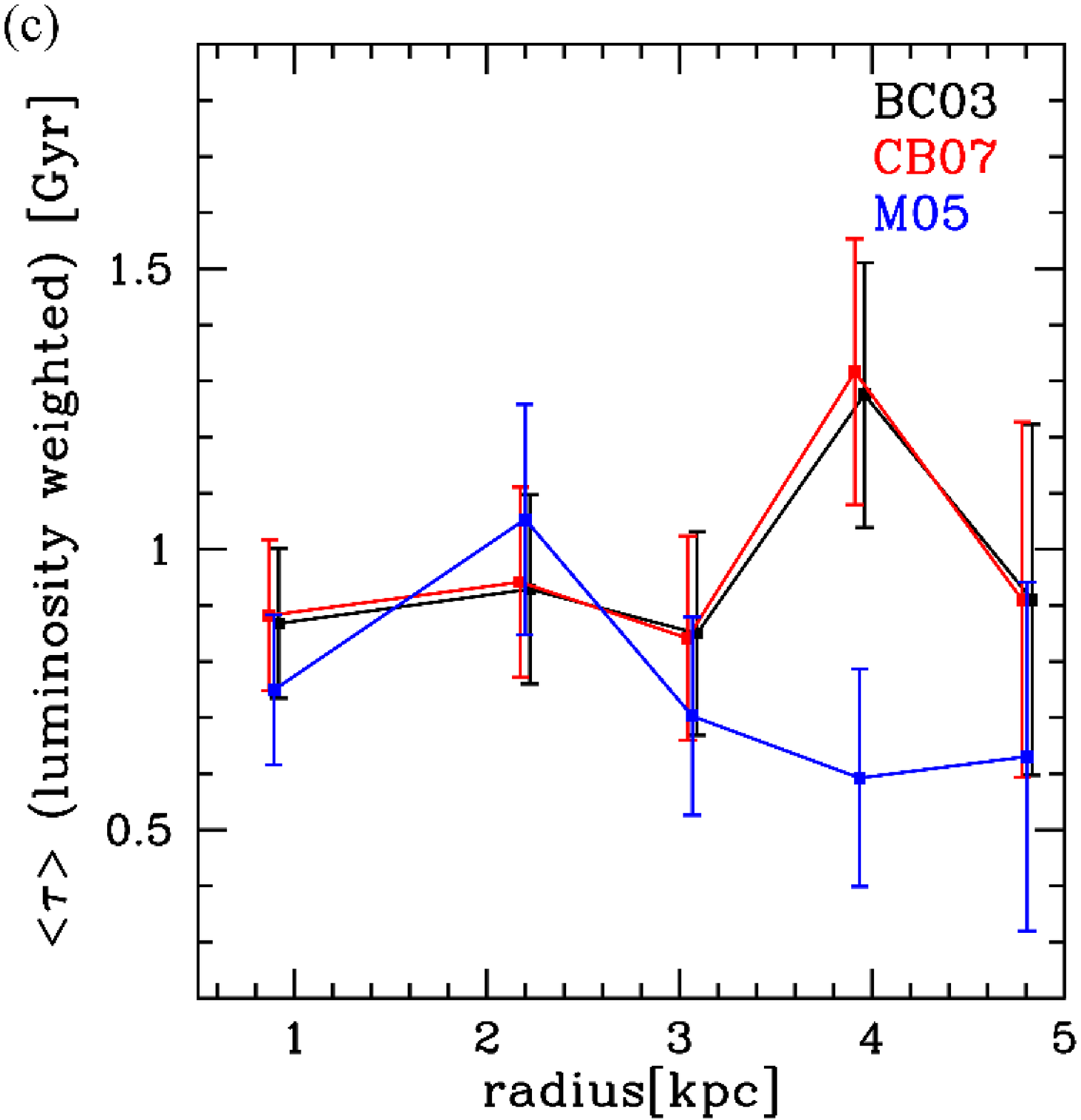}
\hspace{0.1cm}
\includegraphics[width=.35\textwidth]{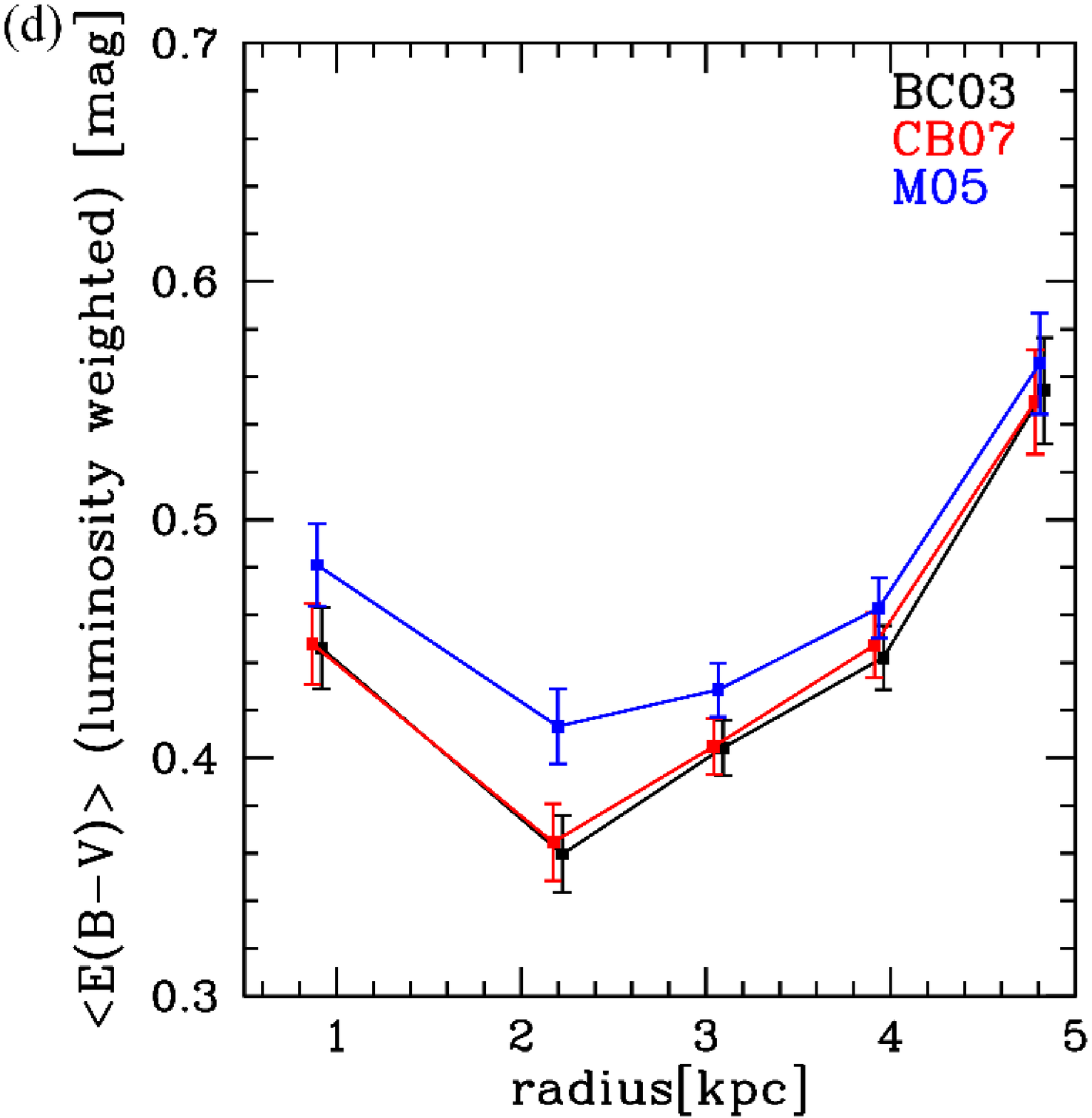}
}
}
\vspace{0.5cm}
\centerline{\rotatebox{0}{
\includegraphics[width=.35\textwidth]{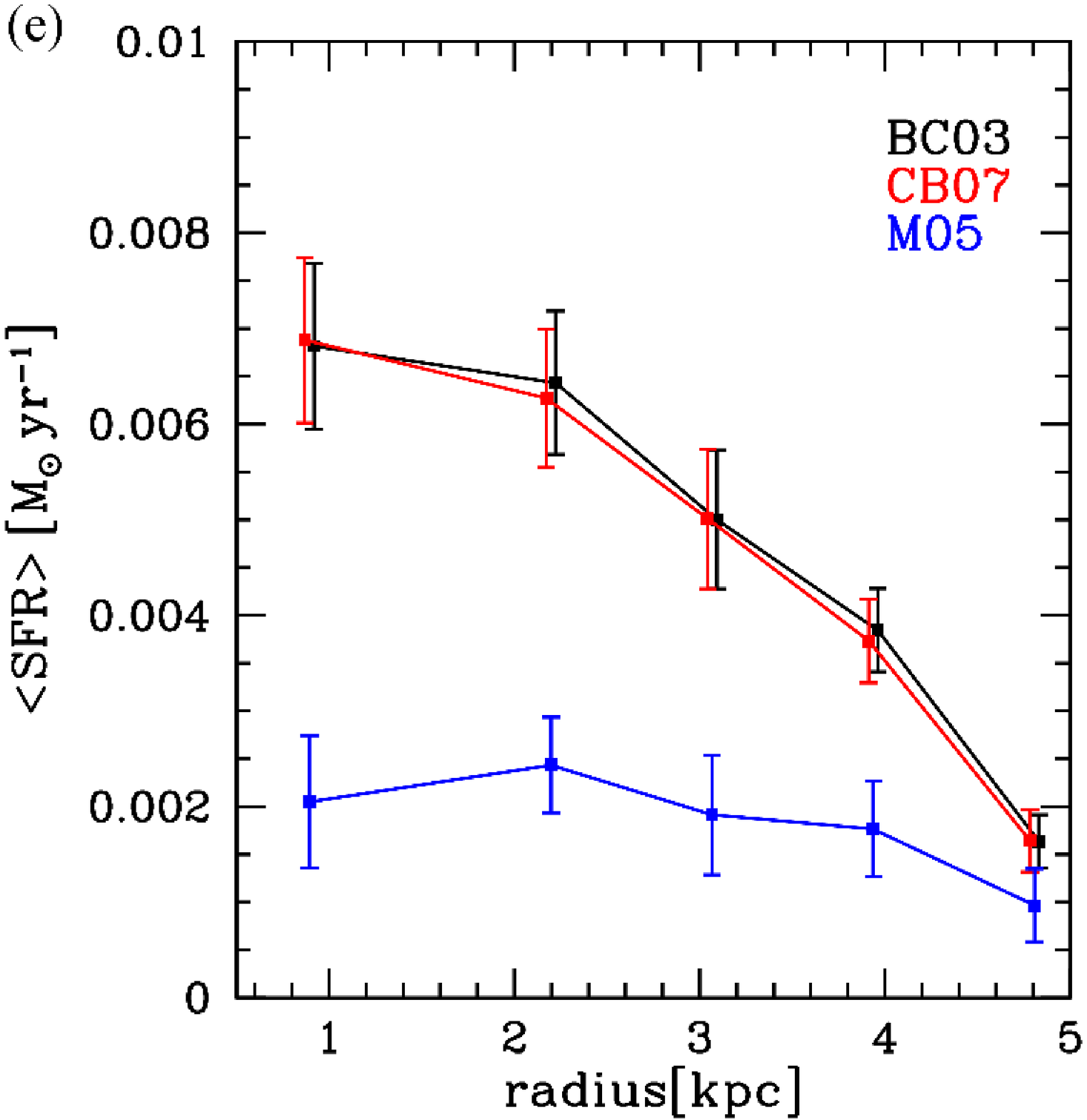}
\hspace{0.1cm}
\includegraphics[width=.35\textwidth]{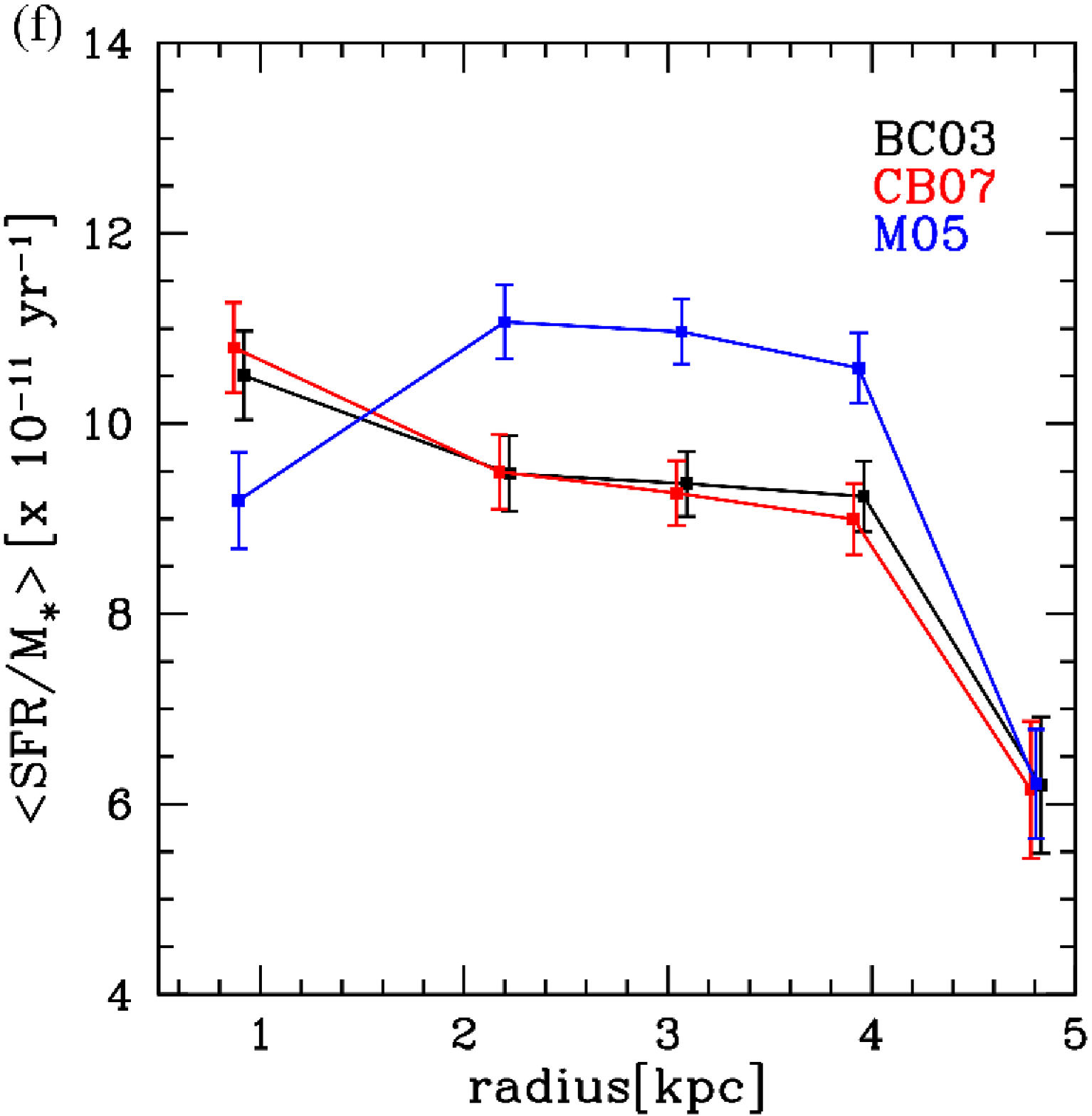}
}
}
\tiny{
\caption{
The effect of stellar population synthesis model differences on the
the radial variation of  stellar population quantities in a disk
galaxy at $z=1.1$. (a) The $i775$ image of the galaxy. The following
panels show the radial variation of (b) the luminosity-weighted
stellar population age (c) the luminosity-weighted SFR e-folding time
$\tau$ (d) the dust obscuration given in terms of $E(B-V)$ (e) the mean SFR and
(f) and the mean specific SFR (sSFR) across the galaxy. The models tested are BC03 (black), CB07 (red)
  and M05 (in blue). The innermost annulus is larger
  than the FWHM of the PSF in the $i775$ image ($0.11''$). 
}   
\label{fig:radialgradient_models}}
\end{figure}

In Figure~\ref{fig:radialgradient_models}, we present the radial
variation of the stellar population age, SFR e-folding time,
obscuration, SFR and sSFR using the $bviz$
passbands for the same galaxy at $z\sim1$ in
Figures~\ref{fig:models_optical_age} and~\ref{fig:models_optical_ebv}. The radial trends
in each quantity are measured according to
the method described in section 2.3.3. All stellar population
synthesis models agree on the
qualitative trends of all the parameters with radius but there are
some quantitative differences that arise, particularly between the
different population synthesis models All three models agree that the stellar populations in the galaxy at all radii are young ($<0.3$Gyr) and that
the stellar population age decreases rapidly from the center to the
outskirts. M05 predicts a mean stellar population age that is somewhat lower in the
innermost part of the galaxy compared to BC03 and CB07, and this
systematic difference is of the order of the statistical uncertainty
in the age measurement. BC03 and CB07 also predict a somewhat sharper drop in the stellar population age
between 2 and 3 kpc, but again, the difference is well within the statistical uncertainties.

The models agree on a range of $\tau\sim0.5-1$ Gyr across the
galaxy. Between 3 and 5 kpc, however, the M05 model prediction
diverges by $1\sigma$ from both BC03 and CB07 models. This region
corresponds to the spiral arm and the clumps of star-forming HII
regions within it. This implies that the different treatment of TP-AGB
stars in M05 compared to the other models has its biggest effect in
the spiral arms of the galaxy. The dust obscuration trends are quite
similar among the models which show that the center of the galaxy is
moderately obscured ($E(B-V)\sim0.45$). The obscuration then decreases
up to 2 kpc and then increases to a maximum ($E(B-V)\sim0.6$) in the
spiral arms. 

The different models all find that the mean SFR in
each annulus peaks in the center of the galaxy. To explain
this, we note that the current SFR is determined not only by the mean age and
$\tau$ of the stellar populations, but also by the total stellar mass
formed to date, which normalizes the SFR. M05 predicts a lower
stellar mass throughout the galaxy than does BC03 or CB07. When the
stellar mass is normalized out to give the specific SFR (sSFR), we see
that the discrepancy between M05 and BC03/CB07 narrows considerably:
$1\sigma$ at most in all radial bins. Comparing the trend in sSFR with the
morphology of the galaxy, we conclude that the bar structure in the
center of the galaxy is actively star-forming. The
models disagree, however, on how much more star-forming the bar is relative to the
spiral arms at a radius of $\sim4$kpc. BC03 and CB07 predict a slightly higher sSFR
(by around $2\times10^{-11}\,yr^{-1}$) in the center of the galaxy
compared to the spiral arms, whereas M05 predicts a somewhat lower sSFR
in the center relative to the spiral arms. Nonetheless, a general picture
emerges for this galaxy, which is that star formation is not simply
confined to the center but takes place throughout the galactic disk.
The results, albeit for a single galaxy, imply that pixel-$z$ is robust to differences 
between various stellar population synthesis codes when it is used 
to measure radial variations across galaxies. 

\begin{figure}
\centerline{\rotatebox{0}{
\includegraphics[width=.4\textwidth]{fg15aarxiv.eps}
}
}
\vspace{0.7cm}
\centerline{\rotatebox{0}{
\includegraphics[width=.4\textwidth]{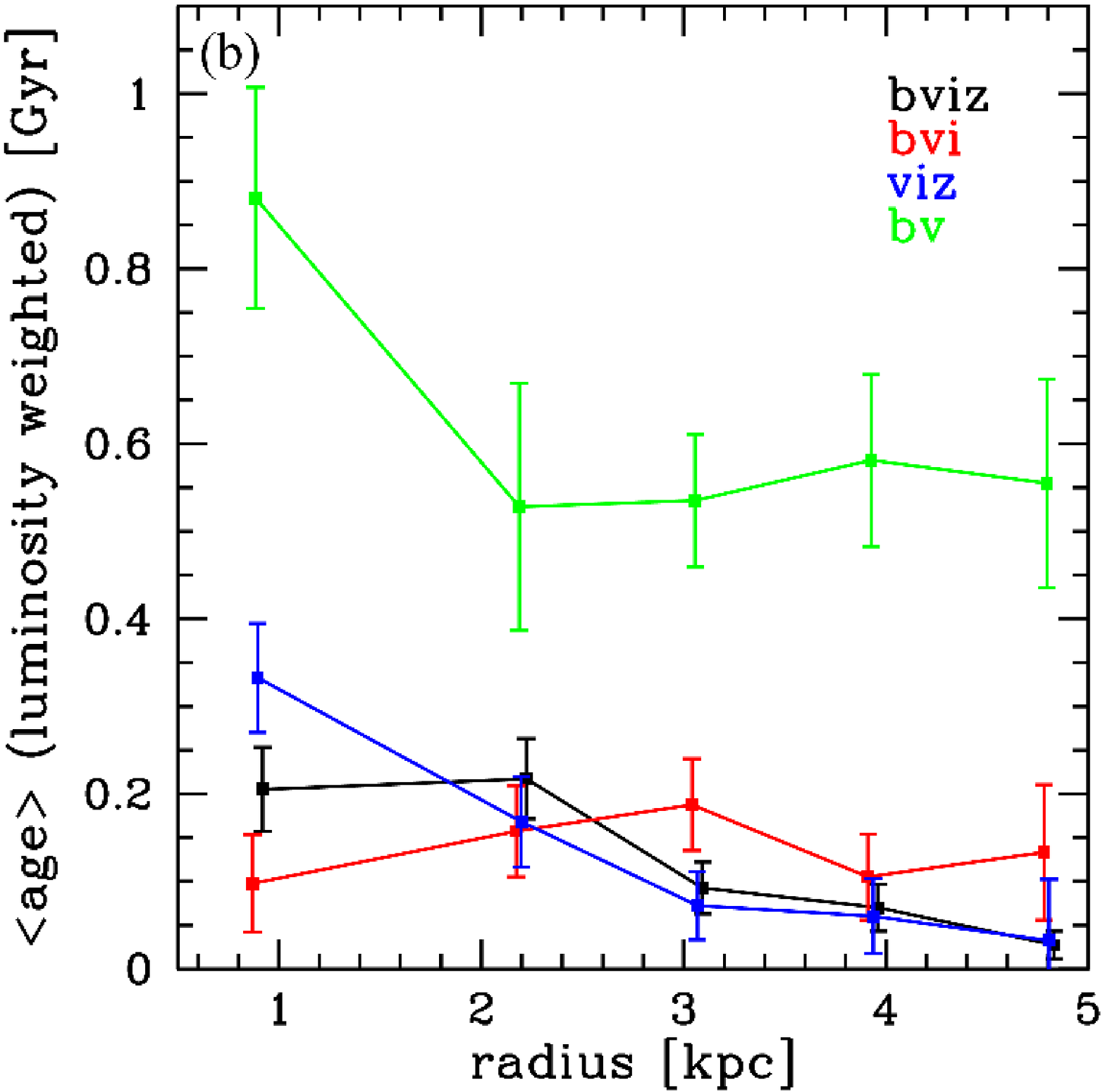}
\hspace{0.1cm}
\includegraphics[width=.4\textwidth]{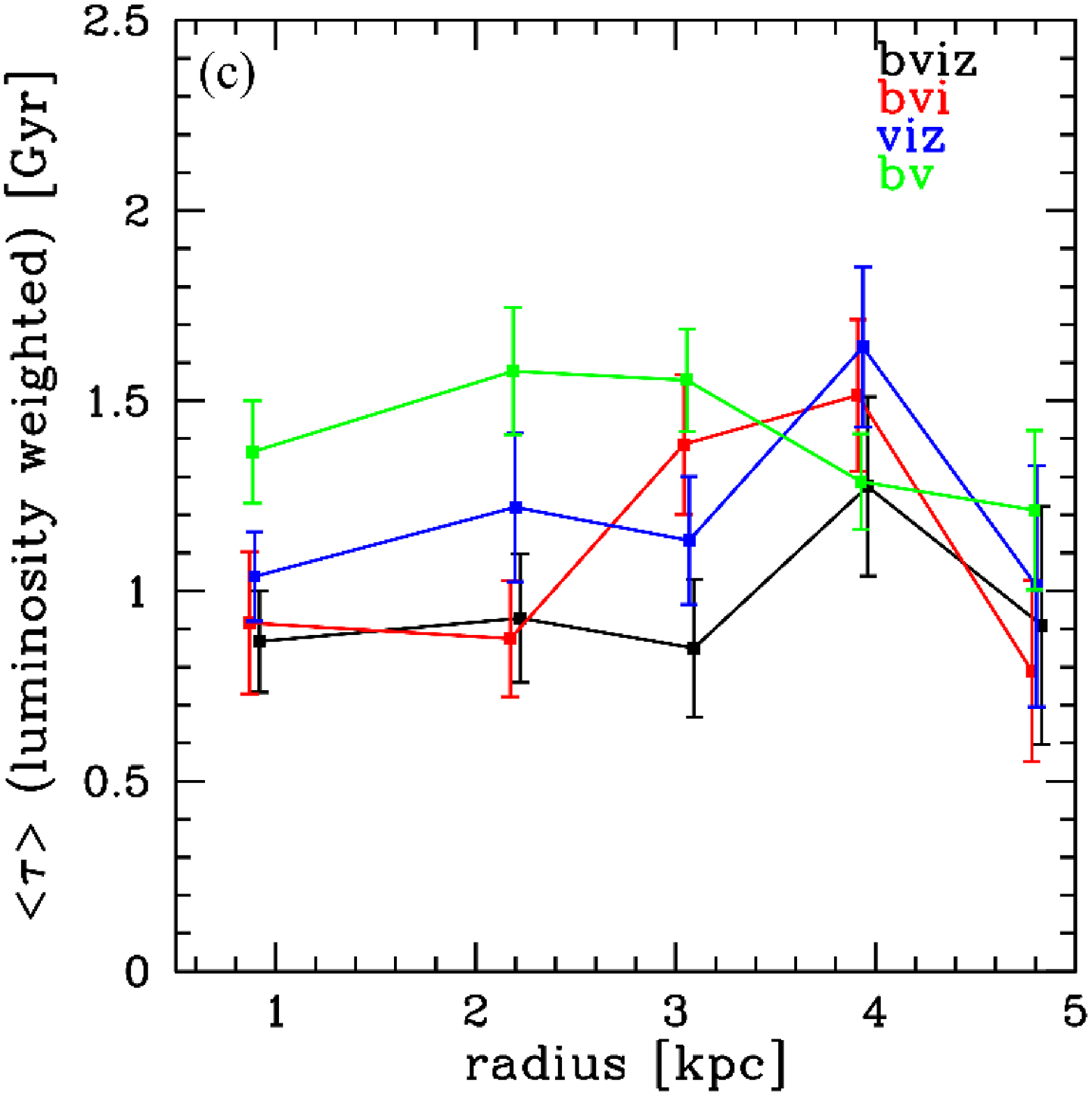}
}
}
\vspace{0.5cm}
\centerline{\rotatebox{0}{
\includegraphics[width=.4\textwidth]{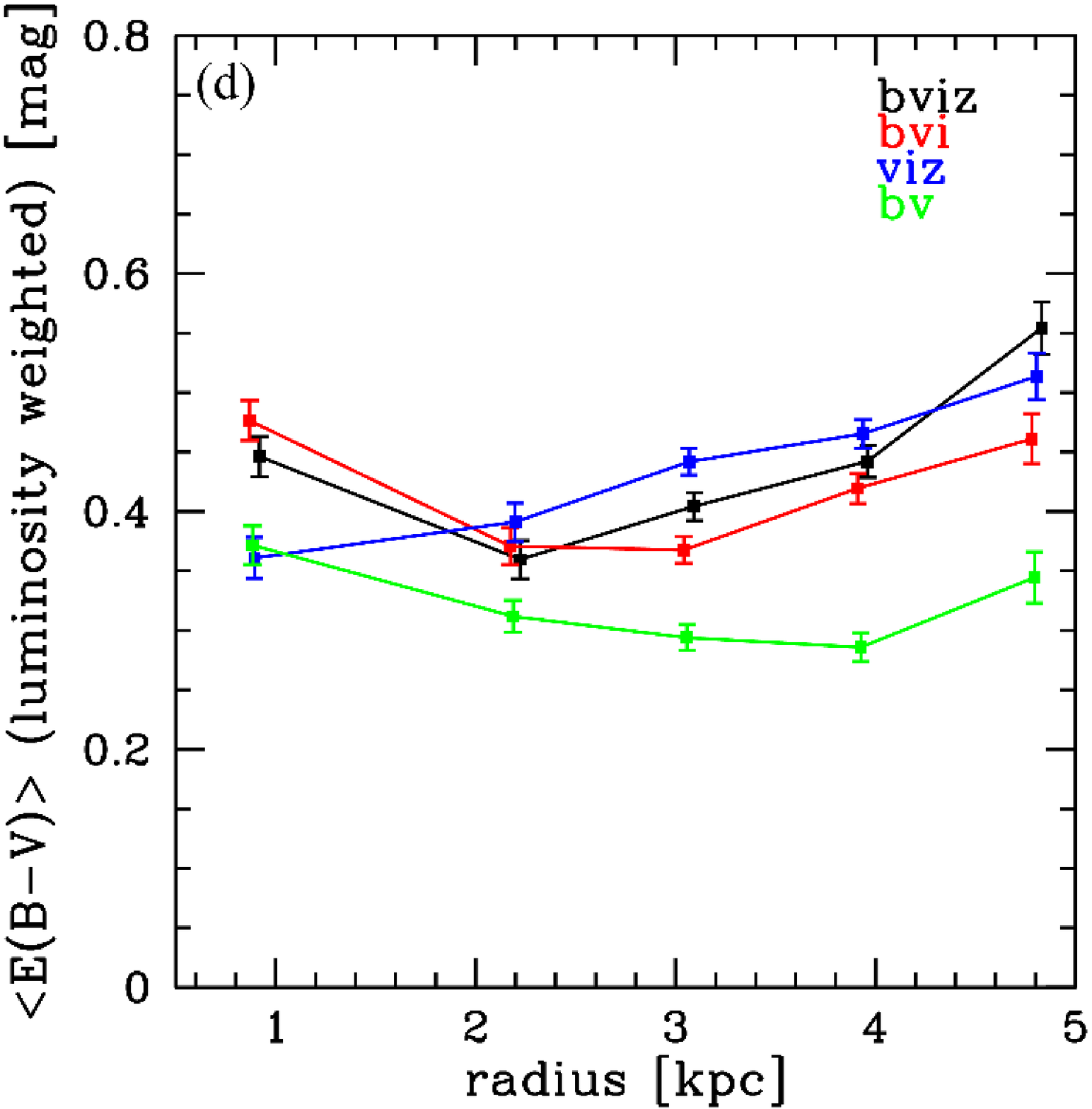}
\hspace{0.1cm}
\includegraphics[width=.4\textwidth]{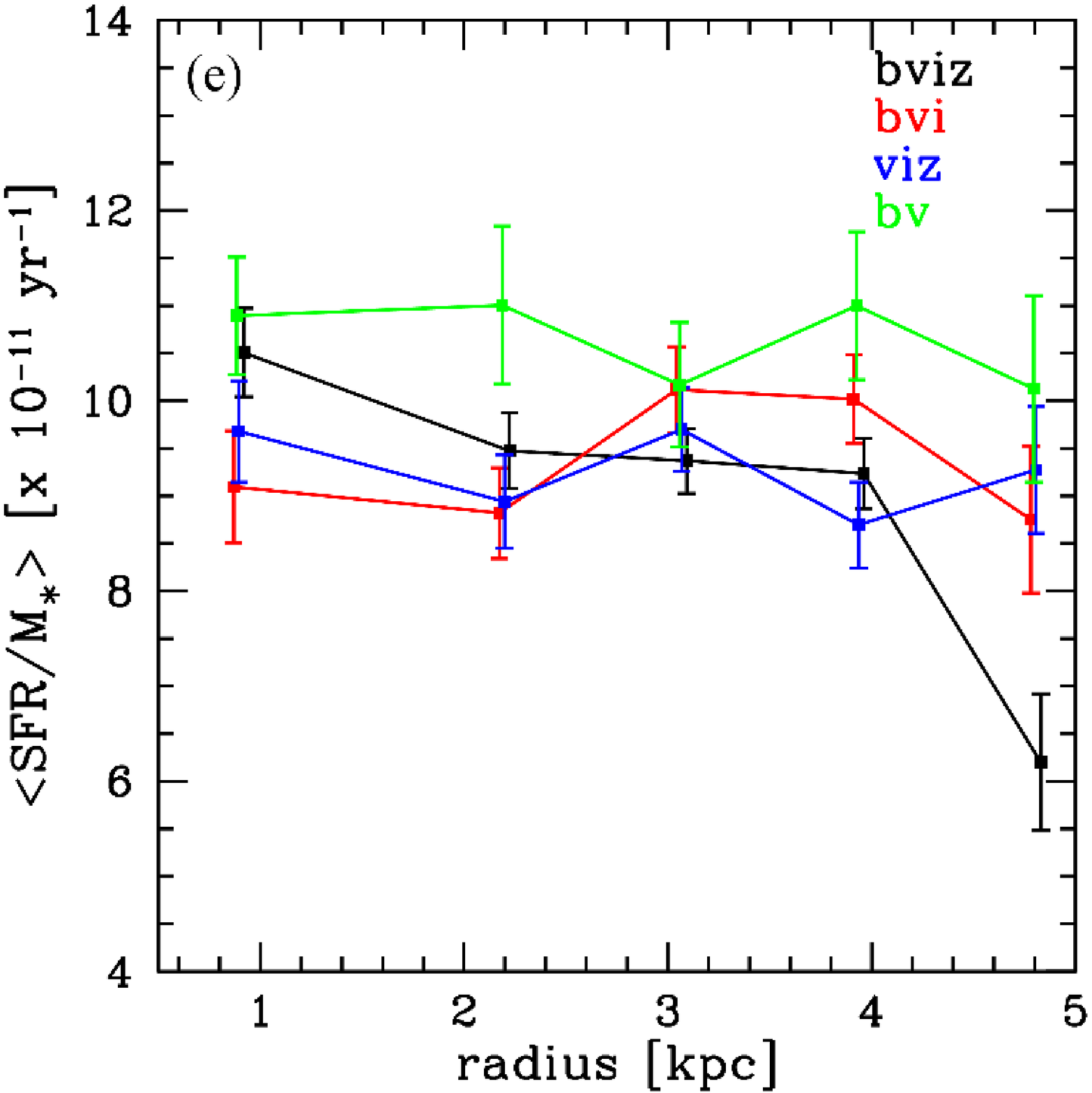}
}
}
\tiny{
\caption{
The effect of passband differences on the
the radial variation of  stellar population quantities in a disk
galaxy at $z=1.1$ for a fixed population synthesis model (BC03). (a)
The $i775$ image of the galaxy. The following panels show the radial variation of (b) the luminosity-weighted
stellar population age (c) the luminosity-weighted SFR e-folding time
$\tau$ (d) the dust obscuration given in terms of $E(B-V)$ (e) the
mean specific SFR (sSFR) across the galaxy. The passbands tested are
$bviz$ (black), $bvi$ (red), $viz$ (blue) and $bv$ (green). The innermost annulus is larger
  than the FWHM of the PSF in the $i775$ image ($0.11''$). 
}   
\label{fig:radialgradient_passbands}}
\end{figure}

Figure~\ref{fig:radialgradient_passbands} shows the effect of passband
limitations on measuring radial variations of stellar population
parameters. The population synthesis model was fixed to BC03. Using only the $b$ and
$v$ bands results in large systematic
uncertainties which affects all the radial quantities but particularly
the mean age and the obscuration
determinations. The result of using only the $b$ and $v$ bands 
is that the mean age is overestimated throughout the galaxy relative
to the $bviz$ filter set, by as much as $4\sigma$. This can be explained by
the fact at $z\sim1$, the 4000\,\AA \ break passes out of the
bluest optical bands and, as a result, the colors predicted by the
model for the pixels are redder, leading to an older stellar
population for the pixels. The obscuration
is underestimated throughout, and at a radius of 4 kpc the difference with respect
to the $bviz$ value is as much as $4\sigma$. The $\tau$ and the sSFR
estimates are less impacted by the loss of 2 passbands. 

The 3-passband combinations $bvi$ and $viz$ both come closest to
reproducing the $bviz$ radial trends in all the parameters. The
systematic difference between the $bvi$ and $viz$ combinations is
relatively small, but some differences do emerge nevertheless. In the
center of the galaxy, the $viz$ filter combination predicts a larger
(by $1\sigma$) mean stellar population age than the bluer $bvi$ bands. Elsewhere in the galaxy, the three-filter
combinations give results which are consistent (within their statistical
uncertainties) with each other and also with the full $bviz$ filter
set. The radial variation of the dust obscuration is consistent among
the $bvi$, $viz$ and $bviz$ filter sets except in the innermost
annulus where the $viz$ filters underpredict the level of obscuration
relative to the other two combinations. The differences in $\tau$ are
small, although there is one important difference at $r=3$kpc where the $bvi$ set predicts a higher $\tau$
compared to the $viz$ and $bviz$ sets. The effect of filter
cominbations on the sSFR is also relatively small throughout the
galaxy except in its outermost part ($r\sim5$kpc) where the three-filter combinations
overestimate the sSFR relative to the full optical set.

If we were to generalize this test, a minimum of 3 optical
passbands are therefore needed for robustly computing radial trends of
stellar populations in these high redshift galaxies.
 
\subsubsection{Disk galaxies at $z\gtrsim1$: spiral arms and clumps}

\begin{figure}
\centerline{\rotatebox{0}{
\includegraphics[width=0.7\textwidth]{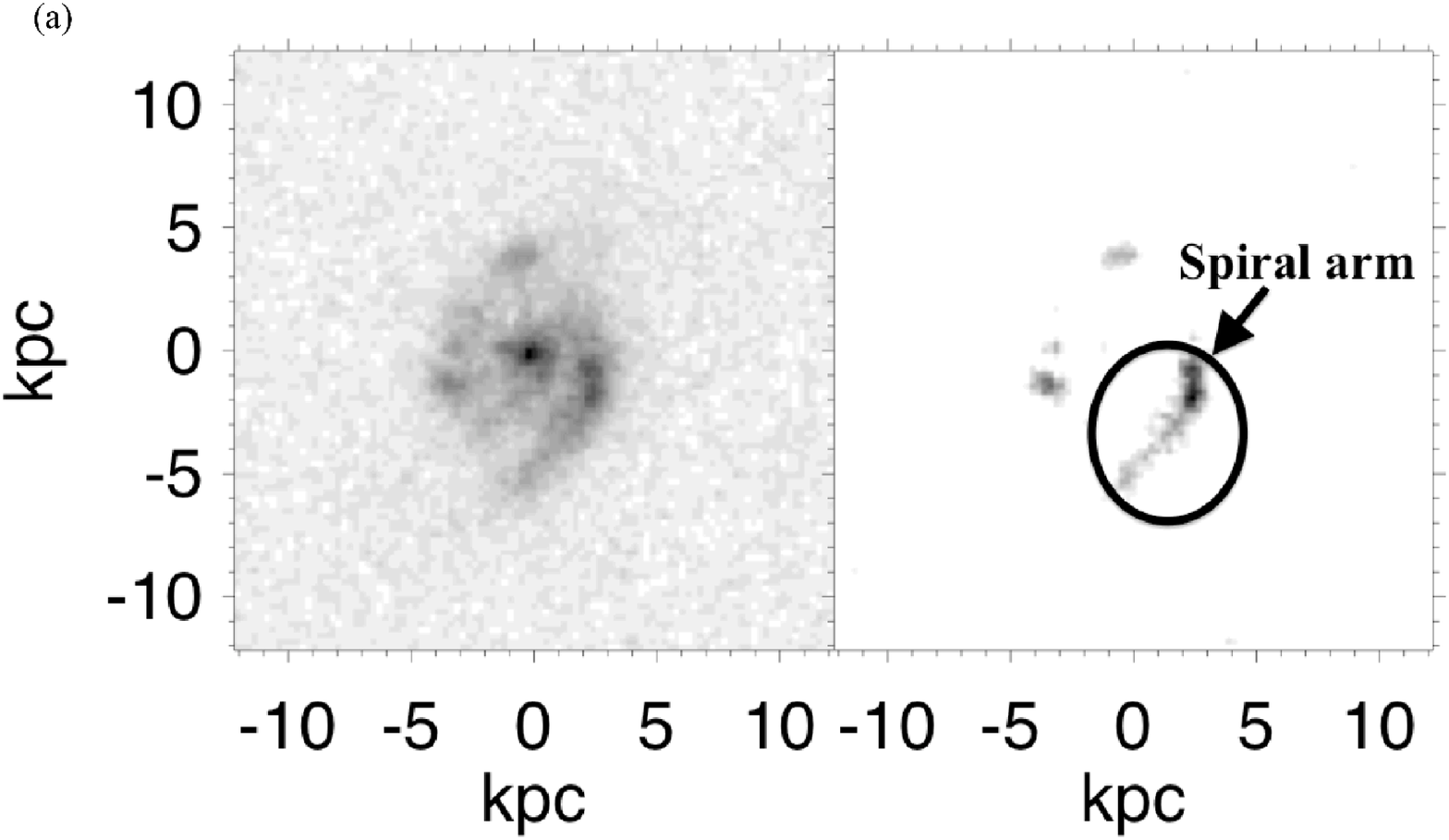}
}
}
\vspace{0.1cm}
\centerline{\rotatebox{0}{
\includegraphics[width=.4\textwidth]{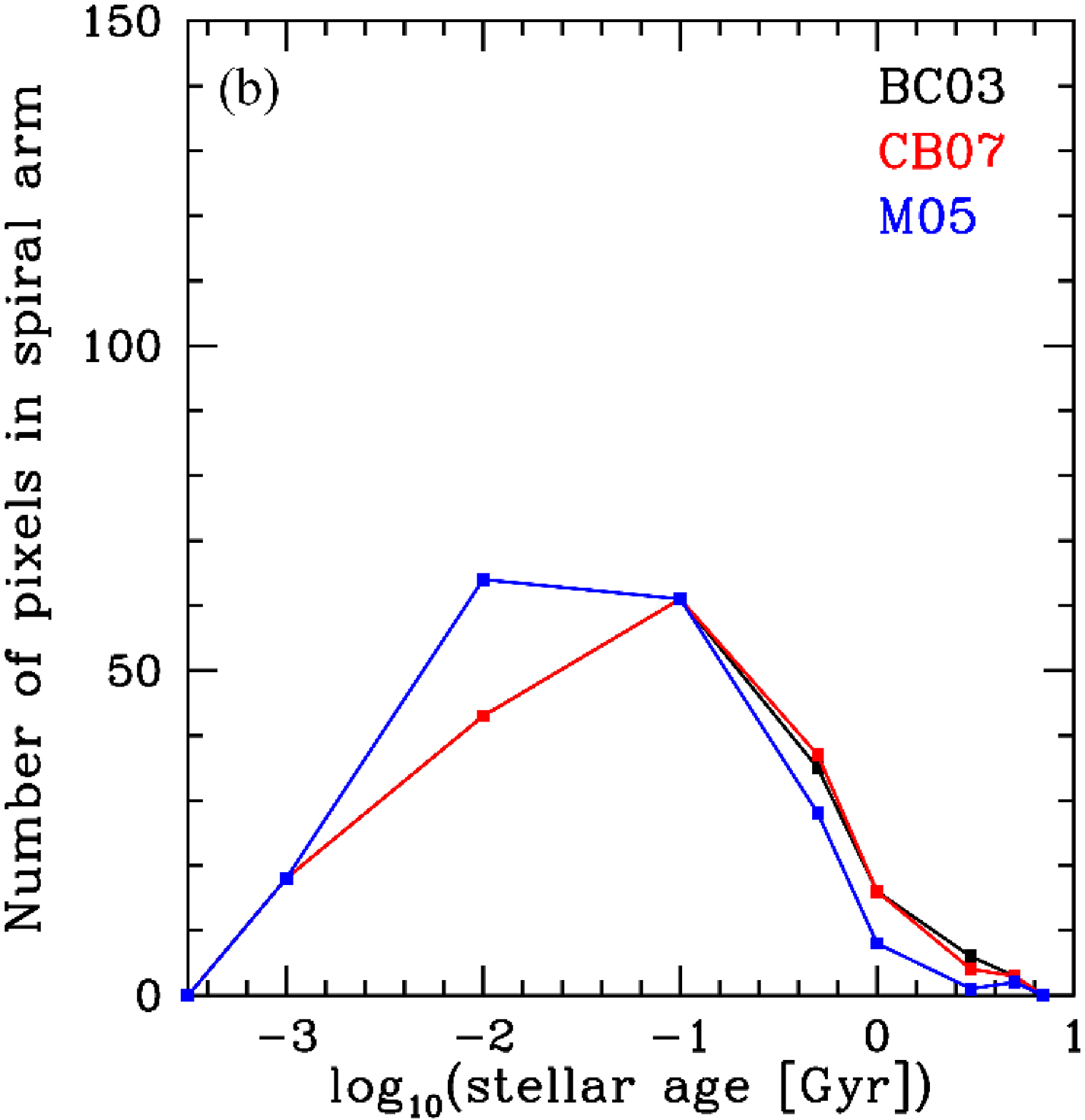}
\hspace{0.1cm}
\includegraphics[width=.4\textwidth]{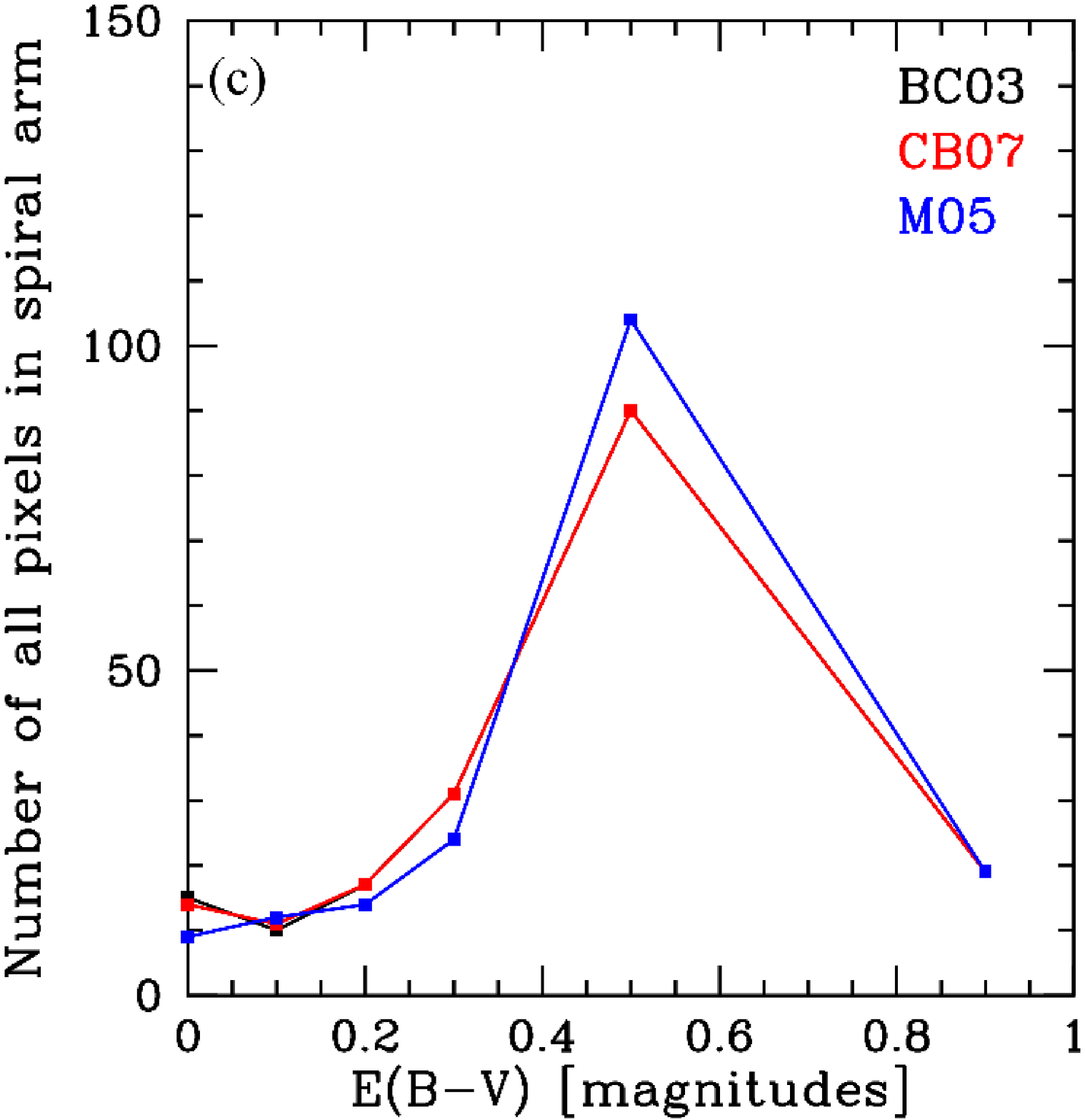}
}
}
\vspace{0.1cm}
\centerline{\rotatebox{0}{
\includegraphics[width=.4\textwidth]{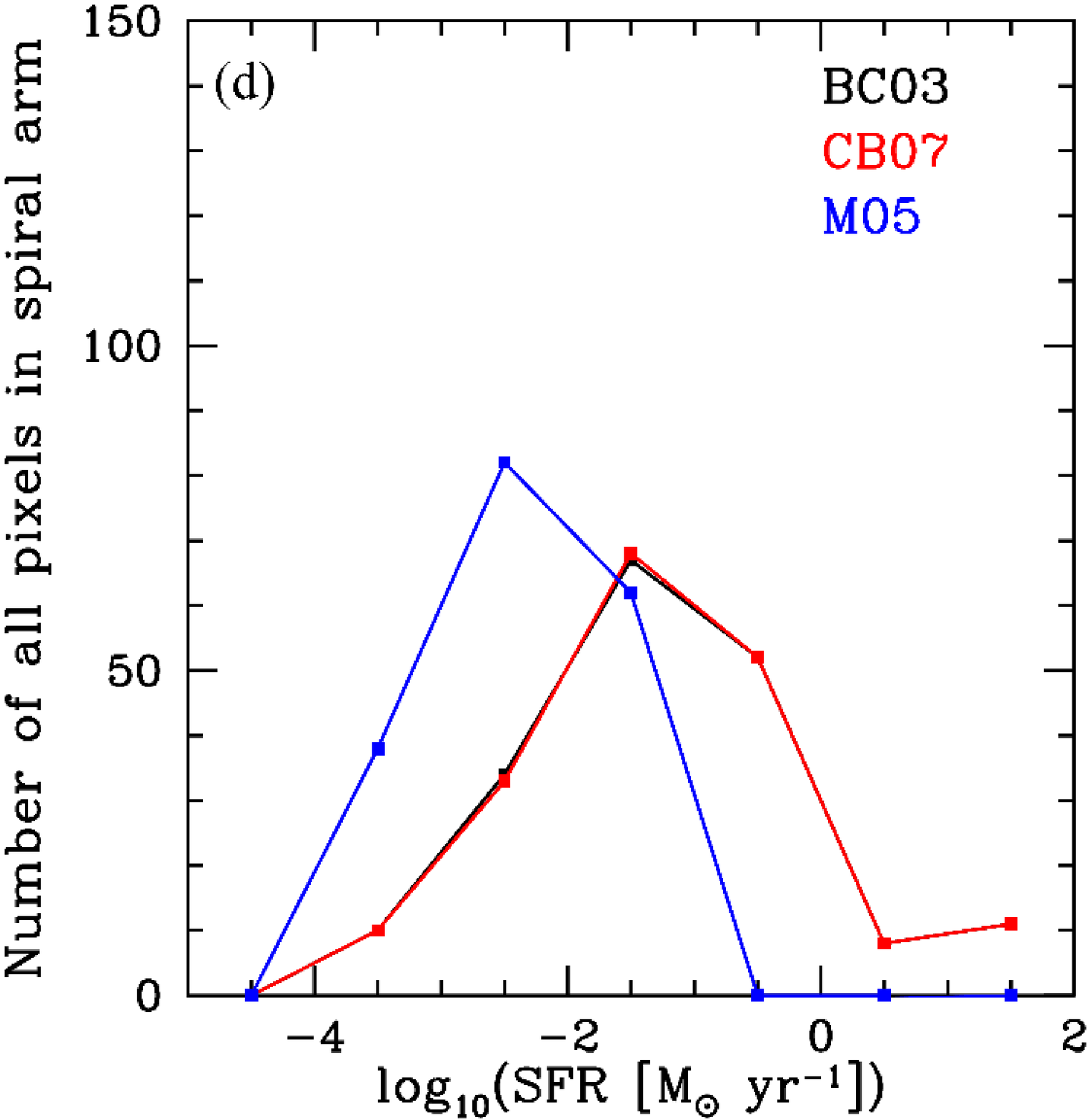}
\hspace{0.1cm}
\includegraphics[width=.4\textwidth]{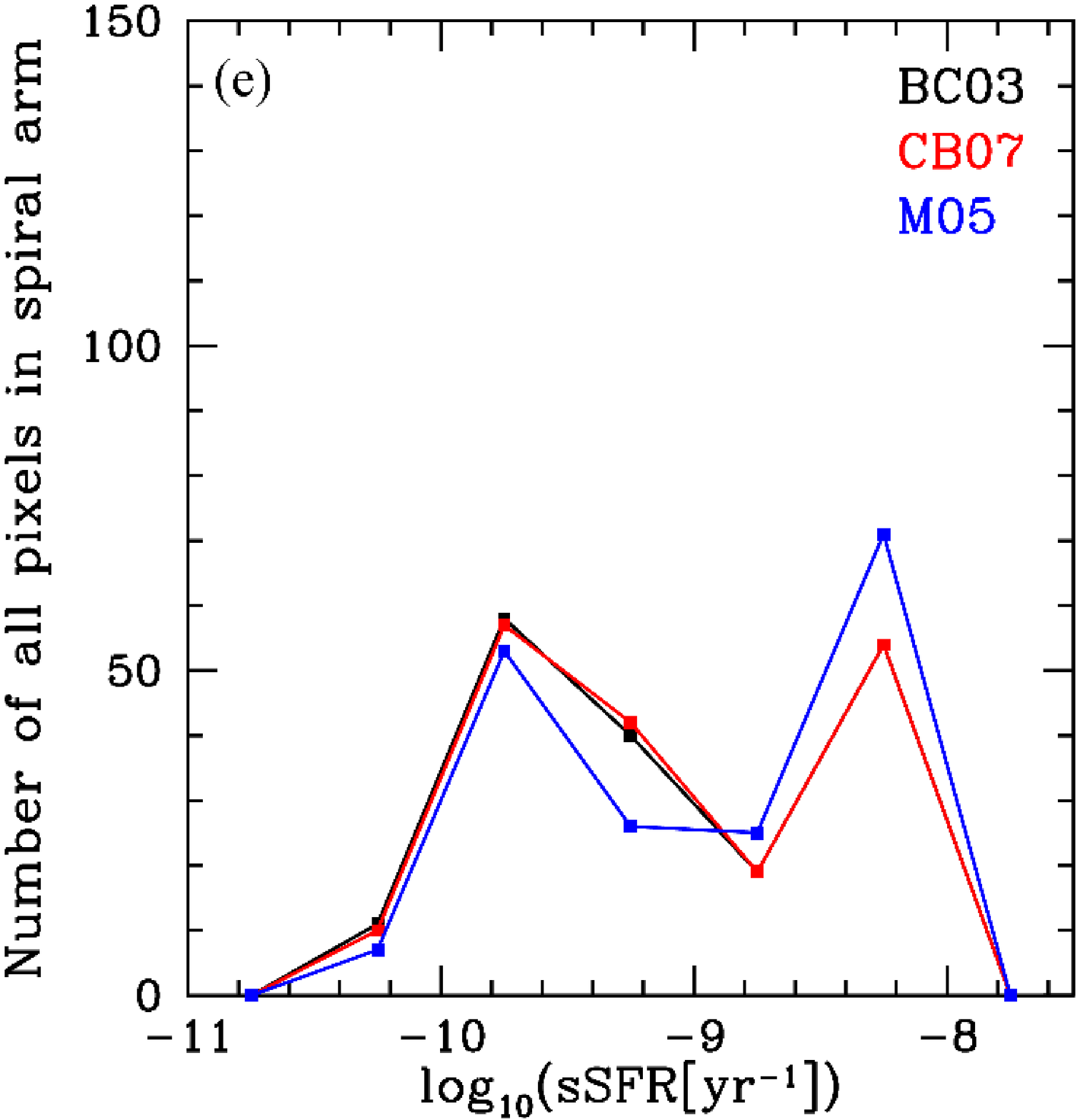}
}
}
\tiny{
\caption{
Spiral arm structure with pixel-$z$: the distribution of
  stellar population properties within a spiral arm. (a) The $i775$
  image of the galaxy (left) and the right spiral arm
  of the galaxy which has been isolated using GALFIT (right). The remaining
  panels show the distribution for the pixels within the spiral arm
  of the following parameters:  (b) the stellar population age (c)
  dust obscuration (d) the SFR and (e) the specific SFR (sSFR). Three population synthesis models are shown in each panel: BC03
(black), CB07 (red) and M05 (blue). 
 }   
\label{fig:spiralarms}}
\end{figure}

\begin{figure}
\centerline{\rotatebox{0}{
\includegraphics[width=.7\textwidth]{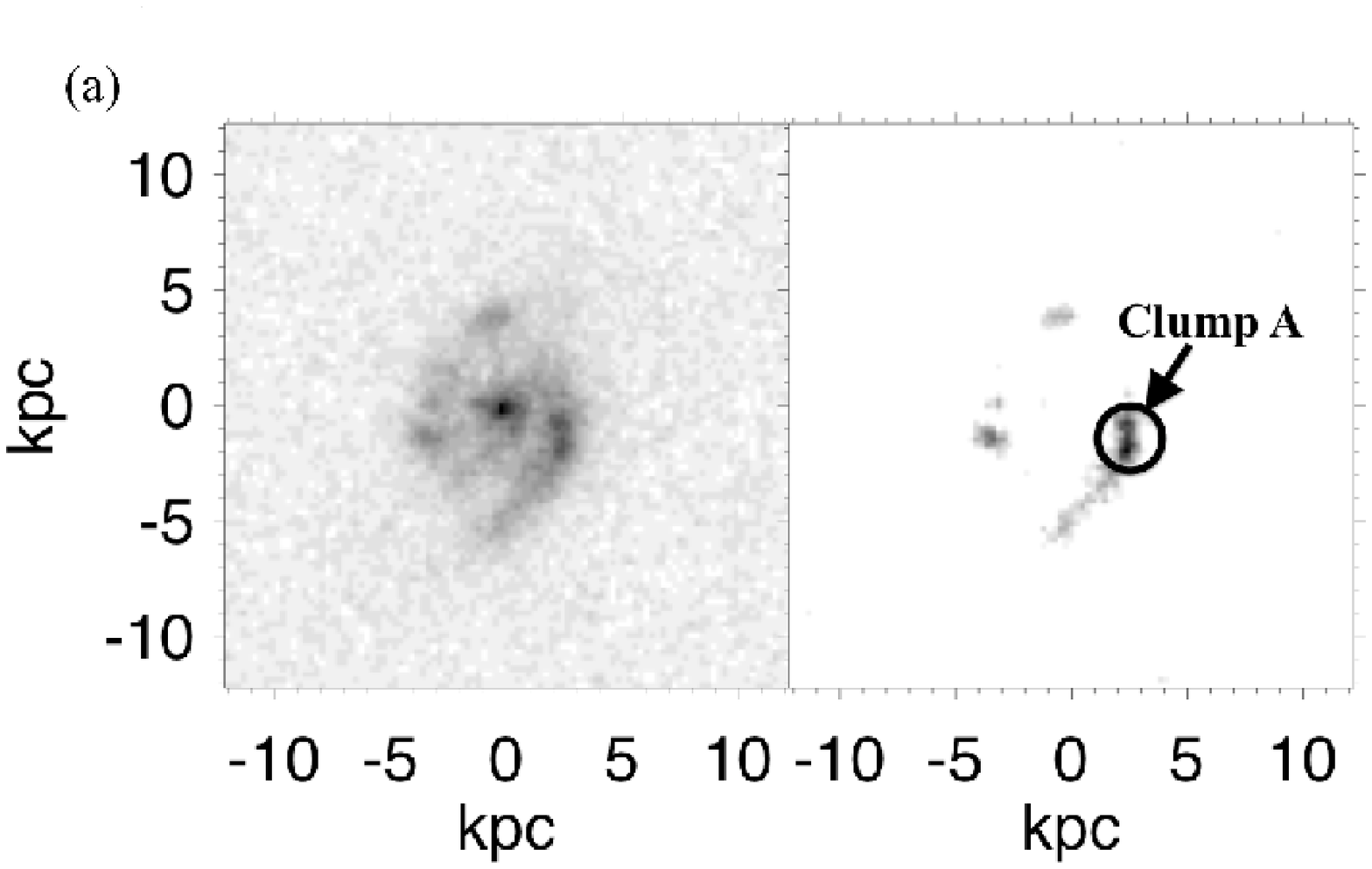}
}
}
\vspace{0.1cm}
\centerline{\rotatebox{0}{
\includegraphics[width=.4\textwidth]{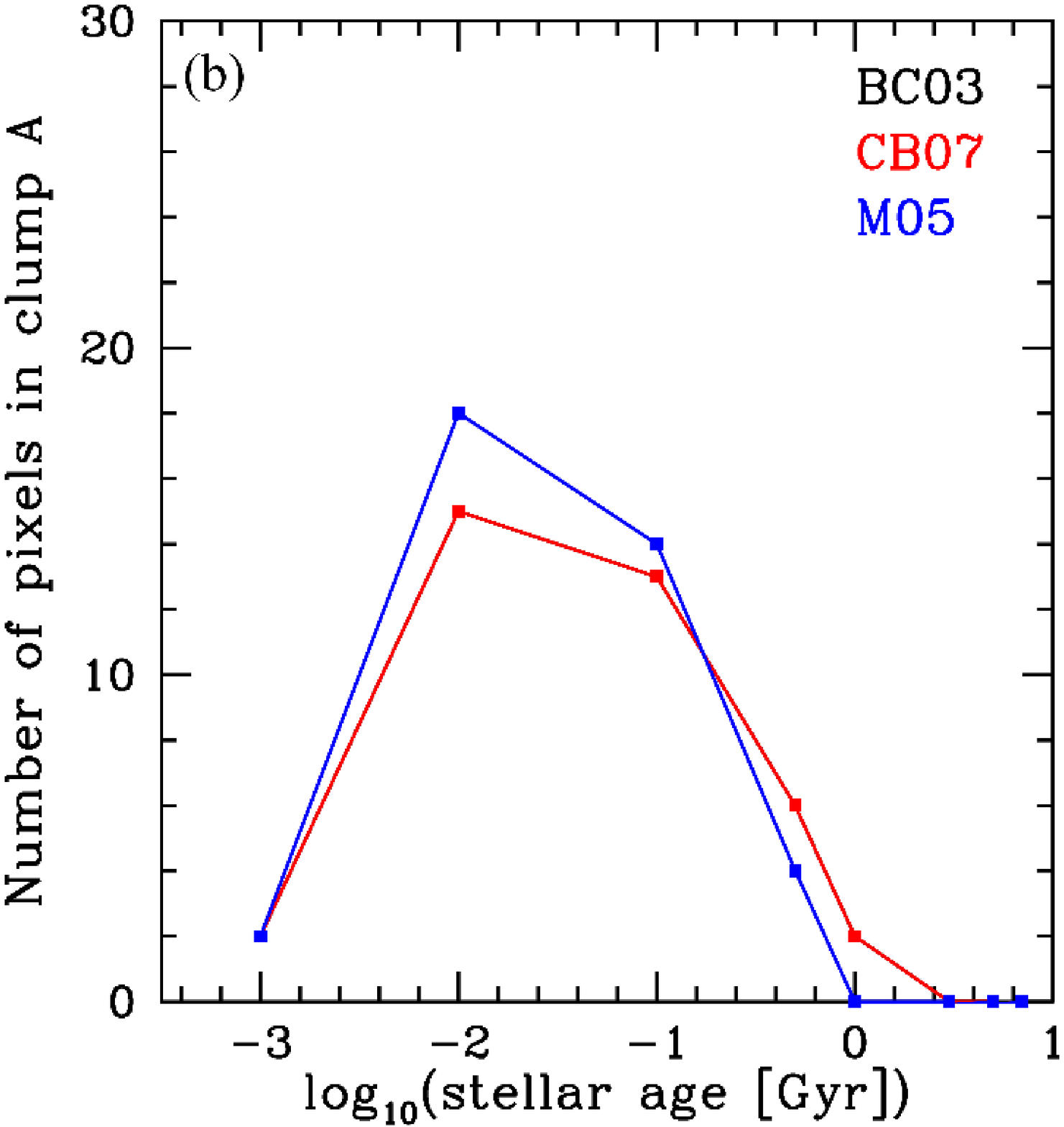}
\hspace{0.1cm}
\includegraphics[width=.4\textwidth]{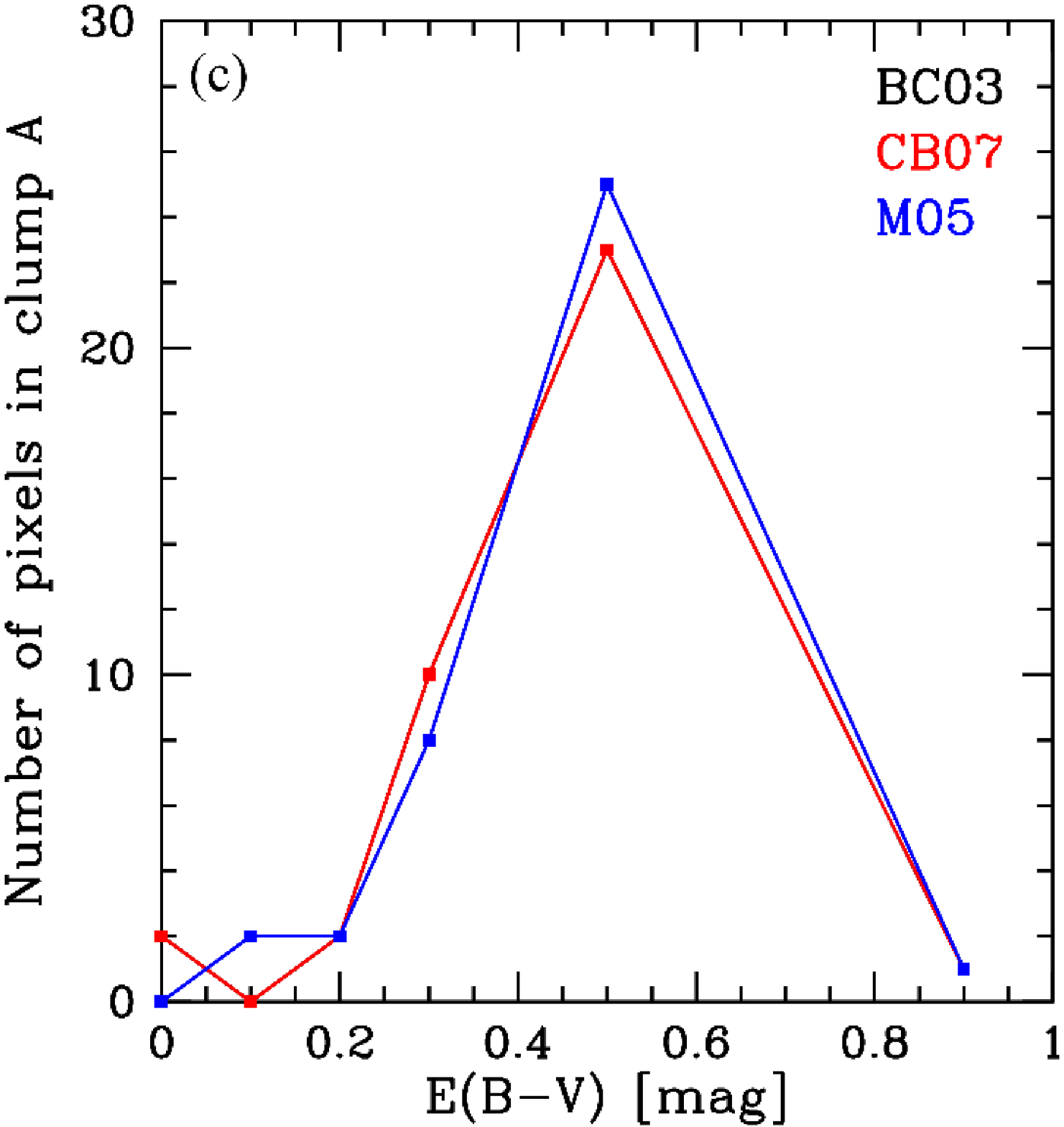}
}
}
\vspace{0.1cm}
\centerline{\rotatebox{0}{
\includegraphics[width=.4\textwidth]{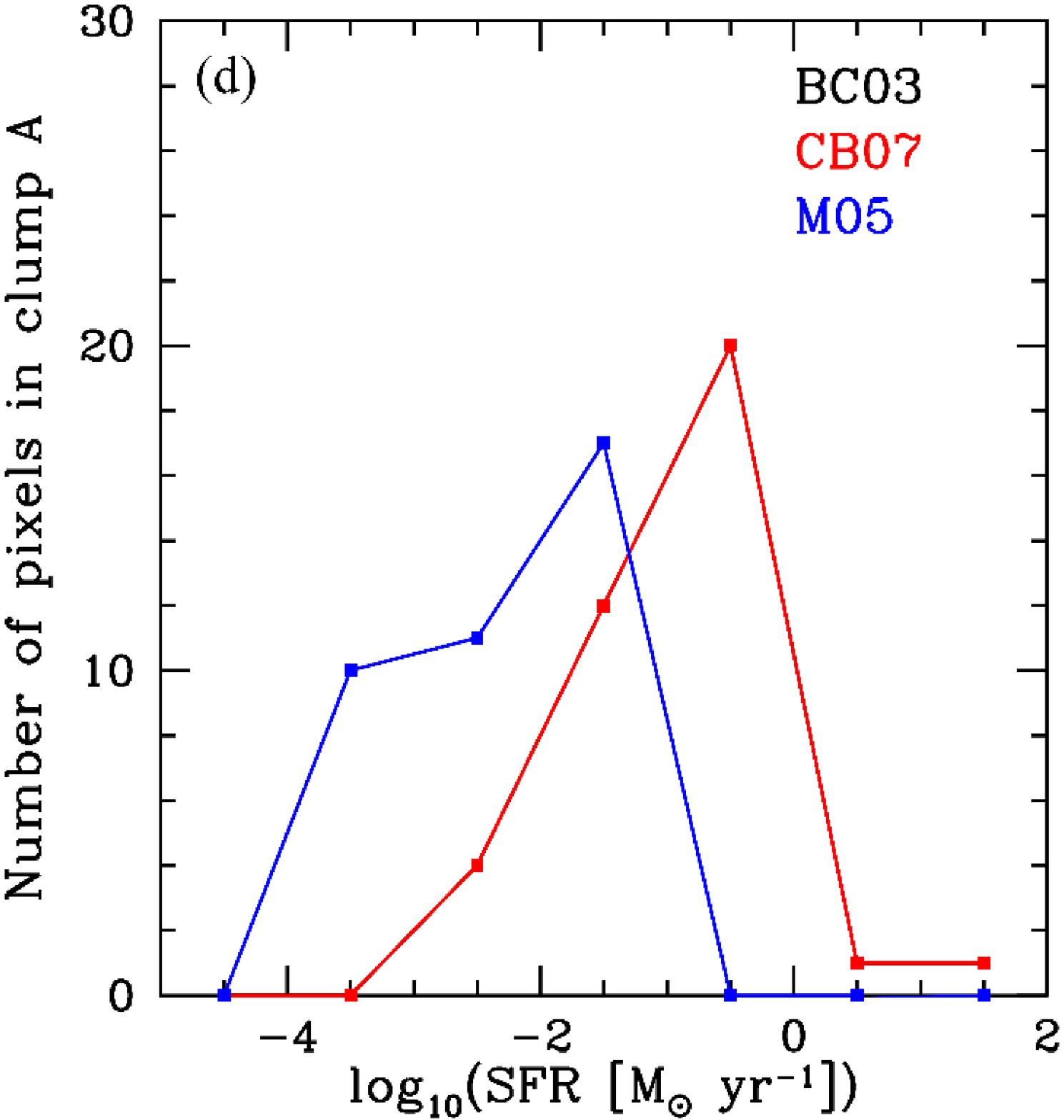}
\hspace{0.1cm}
\includegraphics[width=.4\textwidth]{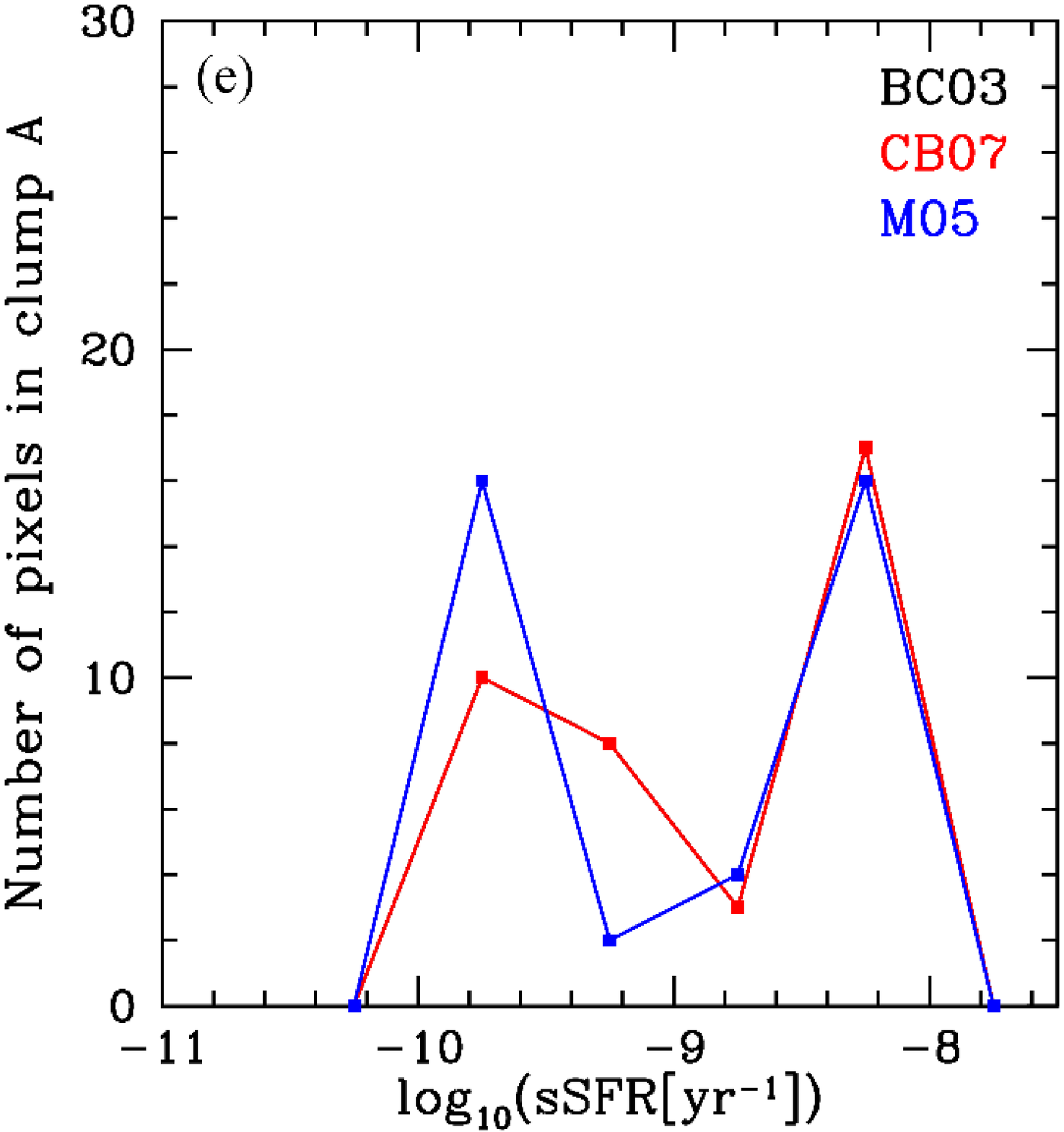}
}
}
\tiny{
\caption{
Clump structure with pixel-$z$: the distribution of
  stellar population properties within clump A of the galaxy. (a) The $i775$
  image of the galaxy (left) and clump A
 which has been isolated using GALFIT (right). The remaining
  panels show the distribution for the pixels within clump A
  of the following parameters:  (b) the stellar population age (c)
  dust obscuration (d) the SFR and (e) the specific SFR. Three population synthesis models are shown in each panel: BC03
(black), CB07 (red) and M05 (blue). 
 }   
\label{fig:clumpA}}
\end{figure}

\begin{figure}
\centerline{\rotatebox{0}{
\includegraphics[width=.7\textwidth]{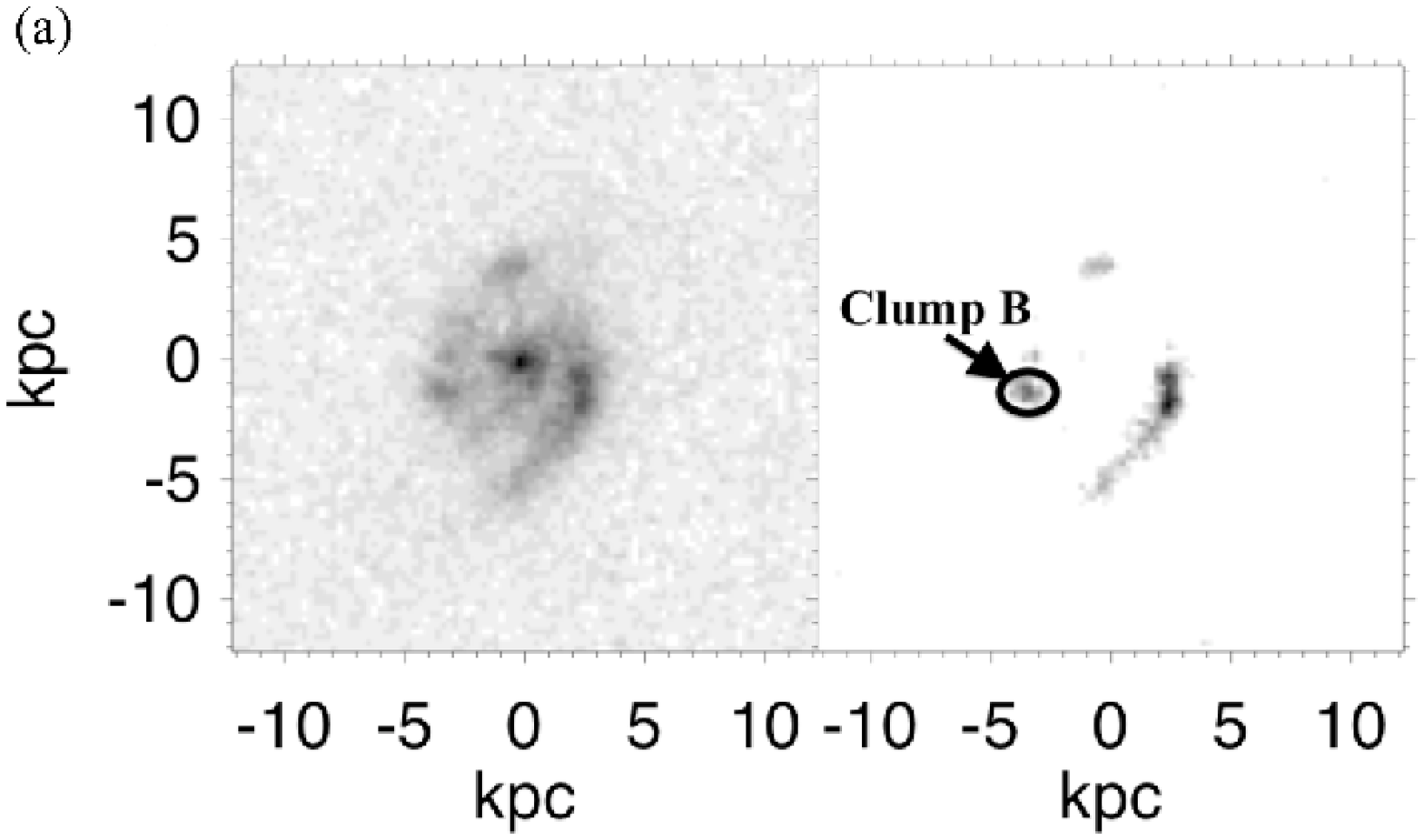}
}
}
\vspace{0.1cm}
\centerline{\rotatebox{0}{
\includegraphics[width=.4\textwidth]{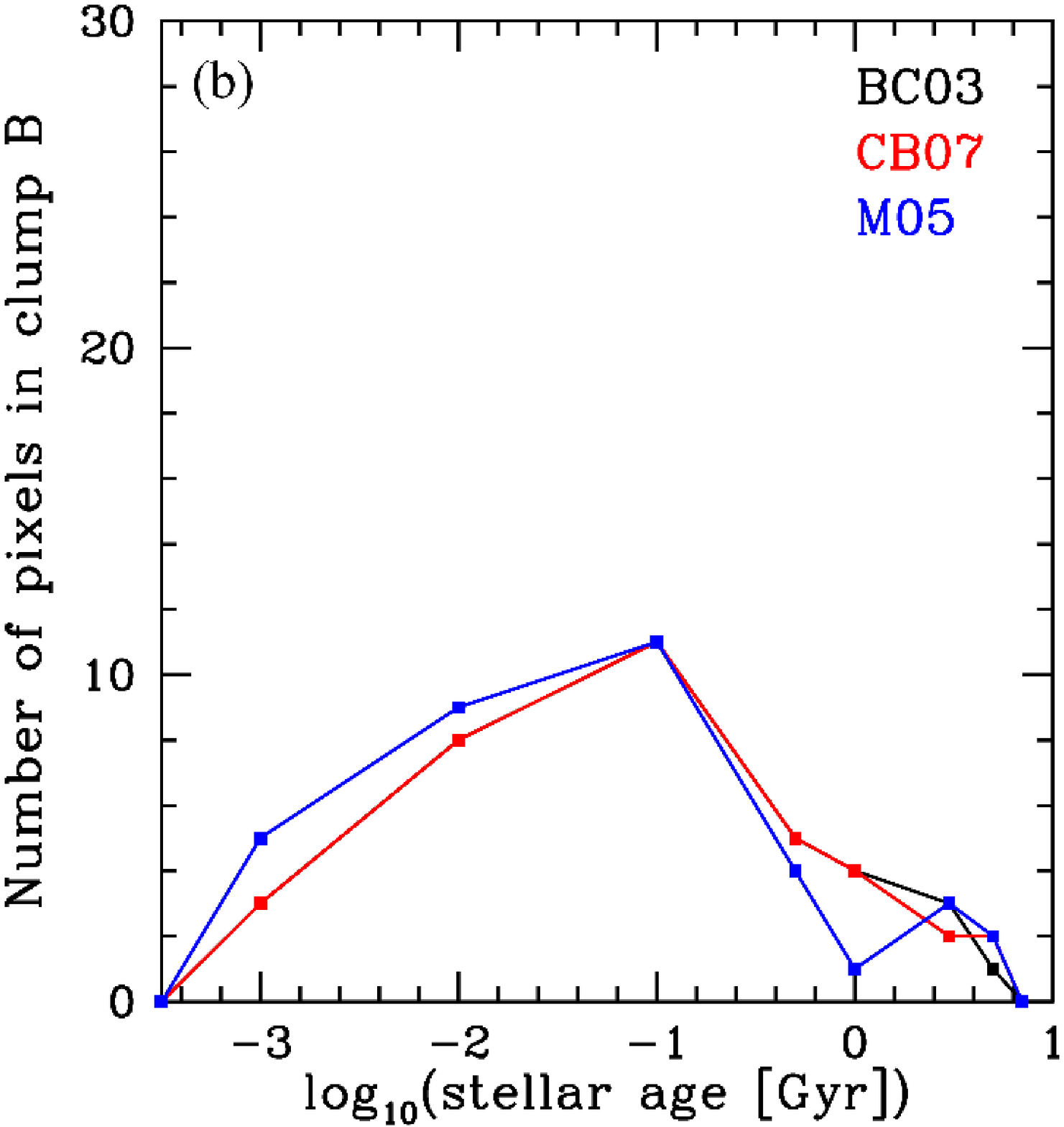}
\hspace{0.1cm}
\includegraphics[width=.4\textwidth]{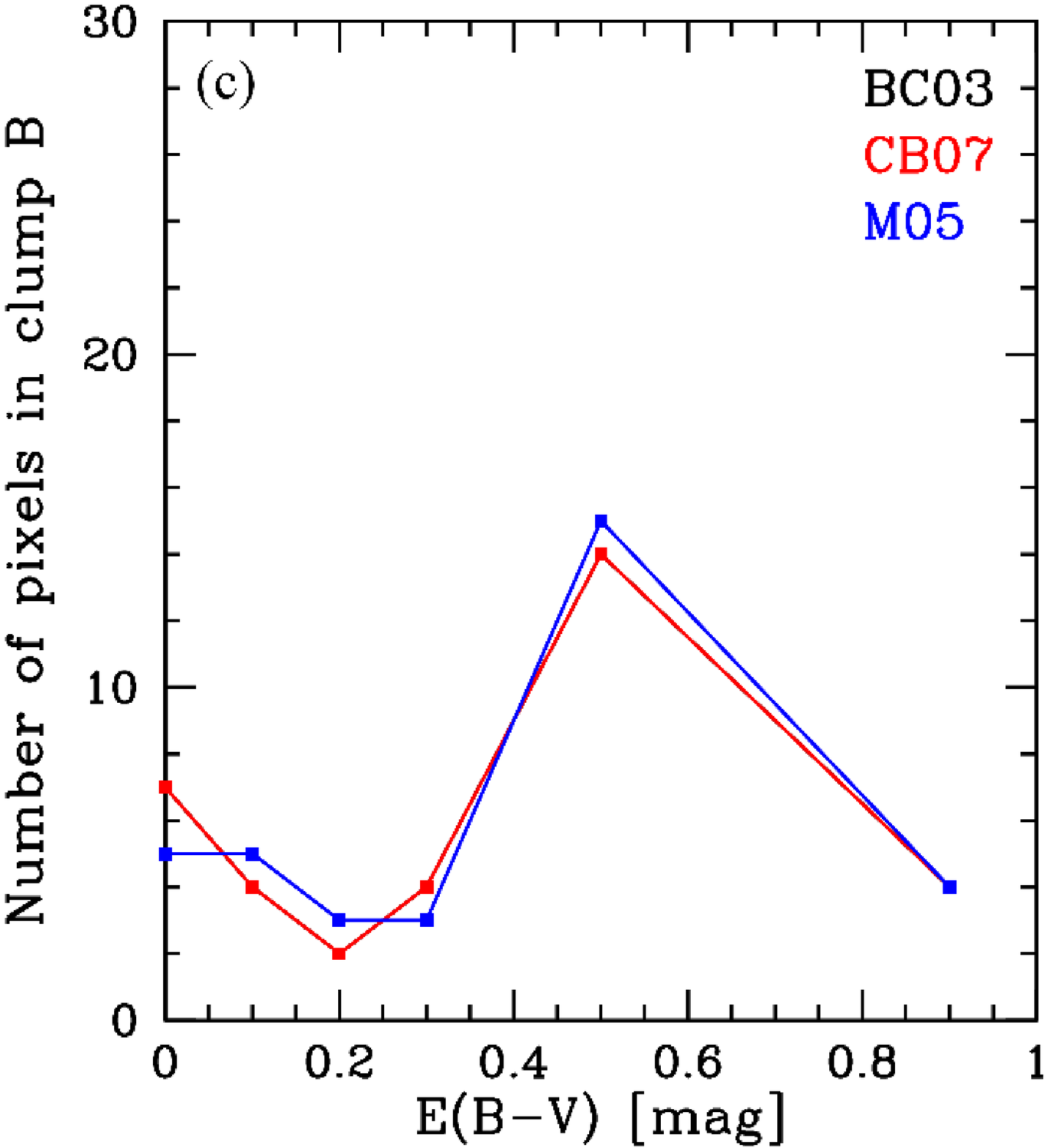}
}
}
\vspace{0.1cm}
\centerline{\rotatebox{0}{
\includegraphics[width=.4\textwidth]{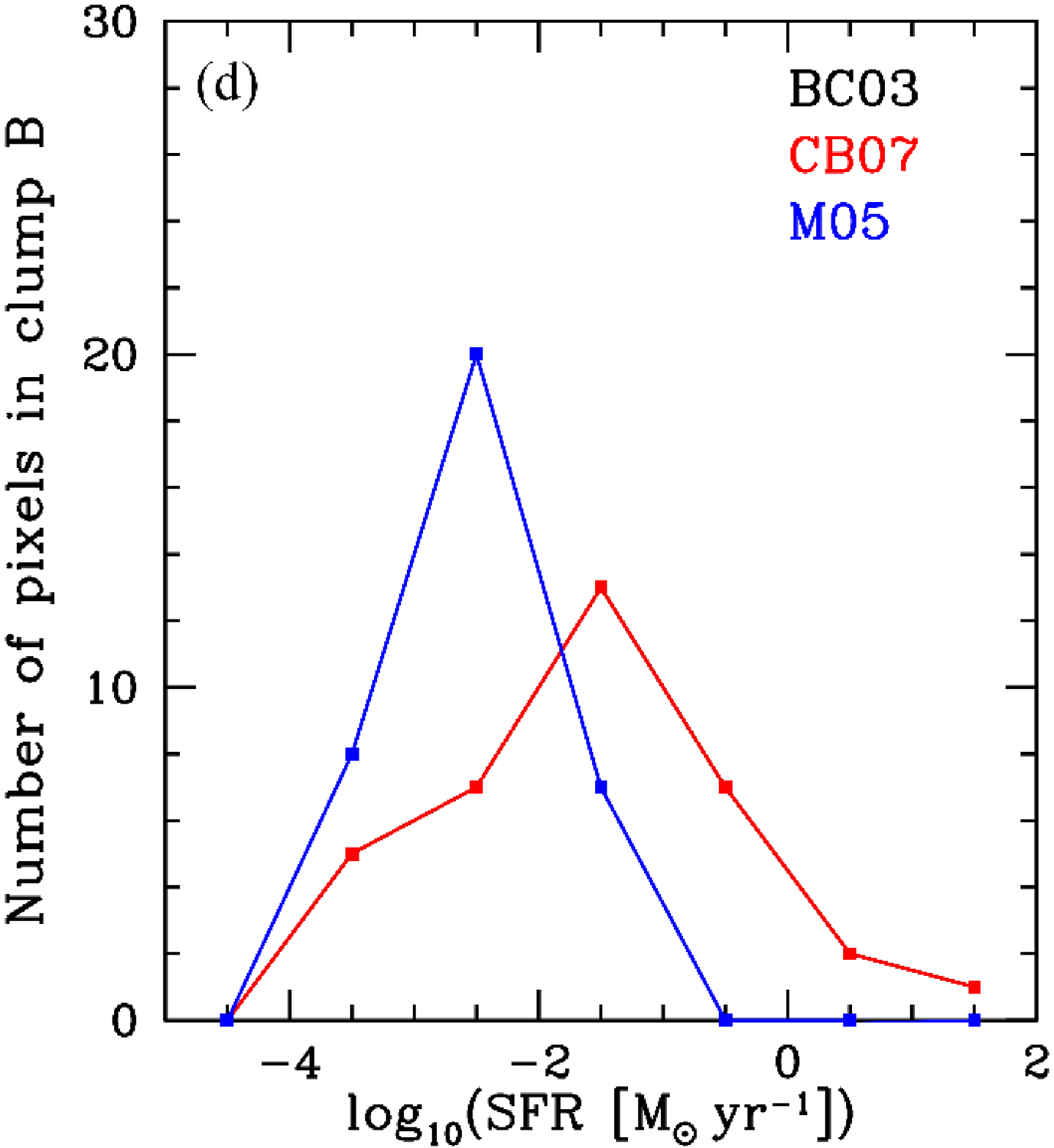}
\hspace{0.1cm}
\includegraphics[width=.4\textwidth]{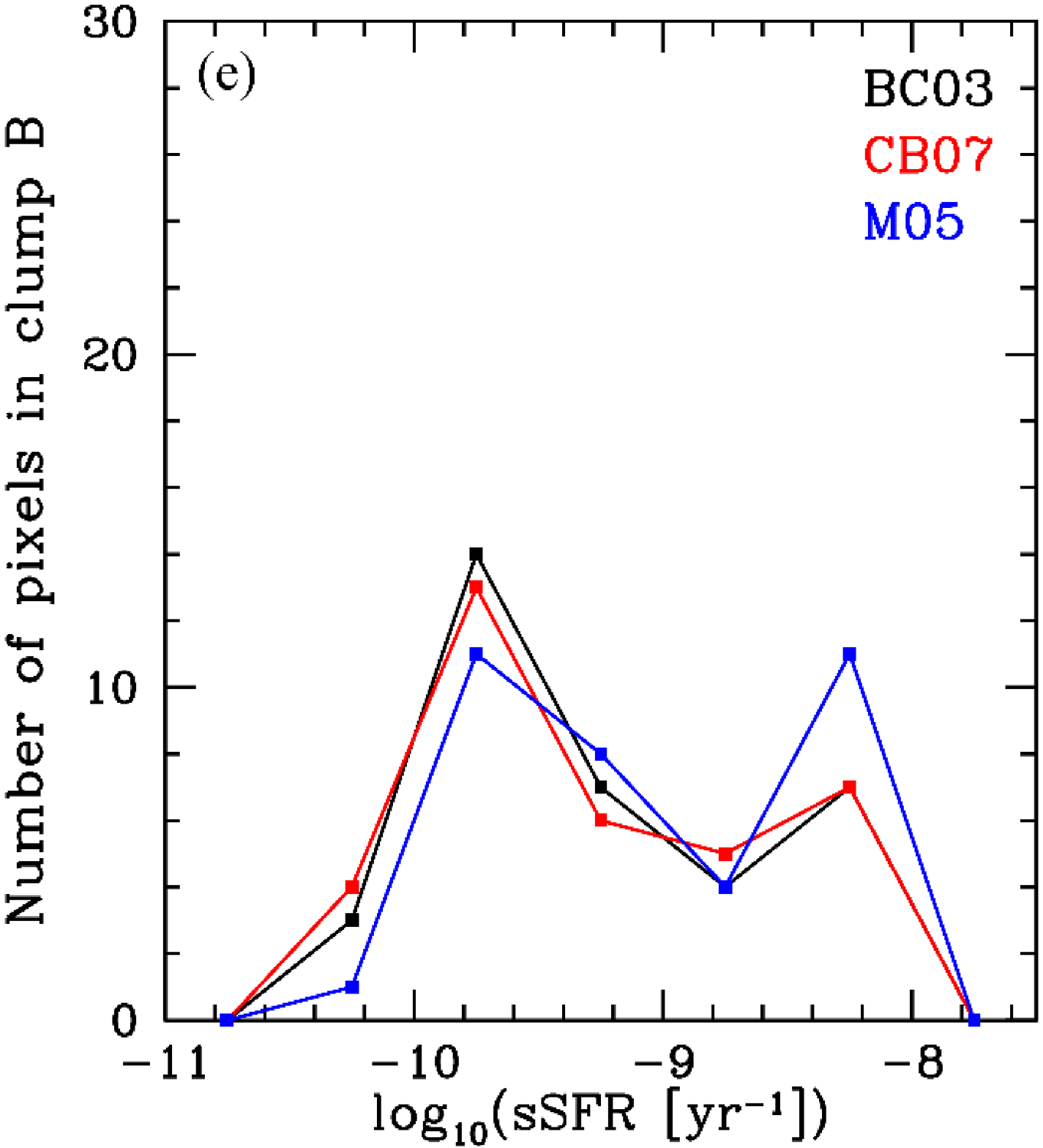}
}
}
\tiny{
\caption{
Clump structure with pixel-$z$: The distribution of
  stellar population properties within clump B of the galaxy. (a) The $i775$
  image of the galaxy (left) and clump B
 which has been isolated using GALFIT (right). The remaining
  panels show the distribution for the pixels within clump B
  of the following parameters:  (b) the stellar population age (c)
  dust obscuration (d) the SFR and (e) the specific SFR. Three population synthesis models are shown in each panel: BC03
(black), CB07 (red) and M05 (blue). 
 }   
\label{fig:clumpB}}
\end{figure}

In Figures~\ref{fig:spiralarms}, \ref{fig:clumpA} and \ref{fig:clumpB}, we illustrate the way pixel-$z$ can be
used to study the stellar populations of subcomponents of galaxies such as
spiral arms and clumps, and we explore how this can be biased by the choice of the SPS model. The optical filter set $bviz$ is used
throughout this analysis. In the spiral arm of the disk galaxy which has
been isolated, the BC03 and CB07 models predict almost identical distributions
for the parameters among the pixels that make up the
spiral arm. In contrast, some differences start to emerge between M05
and the other two models. The age distribution of the
pixels peaks at 0.1 Gyr for the BC03 and CB07 models, and at a lower
age (0.01 Gyr) for the M05 model. All three models show good agreement for
the dust distribution among the pixels in the arm, which peaks at $0.5$
Gyr. There is however a systematic offset in the SFR distributions
which peak at 0.003 $M_{\odot}\,yr^{-1}$ with M05, whereas it peaks at 0.03
$M_{\odot}\,yr^{-1}$ with BC03/CB07. This difference is largely due to
the lower stellar mass predicted for the pixels in the arm by M05
compared to the other models. When stellar mass is normalized,
we see a much closer agreement between M05 and BC03/CB07. All the
models predict a sSFR distribution that is bimodal in the arm, one peaking at 
$10^{-9.8}yr^{-1}$ and the other at $10^{-8.3}yr^{-1}$.

We also investigate clumps in the disk and the impact of model
differences on the inferred properties of clumps in the disk of the
galaxy. Clump A in Figure~\ref{fig:clumpA} is part of the outer arm of the
galaxy. As in the case of the spiral arm, the distribution of the stellar population age and dust
obscuration among the pixels in clump A, is quite similar among the
different population synthesis models. As with the full spiral arm, a substantial
difference exists in the SFR
distributions. The M05 models predict the clump to
have a lower SFR than BC03/CB07. The distribution peaks at
$\sim0.01 M_{\odot}\,yr^{-1}$ with M05 and $0.3 M_{\odot}\,yr^{-1}$
with BC03/CB07, due to a lower stellar mass predicted for the clump by the
M05 model. It is also worth nothing that the peak of the SFR distribution
within the clump occurs at a higher SFR than in the spiral arm. The sSFR distributions
in the clump are far more
consistent among all the models, again showing a bimodal star formation in the clump, with
one population peaking at $10^{-9.8}yr^{-1}$ and the other at
$10^{-8.3}yr^{-1}$ as in the case of the spiral arm. The SFR and sSFR
distributions imply that the star formation process in the spiral arm might be driven by the
star formation in the clump itself. 

Finally, in Figure~\ref{fig:clumpB}, we examine clump B on the left
side of the galaxy. Again, model differences affect primarily the
SFR, with M05 predicting the SFR distribution to peak at
$\sim0.001  M_{\odot}\,yr^{-1}$ while BC03/CB07 predict that it peaks
around 0.03 $M_{\odot}\,yr^{-1}$. Irrespective of the model, however,
it is evident that the mean SFR of clump B is significantly lower than
clump A, indicating that clump A is the primary driver of star
formation in the outskirts of the galaxy. Despite some differences between
M05 and the other models, therefore, pixel-$z$ can be used to
effectively and accurately probe mean properties of stellar populations
within components of galaxies such as spiral arms and clumps. As discussed in section 1.1,
this can be used to probe substructure in $z\sim2$ galaxies in order to
determine if these objects are largely driven by mergers or are disks
formed in-situ.

\subsection{Incorporating WFC3 to Probe Substructure: Expectations from the CANDELS Survey}
\label{subsec:candels}

\begin{figure}
\centerline{\rotatebox{0}{
\includegraphics[width=.4\textwidth]{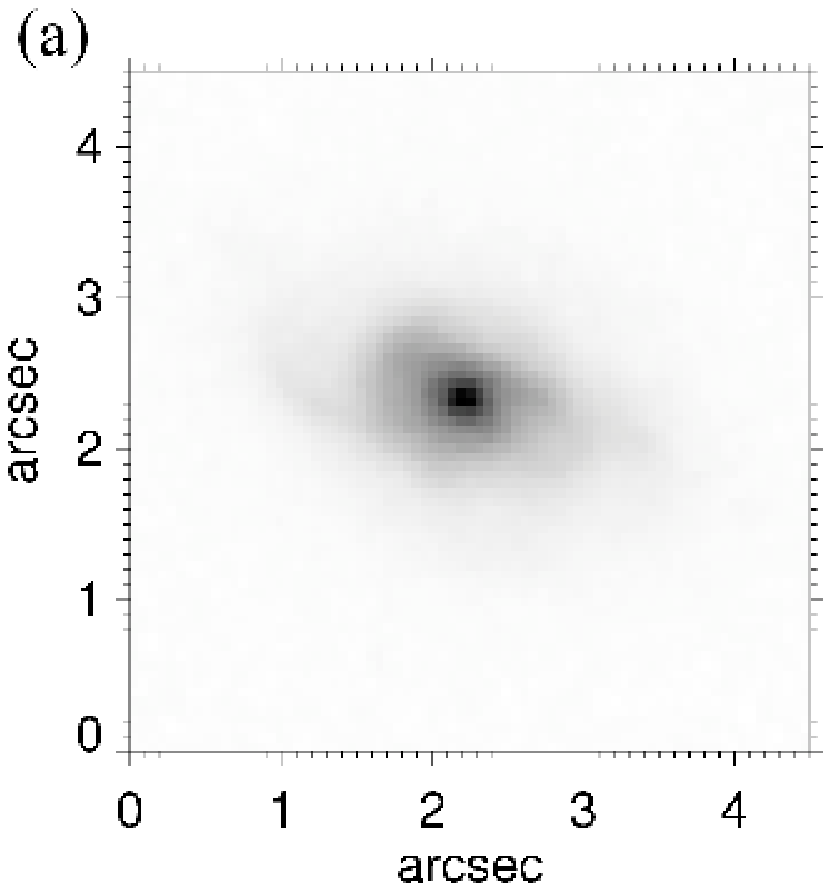}
}
}
\vspace{0.1cm}
\centerline{\rotatebox{0}{
\includegraphics[width=0.7\textwidth]{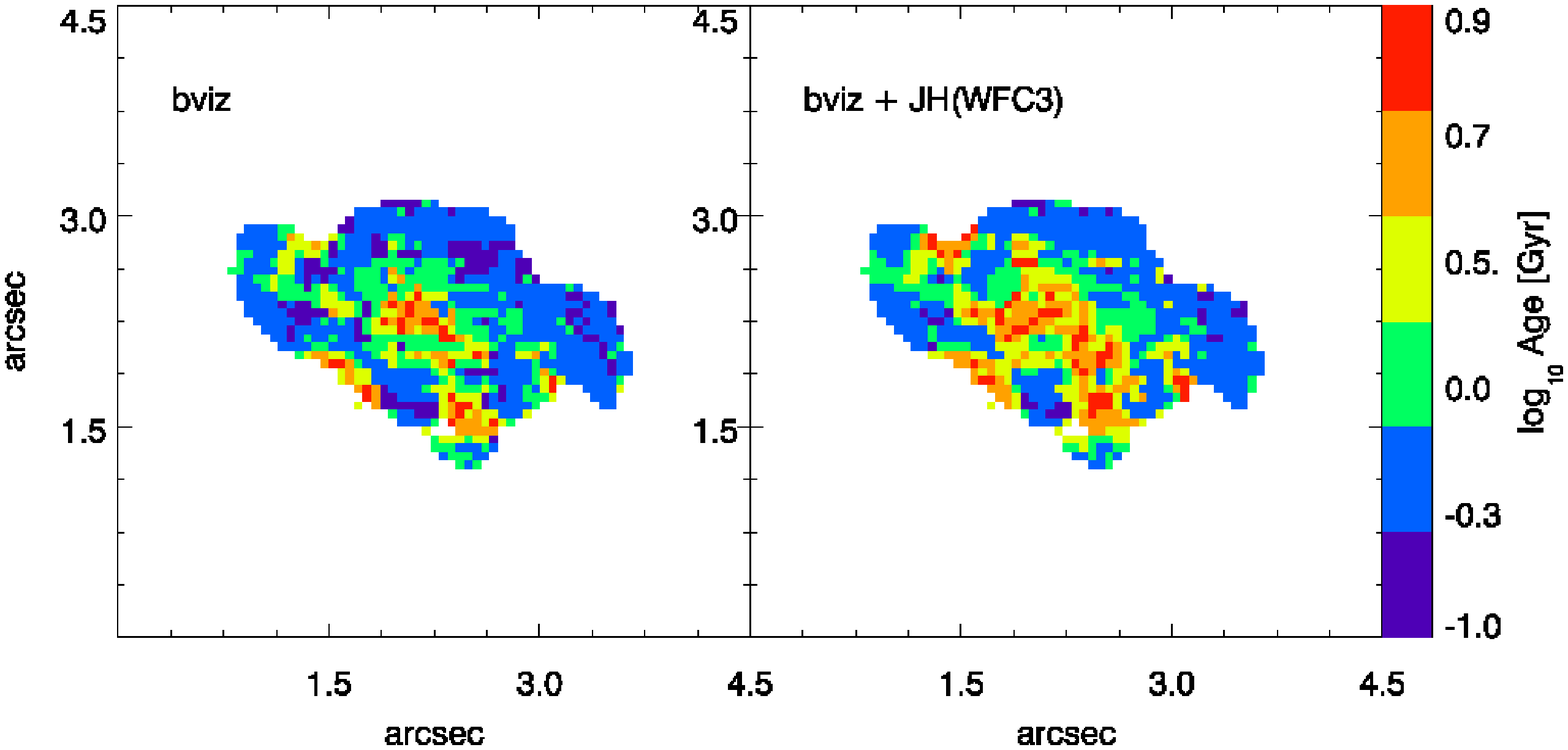}
}
}
\vspace{0.1cm}
\centerline{\rotatebox{0}{
\includegraphics[width=0.7\textwidth]{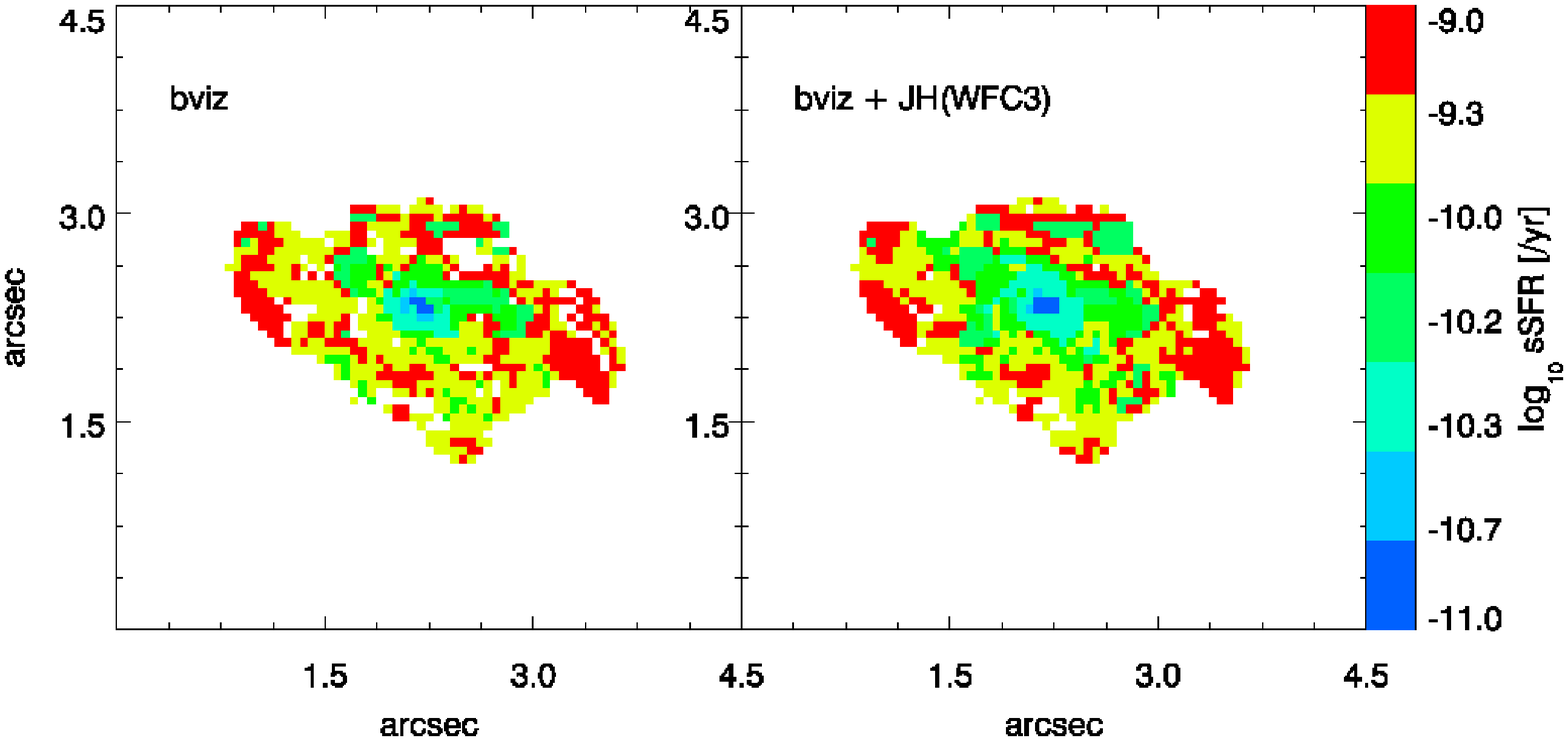}
}
}

\tiny{
\caption{
Effect of adding WFC3 $J$ and $H$ band images from the CANDELS survey on the
derived pixel-$z$ parameters using the Maraston 05 stellar
population synthesis models (a) $H$ band image of a galaxy at $z=0.6$
(b) Stellar population age map of the galaxy obtained using the $bviz$
passbands (left) and the $bvizJH$ bands (right) (c) Specific SFR
(sSFR) map of the galaxy obtained using the $bviz$
passbands (left) and the $bvizJH$ bands (right). 
}   
\label{fig:CANDELS_maps}}
\end{figure}

\begin{figure}
\centerline{\rotatebox{0}{
\includegraphics[width=.45\textwidth]{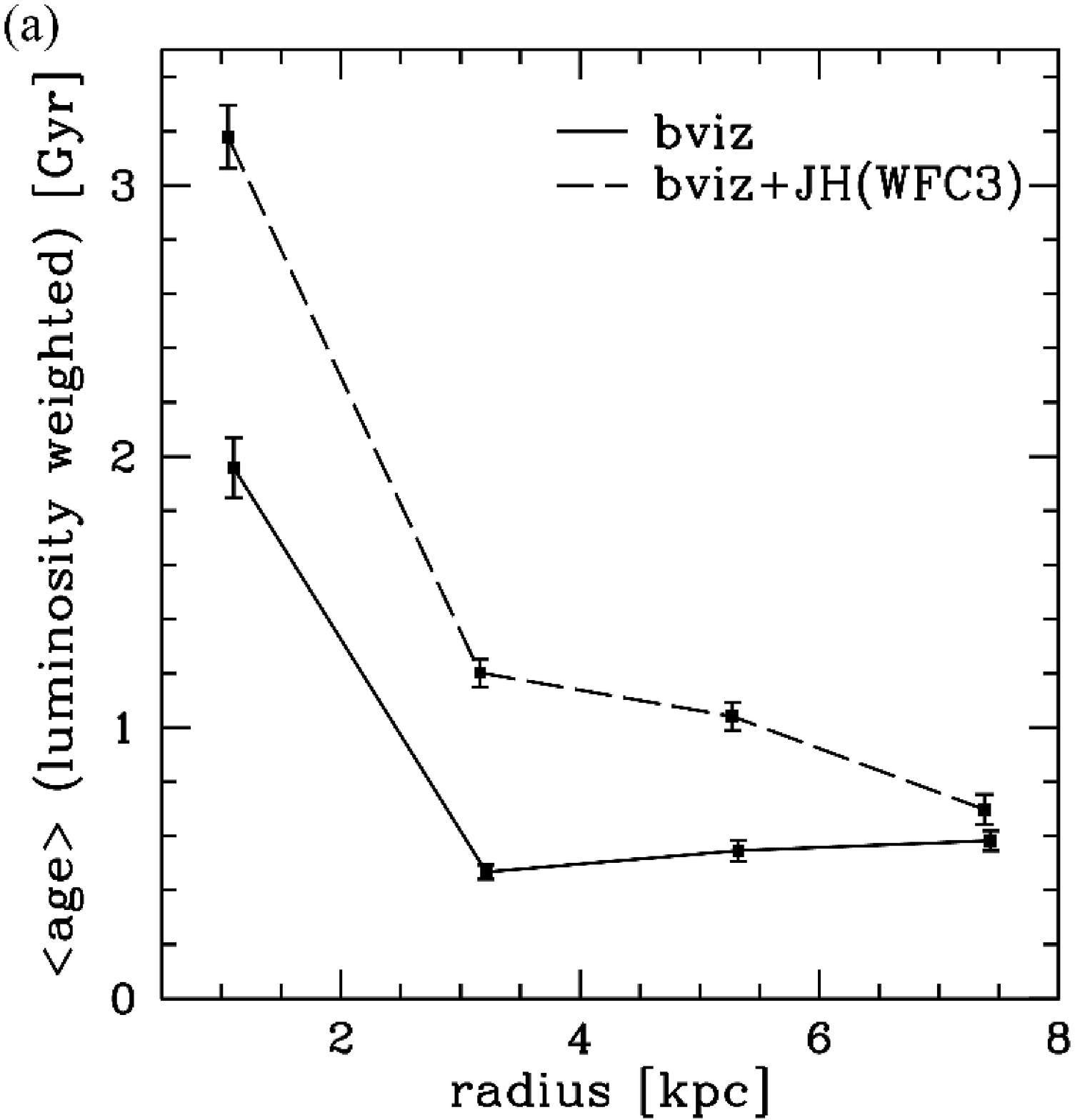}
}}
\vspace{0.1cm}
\centerline{\rotatebox{0}{
\includegraphics[width=.45\textwidth]{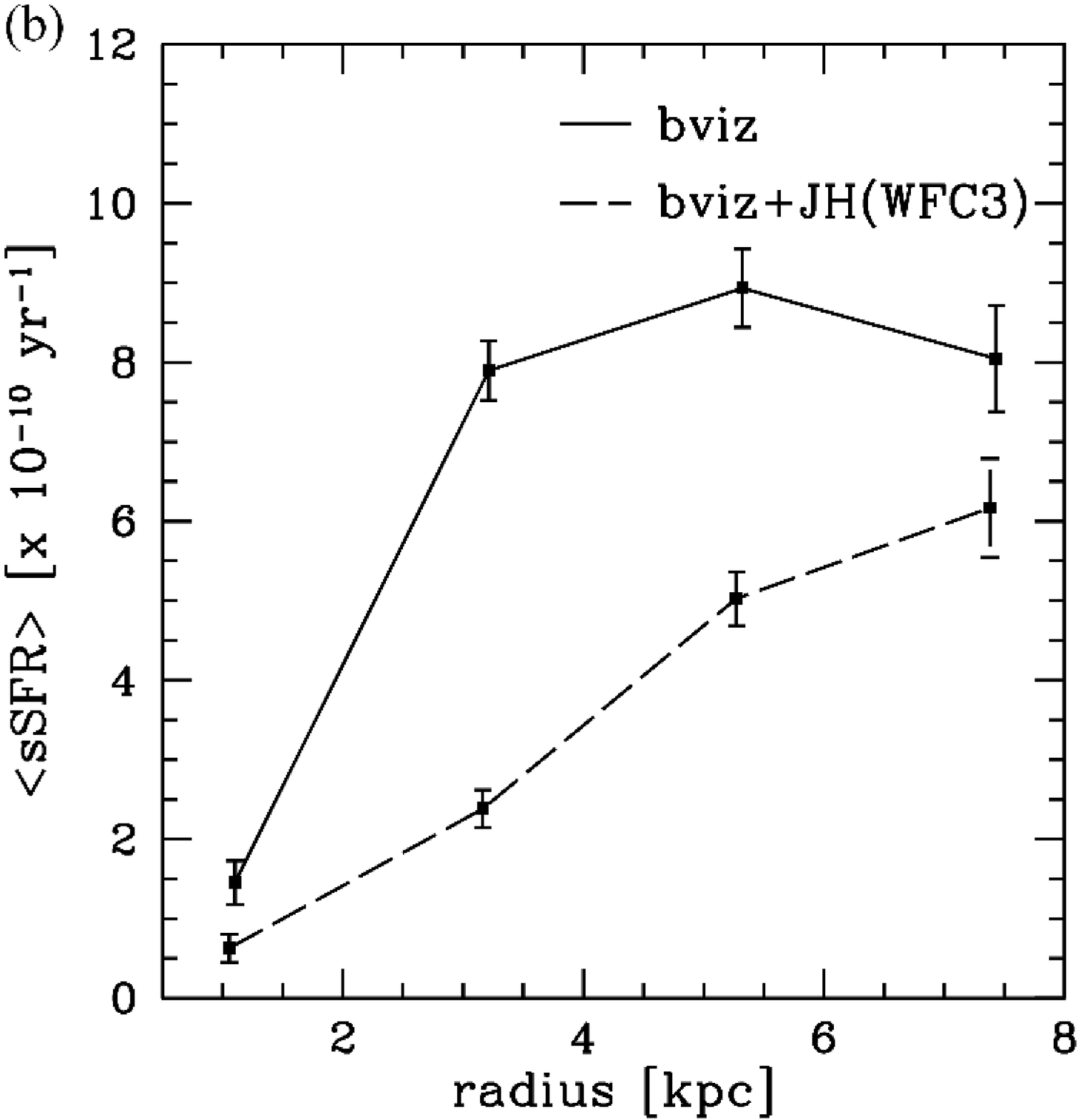}
}
}
\tiny{
\caption{
Effect of adding WFC3 $J$ and $H$ band images from the CANDELS survey
on the radial variation of the pixel-$z$ parameters across the galaxy
in Figure~\ref{fig:CANDELS_maps}. The Maraston 05 stellar
population synthesis model was used.  (a) Radial variation of the stellar
population age using the $bviz$ passbands (solid line) and the $bvizJH$ bands (dashed line) (b)
Radial variation of the specific SFR (sSFR) using the $bviz$
passbands (solid line) and the $bvizJH$ bands (dashed line). 
 }   
\label{fig:CANDELS_radialplots}}
\end{figure}

Figure~\ref{fig:CANDELS_maps} shows the results of adding the NIR WFC3 imaging in the $F125W$
and $F160W$ passbands from the CANDELS survey to the existing $bviz$ bands
for an example disk galaxy at $z=0.6$, with the M05 models. The most
notable difference between using the optical-only passbands and the
combined optical and WFC3/NIR passbands is in the spatial distribution
of the oldest ($t>4$ Gyr) stellar populations in the galaxy. 
In the ACS-only case, the oldest stellar populations in the galaxy
are confined in highly localized regions in the galaxy, with a
concentration in the center and and in pockets in the south-west and sout-east of
the image.  The addition of the WFC3/NIR data, however, suggests that
the oldest (and thus redder) populations are not simply confined in these pockets but
are present more extensively in the disk itself. In the map of the
specific star formation rate in the galaxy, both the optical-only and the combined optical and
NIR cases suggest a qualitatively similar trend in the spatial distribution of the
sSFR i.e., the sSFR is lowest in the center of the galaxy ($\sim10^{-11}\,yr^{-1}$) and
increases towards the outskirts. Away from the center, however, the addition of the WFC3/NIR
passbands, suggests that the disk contains a higher
proportion of stellar populations with a sSFR
($\sim10^{-10}\,yr^{-1}$) that is lower than what is predicted by the optical-only
colors. The differences observed in the maps of the stellar population
age and sSFR are confirmed by the radial variation of these
quantities, shown in Figure~\ref{fig:CANDELS_radialplots}.
The WFC3/NIR colors reveal an older population throughout the disk of
the galaxy relative to the optical-only colors, with
more than a $5\sigma$ difference in the luminosity-weighted mean age
between the $bviz$ and $bvizJH$ colors and up to a radius of 6 kpc from the galaxy center, . 
The radial variation of the sSFR also reflects the fact that the
WFC3/NIR colors are more sensitive to the older stellar population in the disk,
which were otherwise not revealed by the ACS-only passbands. The
WFC3/NIR colors reveal a lower sSFR at all radii compared to the
ACS-only colors, although the difference is most significant between 3 and 5 kpc from the
galaxy center.


\section{SUMMARY}

The pixel-$z$ method performs SED-fitting to individual pixels in
multicolor images of galaxies and obtains the values of stellar
population properties, such as age, SFR, obscuration and metallicity,
in the pixels. We use it to make detailed measurements of the
properties of components in galaxies such as spiral arms and clumps as well as to measure the  
 radial variation of star formation and other properties across galaxies.
In this work, we present the application of the pixel-$z$ technique
to a sample of 467 resolved galaxies at $z\sim1-3$ in the GOODS-South field in order to extract
 spatially resolved properties of these objects from their pixel
 colors. We have focused on the impact of systematic biases on how
 accurately this technique can probe substructure and stellar
 populations in these high redshift objects. We investigate various biases
 that affect the determination of stellar population properties within
 pixels. For each potential source of systematic error, we examine the distribution
 of the systematic error across all pixels in the galaxy sample. We
 also study the effect on the derived properties of components of galaxies. We summarize our findings
 below.

\begin{itemize}
\item Using optical-only passbands, we find that differences
  between population synthesis models can bias the derived stellar population
  properties, although the impact is minimal for the majority of
  pixels. The impact of differences between the BC03 and CB07 models are small, with over 94 percent of pixels in the galaxies
  having resulting systematic errors in their age, dust and $\tau$
  properties that are below the statistical errors. However, a larger systematic uncertainty
  arises from differences between the M05 models and BC03/CB07
  ones. This is likely related not to the differing prescriptions
  for TP-AGB stars but to other fundamental differences between
  the M05 models and the BC03/CB07 models.
 This causes systematic uncertainties in the pixel properties
  that impact a higher proportion of the pixels, resulting in an
  outlier fraction of pixels (with more than twice the statistical error) of
  approximately 0.15, 0.11 and 0.03 for the $\tau$, stellar population age and
  color excess (dust obscuration) respectively.
\item We explore systematic uncertainties in the stellar population
  properties within pixels resulting from a changing parameter space
  used by pixel-$z$. Increasing the range of parameter values that
  pixel-$z$ searches is expected to exacerbate SED
  degeneracies. We find that when using the 4 optical
  bands, this systematic can be well controlled. The fractions of
  pixels whose stellar population parameters are affected in this
  manner are only 0.03, 0.06 and 0.01 for the $\tau$, stellar population age and
  color excess respectively.
\item We explore the effect of limiting the number of
  passbands available for the pixel-$z$ analysis. We find that this
  results in the largest systematic errors in deriving stellar
  population properties in the pixels. Using 2
  passbands alone results in large ($>4\sigma$) systematic
  uncertainties in the parameters in the majority of the
  pixels. As expected, the more passbands that are available to sample
  the spectral range of the pixels, the more accurately
  stellar population parameters can be constrained. We
  find, however, that moderately robust results can be obtained with just three of
  the optical filters within the set of $bviz$, provided they are
  selected to span the 4000\,\AA\ break. We find that omitting one of
  the 4 optical passbands can bias (by more than twice the statistical
  error) the estimates in the stellar population age
  and $\tau$ in more than 20 percent of pixels in the sample. The SFR
  and the stellar population age are most impacted of all the stellar
  population properties by such limitations. Passband limitations are
  thus the largest source of systematic error in the pixel-$z$ method.
\item Using a subsample of 46 galaxies with near-infrared data from
  NICMOS/NIC3 camera, we study how systematic biases in the pixel-$z$
  parameters due to differences among stellar population synthesis
  models, are affected by the addition of the NIR passbands relative
  to the optical-only colors. The most important factor here is that
  the different models predict generally similar rest-frame UV colors
  but show significant differences in their predicted rest-frame NIR
  or optical (red) colors. Thus the impact of adding the NICMOS data
  has a redshift dependence. 
  For the low redshift ($z<1$)
  sample, the addition of the NIR passbands exposes 
differences in the rest-frame NIR colors among the three models being compared, and this impacts the stellar
population age most strongly. For example, for BC03 and CB07, the fraction of pixels showing
$<1\sigma$ differences in age decreases from 0.89 with
optical-only colors to 0.81 with NIR colors. However, for $z>1$, the
bias in the pixel-$z$ parameters due to these model differences that
arise from using NIR colors is not significantly different to those arising from using optical-only colors. 
\item We investigate any systematic difference between summing the
 SED fits to individual pixels in a galaxy (as done by pixel-$z$) and the global SED fit to
 the total flux in the same pixels within a fixed aperture. We find in general a good agreement between the SFR estimated by pixel-$z$ and the one inferred from SED-fitting to the global
  photometry in the galaxy. While the majority of galaxies show
  differences in the SFR determined by the two methods which are
  smaller than the statistical uncertainties of the SFR measurements,
  we detect two small populations of galaxies with significant offsets. The first is in galaxies
  which show low SFR ($<1 M_{\odot}\,yr^{-1}$), where pixel-$z$
  estimates a higher SFR (by more than $2\sigma$) than that
  determined by the global SED fit. A second small outlier population
  is found for galaxies showing SFR$\sim 10 M_{\odot}\,yr^{-1}$, where
  the global SED fit to the total flux produces a higher SFR than that
  estimated by the sum of the individual SED fits.
 \item We show in some individual examples, that pixel-$z$ can be robustly applied to 
  substructure and stellar populations within components in
  galaxies at $z\sim1-3$. We illustrate this with measurements of the radial variations of
  the age, obscuration, SFR and sSFR in these objects. We show that
  systematic uncertainties in these measurements that result from
  using different stellar population synthesis models can generally be well constrained. In these
  examples, we find that the largest systematic differences arise
  between the M05 and BC03/CB07 models, when using
  optical-only passbands, as suggested by our statistical
  study on the full pixel and galaxy sample. 
\item We also show how pixel-$z$ can be used to measure robustly the mean
  properties of components within disk galaxies such as spiral arms
  and clumps. Our results for individual galaxies suggests that
  pixel-$z$ is generally robust to systematic errors from stellar
  population synthesis model differences in determining the distribution of age, dust obscuration
  and sSFR within these clumpy structures in high redshift disk galaxies. 
\end{itemize}

It is important to note that, despite testing the above systematics, there are still several underlying implicit
assumptions within pixel-$z$ which may contribute systematic
effects to inferred results on the stellar population parameters. These assumptions include independent pixels, independent stellar
populations in each pixel and star formation histories that are
described by a single burst and exponential decline. These are likely
to have less impact when the statistical properties of large
populations of galaxies (and pixels) are investigated. In order to
address the impact of these assumptions rigorously, a detailed
comparison of this technique against spatially resolved spectroscopic
observations for a large sample of galaxies (as proposed by Croom et
al.\ 2011) will be required. Comparison with such IFU observations for
a large sample of galaxies will enable a calibration of the
pixel-$z$ method and an eventual refinement of these assumptions.

\section{Acknowledgements}
NW thanks A. Becker and R. Owen at the University of Washington for
help and advice on using code from the LSST Trac, S. Charlot and
G. Bruzual for providing the CB07 models, and R. Griffith and A. Beelen 
for useful discussions. NW acknowledges support from
the Centre National d'\'Etudes Spatiales (CNES) and the Centre National de
la Recherche Scientifique (CNRS), as well as support from NSF AST
0806367 during his time as a graduate student at the University of
Pittsburgh. AJC acknowledges support from NSF award AST-0709394. 
ARZ is funded by the Pittsburgh Particle Physics, Astrophysics, and Cosmology Center (PITT PACC) at
the University of Pittsburgh and by NSF grants AST 0806367 and AST
1108802.

\newpage


\newcommand{\ssps}[1]{\renewcommand{\arraystretch}{#1}}
\begin{table}\centering 
\begin{tabular}{@{}rrrrcrrrcrrr@{}}

\toprule 
&\multicolumn{3}{c}{[BC03,CB07]}& \multicolumn{3}{c}{[BC03,M05]}& \multicolumn{3}{c}{[CB07,M05]} \\

\midrule
  & $<1\sigma$ & $>2\sigma$ && $<1\sigma$ & $>2\sigma$  && $<1\sigma$
  & $>2\sigma$\\ 
\midrule
$\tau$ & 0.94 & 0.03 && 0.71 & 0.15  && 0.80 & 0.17\\
age & 0.94 & 0.04 && 0.86 & 0.11  && 0.87 & 0.11\\
$E(B-V)$ & 0.99 & 0.01 && 0.95 & 0.03  && 0.94 & 0.03\\
\bottomrule 
\end{tabular} 
\caption{Effect of stellar population synthesis models. Shown for each
  pixel-$z$ inferred parameter (row) is the fraction of pixels with systematic differences of (a)
  $<1\sigma$ and (b) $>2\sigma$ for each of the pairs of models that
  were compared.} 
\label{tab:table1}
\end{table}

\newcommand{\metallcities}[1]{\renewcommand{\arraystretch}{#1}}
\begin{table}\centering 
\begin{tabular}{@{}rrrrcrrrcrrr@{}}
\toprule 
&\multicolumn{3}{c}{$<1\sigma$}& \multicolumn{3}{c}{$>2\sigma$}\\
\midrule
$\tau$   & 0.94 & 0.03 \\
age   & 0.92& 0.06 \\
E(B-V)   & 0.99& 0.01 \\
\bottomrule 
\end{tabular} 
\caption{Effect of increasing the metallicity range, and consequently
  the number of SEDs from 2178 to 4356, in the pixel-$z$ grid. Shown for each
  inferred pixel-$z$ parameter is the fraction of pixels with systematic differences of 
  $<1\sigma$ and $>2\sigma$.} 
\label{tab:table1}
\end{table}

\newcommand{\passbands}[1]{\renewcommand{\arraystretch}{#1}}
\begin{table}\centering 
\begin{tabular}{@{}rrrrcrrrcrrr@{}}
\toprule &\multicolumn{3}{c}{[$b+i+z$]}& \multicolumn{3}{c}{[$b+v+i$]}& \multicolumn{3}{c}{[$v+i+z$]} \\
\midrule
  & $<1\sigma$ & $>2\sigma$ && $<1\sigma$ & $>2\sigma$  && $<1\sigma$
  & $>2\sigma$\\ 
\midrule
$\tau$ & 0.72 & 0.20 && 0.65 & 0.26  && 0.73 & 0.20\\
age & 0.70 & 0.28 && 0.71 & 0.26  && 0.76 & 0.22\\
$E(B-V)$ & 0.82 & 0.11 && 0.78 & 0.13  && 0.78 & 0.14\\
\bottomrule 
\end{tabular} 
\caption{Effect of pasband limitations. Shown for each
  pixel-$z$ inferred parameter is the fraction of pixels in the
  full sample of 467 galaxies which have systematic differences of $<1\sigma$ and $>2\sigma$ as a result of different permutations
  of available broadband filters.} 
\label{tab:table1}
\end{table}


\newcommand{\nirlowz}[1]{\renewcommand{\arraystretch}{#1}}
\begin{table}\centering 
\begin{tabular}{@{}rrrrcrrrcrrr@{}}
\toprule &\multicolumn{3}{c}{[BC03,CB07]}& \multicolumn{3}{c}{[BC03,M05]}& \multicolumn{3}{c}{[CB07,M05]} \\
\midrule
  & \textit{bviz} & \textit{bvizJH} && \textit{bviz} & \textit{bvizJH}  && \textit{bviz}
  & \textit{bvizJH}\\ 
\midrule
$\tau$ & 0.98 & 0.88 && 0.84 & 0.81  && 0.83 & 0.81\\
age & 0.98 & 0.78 && 0.89 & 0.80  && 0.89 & 0.80\\
$E(B-V)$ & 0.99 & 0.91 && 0.96 & 0.94  && 0.97 & 0.92\\
\bottomrule 
\end{tabular} 
\caption{Comparing the level of agreement between pairs of stellar population synthesis models when using
  optical colors only with those arising from using optical and NIR
  colors, for a sample of galaxies at $z \le 1$ with \textit{bviz} and
  \textit{JH} imaging. Listed are the fraction of all pixels in this low
  redshift galaxy sample which show systematic differences smaller
  than $1\sigma$ in their pixel-$z$ parameters.
} 
\label{tab:table4}
\end{table}

\newcommand{\nirhighz}[1]{\renewcommand{\arraystretch}{#1}}
\begin{table}\centering 
\begin{tabular}{@{}rrrrcrrrcrrr@{}}
\toprule &\multicolumn{3}{c}{[BC03,CB07]}& \multicolumn{3}{c}{[BC03,M05]}& \multicolumn{3}{c}{[CB07,M05]} \\
\midrule
  & \textit{bviz} & \textit{bvizJH} && \textit{bviz} & \textit{bvizJH}  && \textit{bviz}
  & \textit{bvizJH}\\ 
\midrule
$\tau$ & 0.95 & 0.96 && 0.78 & 0.80  && 0.80 & 0.82\\
age & 0.99 & 0.97 && 0.95 & 0.93  && 0.96 & 0.92\\
$E(B-V)$ & 0.99 & 0.99 && 0.95 & 0.90  && 0.95 & 0.91\\
\bottomrule 
\end{tabular} 
\caption{Comparing the level of agreement between pairs of stellar population synthesis models when using
  optical colors only with those arising from using optical and NIR
  colors, for a subsample of galaxies at $z>1$ with \textit{bviz} and
  \textit{JH} imaging. Listed are the fraction of all pixels in the low
  redshift galaxy sample which show systematic differences smaller
  than $1\sigma$ in their pixel-$z$ parameters.} 
\label{tab:table1}
\end{table}

\newpage


\clearpage


\clearpage

\end{document}